\documentclass[preprint,3p]{elsarticle}      

\usepackage{amsmath}
\usepackage{natbib}
\usepackage{amssymb}
\usepackage{graphicx}
\usepackage{dcolumn}
\usepackage{bm}
\usepackage{mathrsfs}
\usepackage{color}
\usepackage{psfrag}
\usepackage{url}
\usepackage{soul}
\usepackage{float}  

\usepackage{cancel}  

\usepackage[normalem]{ulem}
\usepackage{color}

\definecolor{carrot}{rgb}{1.0, 0.44, 0.26}
\newcommand{\tim}[1]{{\color{black}#1}}  

\newcommand{\rev}[1]{{\color{black}#1}}  
\newcommand{\reva}[1]{{\color{black}#1}}  

\definecolor{revColor}{rgb}{0.12,0.45,0.0}
\newcommand{\revCom}[1]{\color{black}#1}    
\definecolor{revdColor}{rgb}{0.8,0.15,0.25}

\usepackage{booktabs}
  
\newcommand{\et}{et al.}
\newcommand{\eg}{e.g.}
\newcommand{\ie}{i.e.}
\newcommand{\fig}{Figure}
\newcommand{\figs}{Figures}
\newcommand{\eq}{Eq.}
\newcommand{\eqs}{Eqs.}
\newcommand{\sect}{Section}
\newcommand{\sects}{Sections}
\newcommand{\app}{}    
\newcommand{\tab}{Table}
\newcommand{\tabs}{Tables}
\newcommand{\of}{OpenFOAM}
\newcommand{\dakota}{Dakota}
\newcommand{\vk}{von K\'arm\'an}
\newcommand{\opt}{{\rm opt}}

\newcommand{\var}{\text{var}}
\newcommand{\BE}{\mathbb{E}}
\newcommand{\BR}{\mathbb{R}}
\newcommand{\BQ}{\mathbb{Q}}
\newcommand{\cR}{\mathcal{R}}
\newcommand{\cU}{\mathcal{U}}
\newcommand{\cN}{\mathcal{N}}
\newcommand{\pp}{\partial}
\newcommand{\dd}{\mbox{d}}
\newcommand{\dash}{\mbox{-}}
\newcommand{\bx}{\mathbf{x}}
\newcommand{\fq}{\mathbf{q}}
\newcommand{\fQ}{\mathbf{Q}}

\newcommand{\nd}{n/\delta}
\newcommand{\nx}{n_x/\delta}
\newcommand{\ny}{n_y/\delta}
\newcommand{\nz}{n_z/\delta}
\newcommand{\btw}{\bar{\tau}_w}

\newcommand{\yd}{y/\delta}
\newcommand{\hd}{h/\delta}
\newcommand{\rey}{\mbox{Re}}
\newcommand{\reyt}{\mbox{Re}_\tau}
\newcommand{\dx}{\Delta x}
\newcommand{\dy}{\Delta y}
\newcommand{\dz}{\Delta z}

\newcommand{\dxp}{\Delta x^+}
\newcommand{\dyp}{\Delta y^+}
\newcommand{\dzp}{\Delta z^+}
\newcommand{\Qk}{\mathbb{Q}_{\kappa}}
\newcommand{\QB}{\mathbb{Q}_{B}}
\newcommand{\ut}{u_\tau}
\newcommand{\lut}{\langle u_\tau \rangle}
\newcommand{\but}{\bar{u}_\tau}
\newcommand{\lbut}{\langle \bar{u}_\tau \rangle}
\newcommand{\bp}{\bar{p}}

\newcommand{\bu}{\bar{u}}
\newcommand{\U}{\langle \bar{u} \rangle}
\newcommand{\lu}{\langle u \rangle}

\newcommand{\uv}{\langle u'v'\rangle}
\newcommand{\buv}{\langle \bar{u}'\bar{v}'\rangle}
\newcommand{\bv}{\bar{v}}
\newcommand{\bk}{\bar{\mathcal{K}}}

\newcommand{\sgs}{\rm sgs}
\newcommand{\bD}{\bar{\Delta}}
\newcommand{\rms}{\rm rms}

\newcommand{\einf}{\epsilon_{\infty}}
\newcommand{\rk}{{\rm k}}

\newcommand{\frst}{1^{\rm st}}
\newcommand{\scnd}{2^{\rm nd}}
\newcommand{\thrd}{3^{\rm rd}}
\newcommand{\frth}{4^{\rm th}}

\newcommand{\pdiff}[2]{\frac{\partial #1}{\partial #2}}

\begin{document}

\title{Systematic Study of Accuracy of Wall-Modeled Large Eddy Simulation using Uncertainty Quantification Techniques}

\author[focal1]{S.~Rezaeiravesh\corref{cor1}}
\ead{saleh.rezaeiravesh@it.uu.se}
\author[focal1]{T.~Mukha}
\ead{timofey.mukha@it.uu.se}
\author[focal1,focal2]{M.~Liefvendahl\corref{cor2}}
\ead{mattias.liefvendahl@foi.se}
\cortext[cor1]{Principal Corresponding Author}
\cortext[cor2]{Corresponding Author}
\address[focal1]{Uppsala University, Department of Information Technology, Box 337, SE-751 05 Uppsala, Sweden}
\address[focal2]{FOI, Totalf\"{o}rsvarets forskningsinstitut, 164 90 Stockholm, Sweden}



\begin{abstract}
The predictive accuracy of wall-modeled large eddy simulation is studied by systematic simulation
campaigns of turbulent channel flow.
The effect of wall model, grid resolution and anisotropy, numerical convective scheme and subgrid-scale modeling is investigated.
All of these factors affect the resulting accuracy, and their action is to a large extent intertwined.
The wall model is of the wall-stress type, and its sensitivity to location of velocity sampling, as
well as law of the wall's parameters is assessed.
For efficient exploration of the model parameter space (anisotropic grid resolution and wall model parameter values),
generalized polynomial chaos expansions are used to construct metamodels for the responses which are taken to be measures of the predictive error in quantities of interest (QoIs).
The QoIs include the mean wall shear stress and profiles of the mean velocity, the turbulent kinetic energy, and the Reynolds shear stress.
DNS data is used as reference.
Within the tested framework, a particular second-order accurate CFD code (\of), the results provide ample support for grid and method parameters recommendations which are proposed in the present paper,
and which provide good results for the QoIs.
Notably, good results are obtained with a grid with isotropic (cubic) hexahedral cells, with $15\, 000$ cells per $\delta^3$, where $\delta$ is the channel half-height (or thickness of the turbulent boundary layer).
The importance of providing enough numerical dissipation to obtain accurate QoIs is demonstrated.
The main channel flow case investigated is  $\reyt=5200$, but extension to a wide range of $\rey$-numbers is considered.
Use of other numerical methods and software would likely modify these recommendations, at least slightly, but the proposed framework is fully applicable to investigate this as well.\footnote{\textcopyright 2019. This manuscript version is made available under the CC-BY-NC-ND 4.0 license.\\ https://doi.org/10.1016/j.compfluid.2019.03.025}
\end{abstract}

\maketitle


\section{Introduction}\label{sec:intro}
Wall-bounded turbulent flows occur widely in many engineering \tim{applications as well as in natural phenomena}.
\rev{It is, however, well-known that accurate simulation of such flows meets several severe challenges.}
The main obstacle is the prohibitive computational cost of \tim{applying} scale-resolving \tim{turbulence modeling approaches, such as large eddy simulation (LES), to} turbulent boundary layers (TBLs)\tim{.} \tim{The latter is due to} \tim{the} wide ranges of temporal and spatial scales \tim{that a TBL contains}. 
Adjacent to the wall, in the inner layer of the TBL, \rev{the most} energetic structures are strongly anisotropic, with sizes of the order of the local viscous length scale $\delta_\nu=\nu/u_\tau$, see~\cite{jimenez:13,pope}.
Here, $\nu$ denotes the kinematic viscosity and $u_\tau$ is the wall friction velocity defined as~$u_\tau=\sqrt{\tau_w/\rho}$, with $\tau_w$ and $\rho$ being magnitude of the wall shear stress and fluid density, respectively. 
In contrast, the size of the energetic structures in the outer layer of the TBL \tim{scales with the TBL thickness }\revCom{$\delta$, where}~$\delta \gg \delta_\nu$.
The ratio of $\delta_\nu/\delta$ decreases with \tim{the} Reynolds ($\rey$-)number. 
This results in the number of grid cells required for \tim{LES approximately scaling as $\rey^{1.85}$} \rev{for a TBL affected by a weak pressure gradient,} see \cite{chapman:79,choi:12,saleh:17}.

A possible strategy to circumvent this restrictive computational cost is to only aim for resolving the structures in the outer layer of TBL, and approximate the uncaptured effects in the near-wall region by some type of modeling.
The resulting approach is referred to as wall-modeled (WM)LES, for which the required number of grid cells increases linearly with \tim{the} $\rey$-number \tim{\cite{chapman:79,choi:12}}, and hence \rev{leads to a large reduction in the computational cost as compared to} \tim{classical LES, which is from hereon referred to as wall-resolving (WR)LES}.

Different methodologies have been proposed for wall modeling, see the reviews~\cite{bose:18,larsson:16,moin:16,piomelli:08} and the references therein.
A particular class referred to as wall-stress modeling is considered in the present study. 
The near-wall treatment in this type of modeling \tim{consists of} accurately predicting the local value of \rev{the filtered} wall shear stress, given information from the outer layer of the TBL.
\tim{The large size of the WMLES grid with respect to $\delta_\nu$ justifies using} a RANS (Reynolds-averaged Navier-Stokes)-based model to \tim{make the} wall shear stress \tim{predictions}.

\rev{It is possible to, in each time step, solve the} full RANS equations on an auxiliary grid \tim{extended from the wall up to some height $h$ in the outer layer}, see \eg~\cite{kawai:13}. 
\tim{At this height, the LES velocity field is imposed as a boundary condition.}
As slightly simpler wall models, turbulent boundary layer equations (TBLE) have been widely used by several authors for WMLES of \tim{both} attached and separated flows, see \eg~\cite{cabot:96,balaras:96,cabotMoin:00,wangMoin:02,piomelli:02}.
In their most complete form, the TBLE are partial differential equations (PDEs) containing the convection \tim{and wall-normal diffusion} of the averaged \tim{wall-parallel} velocities, and the wall-parallel components of averaged pressure gradient.

\tim{By neglecting the variation of the different terms in the TBLE in the wall-parallel directions, a wall model based on ordinary differential equations (ODEs) is obtained.}
\tim{The neglected terms can be modeled, however,} the \tim{simplest} ODE\tim{-based wall model}\tim{, accounting only for} the viscous and turbulent velocity diffusion terms \tim{has also been applied}, see \eg~\cite{wangMoin:02,kawai:12,bermejo:14}.
Exactly equivalent to such \tim{an} ODE\tim{-based model}, are the \tim{models based on} laws of the wall, see \eg~\cite{pope}.
Such laws algebraically express the dependency between the inner-scaled mean velocity and the distance from the wall for equilibrium TBLs.
This type of wall modeling dates back to the pioneering works of Schumann \cite{schumann:75} and Gr{\"o}tzbach \cite{grotzbach:87}, but has been extensively used in more recent works as well,~\eg~\cite{lee:13}. 
In the present study, a wall model of this type is used. 
It should be noted that the PDE- and ODE-based models do not necessarily lead to more accurate results compared to the algebraic models, see~\cite{larsson:16,wangMoin:02,lee:13}.

Using different classes of wall models, there have been successful attempts to overcome some known issues in WMLES practices. 
A long-standing problem, referred to as log(arithmic)-layer mismatch (LLM), appears due to miscalculating the wall shear stress and manifests itself in a vertical shift in the predicted inner-scaled mean velocity profile plotted against the inner-scaled off-wall distance.
To remove the issue, several different remedies have been proposed, some of which are shortly reviewed here. 
Among different factors which can cause the LLM, the role of the numerical errors and the subgrid-scale (SGS) model are prominent, as pointed out, for instance, in Ref. \cite{larsson:16}.
In particular, due to the coarseness of the WMLES grid in the near-wall region \rev{relative to the energetic flow scales}, the employed discretization and SGS modeling practices are unable to represent accurate velocity values in the first few off-wall cells, see \eg~\cite{cabotMoin:00,nicoud:01,kawai:12}.
Therefore, as first proposed by Kawai and Larsson \cite{kawai:12}, the sampling height $h$ should be chosen \rev{to allow for a certain spatial (grid) resolution bellow it.}
As a consequence, the prediction of the wall shear stress by the wall model is \rev{often} improved, as has been confirmed also in succeeding studies, for instance, see \cite{lee:13}. 
As another remedy to the LLM, the parameters in the turbulent eddy viscosity appearing in the ODE- and PDE-based wall models were suggested to be dynamically adjusted, see \cite{cabotMoin:00,wangMoin:02,kawai:13,park:14}.
Similarly, different approaches for adjusting the SGS viscosity have been proposed, see \cite{wu:13,templeton:05} and the references therein. 
A less complicated yet effective modification was proposed by Yang \et~\cite{yang:15}.
According to them, the use of temporally-filtered velocity samples in the wall model (instead of instantaneous ones), results in more accurate predictions of wall shear stress even for sampling from the wall adjacent cell. 
More recently, the improving effect of averaging the sampled velocity over the wall-parallel planes has been demonstrated by Yang \et~\cite{yang:17}.
They, however, emphasize that due to probable issues with unstructured grids, temporal filtering is recommended. 
On another track, Nicoud \et~\cite{nicoud:01} used techniques from optimal control theory to optimize the predicted wall shear stress so that the LES velocity in the outer layer is forced to converge to an a-priori known target reference. 
To make this strategy predictive, Templeton~\et~\cite{templeton:08} later suggested to couple the LES with a near-wall RANS model that provides the target mean velocity profile.

Wall modeling is not the only influential factor in determining the quality of WMLES. 
This is better understood noting that wall modeling depends on the information imported from the LES solution.
The errors induced by the numerics and modeling are \rev{coupled} together and also the grid resolution naturally affects the simulations.
\rev{The interaction of the errors pertaining to the wall model, numerics, and SGS modeling is quite complex. 
The main objective of the present study is to investigate in detail how these factors affect the predictive accuracy of WMLES.}

To this end, two approaches\revCom{, a-priori analyses and systematic simulations,} are considered.
In a-priori analyses, the connection between the LES solver and the wall model is removed allowing to scrutinize how the wall model predictions are affected by different factors. 
To address the sensitivity of WMLES of a canonical wall-bounded turbulent flow to several factors, a large simulation campaign of several hundred WMLES has been carried out.
Specifically, the influence of the following \tim{is} considered: the numerical scheme used to discretize the non-linear convective term in the Navier-Stokes equations, the SGS modeling, the grid resolution and anisotropy, the distance from the wall of the cell whose velocity is imported into the wall model, and, finally, the parameters of the law of the wall.
To the authors' knowledge, the present study is one of the most complete and detailed attempts, see~also~\cite{temmerman:03,pantano:08}, to investigate the influence of these factors on the predictive accuracy of WMLES.

\rev{Specifically}, an algebraic wall model based on the Spalding law of the wall \cite{spalding} is considered for WMLES of fully developed turbulent channel flow at target friction Reynolds number, $\reyt=\delta/\delta_\nu$, equal to $5200$, where $\delta$ is the channel half-height. 
The quantities of interest (QoIs) include the mean wall shear stress, and the cross-channel profiles of the mean velocity, the resolved Reynolds shear stress, and \tim{the resolved} turbulent kinetic energy (TKE).
The corresponding error in these quantities is defined as the deviation from the reference direct numerical simulation (DNS) data of Lee and Moser \cite{lee-moser:15}. 
The simulations are carried out with the finite-volume (FV)-based open-source software~\of\footnote{\texttt{www.openfoam.com}}, \cite{weller:98,meric:14}. 
The discretization of the governing equations in both space and time is second-order accurate.
For the purpose of wall modeling, the recently developed library by Mukha \et~\cite{mukha:18} based on \of~technology is employed, which provides great flexibility \tim{in the choice of} the controlling parameters of WMLES.

A distinctive aspect of the present study is making use of the relevant tools developed in the framework of uncertainty quantification (UQ), see \eg~\cite{smith,uqHandbook}, in line with the main goals formulated above.
In particular, two different types of sensitivity analysis techniques are applied in order to a-priori study the sensitivity \tim{of} the wall model to the input velocity and the model parameters.
\tim{Also}, metamodels based on generalized polynomial chaos expansion (gPCE) \cite{ghanem:91,xiu_gPCE} and collocation method \cite{xiu:07} are used to reduce the number of simulations required to construct the response surfaces of \tim{the QoIs with respect to the variation of various simulation parameters.}

An outcome of the present study is a set of guidelines for accurate WMLES of turbulent channel flow, the validity of which is demonstrated at Reynolds numbers other than $5200$. 
Despite being developed for channel flow, these best practice guidelines are expected to be \rev{useful as a good starting point for investigating more complex flows, e.g.~involving curved surfaces and large pressure gradients.}
Hence, the guidelines along with other materials of the present study are believed to be of interest for the 
CFD (computational fluid dynamics) community, in general, and for the community of \of~users, in particular. 
\rev{The computational methodology used here has also been employed for WMLES of other flows, including flow around a streamlined body \cite{timECCOMAS:18,matR2} and flow around a model ship hull \cite{matR3}.}

The paper is organized as follows. 
In \sect~\ref{sec:computMethod}, the computational methodology is explained. 
This includes details \tim{regarding} the CFD methods \tim{used} for LES\tim{,} a summary of the wall modeling approach, and a review of the UQ techniques employed. 
The details of the channel flow simulations including the case setup and the definition of the error measures utilized to evaluate the accuracy of the QoIs are presented in \sect~\ref{sec:simDetails}.
Two important preliminary discussions are made in \sect~\ref{sec:prelimDisc}. 
First, by using the sensitivity analysis techniques provided in the UQ framework and making use of available DNS datasets for velocity, the influential factors on the accuracy of algebraic wall-stress models are a-priori investigated. 
The next discussion \tim{concerns} the connection between different QoIs of turbulent channel flow \tim{that} helps analyze the results of the WMLES presented in \sect~\ref{sec:results}.
The latter comprises of several subsections each being focused on one specific group of factors influencing the QoIs. 
In \sect~\ref{sec:numSGSwm}, it is \tim{first} shown what the footprints of the numerical scheme and SGS model on the QoIs \tim{are}, \tim{when} a grid resolution typical of WMLES is used. 
Then, the gains by including wall modeling are explored. 
The impact of grid resolution and the distance to the sampling point, $h$, \tim{are} investigated in \sects~\ref{sec:nTests} and \ref{sec:hTests}, respectively.
The influence of the wall model parameters on the QoIs is thoroughly explored in \sect~\ref{sec:wmParams}. 
\tim{How the anisotropy of the grid affects the QoIs} is then examined in detail in \sect~\ref{sec:gridAnisot}.
Based on the observations made through the simulation campaign, best practice guidelines for WMLES of turbulent channel flow \tim{are} provided in \sect~\ref{sec:guidelines}, and applied to WMLES of channel flow at different $\rey$-numbers in \sect~\ref{sec:ReEffect}.
\tim{Concluding remarks are given in \sect~\ref{sec:conclusions}.}

\section{Computational Methodology}\label{sec:computMethod}

\subsection{CFD methods}\label{sec:cfd}
The governing equations for LES of an incompressible Newtonian flow are derived by applying spatial filtering to the Navier-Stokes equations, see~\cite{sagaut}. 
Assuming that the filtering and differentiation operations commute with each other, the obtained filtered mass and momentum conservation laws are as follows, 
\begin{eqnarray}
\frac{\partial \bu_j}{\partial x_j} &=& 0 \,, \nonumber \\
\frac{\partial \bu_i}{\partial t} + \frac{\partial \bu_i \bu_j}{\partial x_j} &=&
- \frac{1}{\rho}\frac{\partial \bp}{\partial x_i} + \pdiff{}{x_j} \left( 2\nu \bar{S}_{ij} \right)- \frac{\partial B_{ij}}{\partial x_j}\,.\quad (i=1,2,3) \label{eq:LESmomentum}
\end{eqnarray}
Here, summation over repeated indices is implied, filtered quantities are specified by an overbar,~$u_i$ is the velocity component in the $i$-th direction, $p$ specifies pressure, $\bar{S}_{ij} = \left( {\partial \bu_i}/{\partial x_j} + {\partial \bu_j}/{\partial x_i} \right)/2$ is the filtered rate of strain tensor, and $B_{ij}=\overline{u_i u_j}-\bu_i \bu_j$ is the SGS stress tensor. 
In order to close (\ref{eq:LESmomentum}), the SGS stress tensor $B_{ij}$ must be modeled. 
To this end, a wide range of approaches employ the Boussinesq approximation, which is the hypothesis that the anisotropic part of $B_{ij}$ is structurally similar to the viscous stress tensor, 
\begin{equation}\label{eq:sgsBoussinesq}
B_{ij} = -2\nu_{\sgs} \bar{S}_{ij} +  \frac{1}{3}B_{kk} \delta_{ij} \,,
\end{equation}
where $\nu_{\sgs}$ is the so-called eddy viscosity, and $\delta_{ij}$ denotes the Kronecker delta. 
The isotropic part,~$B_{kk}$, can be lumped into the pressure term in~\eqref{eq:LESmomentum}.
For the eddy viscosity, dimensional analysis can be used to express~it~as, 
\begin{equation}\label{eq:nuSGS}
\nu_{\sgs} = C \bD u_{\sgs},\,
\end{equation}
with the filter width $\bD$ defining a characteristic length scale, $C$ denoting a model constant, and~$u_{\sgs}$ representing a characteristic velocity scale of the subgrid motions.
The choice of $u_{\sgs}$ thus defines the choice of the eddy viscosity model.

In the present study, three wide-spread models are considered.
The first is the dynamic~$k_{\sgs}$-eqn (equation) model, where~\mbox{$k_{\sgs}=B_{kk}/2$} is the SGS kinetic energy, for which a transport PDE is solved, coupled with a dynamic procedure to determine the associated model coefficients, see Ref.~\cite{kim:95}.
Given $k_{\sgs}$, the velocity scale $u_{\sgs}$ is defined as $k_{\sgs}^{1/2}$. 
The second considered closure is the Smagorinsky model for which $u_{\sgs}$ is defined as the magnitude of the rate of strain tensor multiplied by the filter width $\bD$, 
\begin{equation}\label{eq:nuSmag}
\nu_{\sgs} = C_s^2 \bD^2 (2 \bar{S}_{ij} \bar{S}_{ij})^{1/2} \,,
\end{equation} 
where, the model coefficient $C_s$ is \rev{taken to be} $0.17$. 
Finally, the Wall-Adapting Local Eddy (WALE) model is also considered, see~\cite{nicoud:99} \rev{for the expression for $\nu_{\sgs}$, in which the value of the associated coefficient $C_w$ is here taken to be $0.325$.}
In this model, both the rate of strain and the vorticity tensors are used to determine~$u_{\sgs}$, leading to a more accurate behavior of $\nu_{\sgs}$ close to the wall.

To solve the governing equations, version \texttt{3.0.1} of the open-source CFD software package \of~\cite{weller:98,meric:14}, is employed.
For discretization, the solver uses the collocated finite-volume method (FVM).
In FVM, the domain is decomposed into cells, or control volumes.
In the collocated formulation, the unknown solution variable $\varphi$ (pressure or velocity component) is stored in the centroids of the cells and its value can be shown to be a second-order accurate approximation of the average of the respective quantity over the volume of the cell.
Since it is impossible to resolve turbulent motion of scales smaller than the local cell size, the FVM defines an implicit filtering procedure.
The corresponding filter width can be defined as $V_c^{1/3}$, where $V_c$ is the local cell volume.

In FVM, see \eg~\cite{ferziger,wmReport:17}, by application of Gauss's theorem to the momentum equation (\ref{eq:LESmomentum}) in its integral form, a set of volume integrals are converted to surface integrals. 
The latter are then decomposed into the sum of the contributions coming from each face bounding the cell.
In turn, the contribution of each face is approximated as $S_f \varphi_f$, where $S_f$ is the area of the face and $\varphi_f$ is the value of the unknown variable at the face's centroid.
Consequently, interpolation between cell center values is required to obtain $\varphi_f$. 
For the convective term in (\ref{eq:LESmomentum}), there are different options available for this purpose among which two schemes are employed in the present study.

If $\varphi_R$ and $\varphi_L$ represent the values of $\varphi$ at two cell centers $R$ and $L$ ($L$ is upwind to $R$), then $\varphi_f$ at the shared face can be constructed via linear interpolation,
\begin{eqnarray*}
\varphi^{\rm Lin}_f = \alpha \varphi_L + (1-\alpha) \varphi_R \,,
\end{eqnarray*}
where, $\alpha := \|\bx_R-\bx_f\|_2/\|\bx_R - \bx_L \|_2$ with vector $\bx$ denoting the spatial coordinates of the cell and face centers\revCom{, and $\|\cdot\|_2$ representing $L_2$ norm}.
This approximation that is referred to as the Linear scheme in the present study is second-order accurate, but \rev{can be} unbounded, \rev{see~\cite{versteeg}}. 

Alternatively, the LUST (Linear Upwind Stabilized Transport) scheme \cite{weller:12} can be used which is a blending of the linear and second-order upwind interpolations:
\begin{eqnarray*}
\varphi^{\rm LUST}_f =  \beta \varphi^{\rm Lin}_f + (1-\beta) \varphi^{\rm Up}_f \,,
\end{eqnarray*}
where \rev{$\beta=0.75$,} and for the upwind scheme, 
$$
\varphi^{\rm Up}_f  = \varphi_L + (\bx_f-\bx_L)\cdot \nabla \varphi |_L\,.
$$
The \rev{$i$-th component of} the gradient $\nabla \varphi |_L$ is computed at the center of the upwind cell, here at $L$, using Gauss's theorem, 
\rev{
\begin{eqnarray*}
\frac{\pp \varphi}{\pp x_i}\big|_L = \nabla \varphi |_L \cdot \hat{\mathbf{i}} \approx \frac{1}{V_L}\sum_{S_L} S_f \varphi_f^{\rm Lin} \,\hat{n}_f\cdot \hat{\mathbf{i}} \,,
\end{eqnarray*}
where, $\hat{n}_f$ denotes the unit vector normal to $S_f$ pointing outward of $V_L$, and $\hat{\mathbf{i}}$ is a unit vector in the $i$-th direction.}
Since the upwind component of the LUST scheme introduces numerical dissipation~\cite{ferziger}, it is expected to mitigate the oscillations introduced into the solution by the unbounded linear scheme.
Note that LUST is second-order accurate as a linear combination of two second-order accurate schemes.

The discretization in time is performed using a second-order accurate implicit backward-differencing scheme, see~\cite{jasak:96}.
For the discretization of the viscous term, the Linear scheme is used.
The pressure gradient is evaluated using Gauss's theorem as shown above for $\nabla \varphi |_L $. 
To handle the coupling between the pressure and velocity in the momentum equation (\ref{eq:LESmomentum}), the PISO (Pressure-Implicit with Split Operator) method~\cite{issa:86} is employed.

\subsection{Wall modeling}\label{sec:wallModels}
The role of the wall model is to accurately predict the filtered wall shear stress $\btw$, which otherwise is not possible due to the coarseness of computational grid \rev{in the near-wall region}. 
\reva{In the present study, by the near-wall region a part of the TBL's inner layer is meant that includes the whole viscous sublayer and a part of the overlap region.}
As input, the wall model takes the solution to the LES equations~\eqref{eq:LESmomentum} at the sampling point positioned in the outer layer of the TBL, where the grid resolution is fine enough to capture the correct dynamics of the flow.
The predicted wall shear stress is then imposed via a boundary condition at the wall.

As already pointed out in \sect~\ref{sec:intro}, a diverse set of approaches for wall-stress modeling has been proposed, see reviews~\cite{larsson:16,bose:18}. 
In the current study, an algebraic wall model based on a law of the wall is used.
The particular law of the wall employed here is that of Spalding~\cite{spalding}, 
\begin{equation}\label{eq:spalding}
y^+=\lu^+ + e^{-\kappa B}\left[e^{\kappa \lu^+} - \sum_{m=0}^3 \frac{(\kappa \lu^+)^m}{m!}\right].
\end{equation}
Here, $y^+=y\lut/\nu$ and $\langle u \rangle^+ = \lu/\lut$ respectively denote the inner-scaled distance from the wall and the mean velocity in the RANS sense. 
Further, $\kappa$ and $B$ are model parameters with original values $0.4$ and $5.5$ \cite{spalding}, respectively. 
However, in the present study unless specified otherwise, $\kappa=0.395$ and $B=4.8$ are adopted, see the thorough discussion in \sect~\ref{sec:wmParams}. 
A favorable property of the Spalding law is that it is valid from the wall up to the wake region in the outer layer of the TBL.
However, other such laws have also been proposed, for example, the Reichardt law~\cite{reichardt}.
Generally, the conclusions drawn here regarding wall modeling should be applicable to any algebraic or ODE-based model based on the assumption that the mean velocity adheres to the canonical profile of an \rev{equilibrium} TBL.

When used as a wall model, \eq~\eqref{eq:spalding} is assumed to be valid \rev{for the filtered velocity} locally in time, \rev{along a wall-normal line in the inner part of the TBL.}
Thus, the averaged quantities can be exchanged to the corresponding instantaneous filtered ones.
Given the instantaneous wall-parallel LES velocity at a distance~$h$ from the wall, the law is used to predict the instantaneous filtered wall shear stress $\btw=\rho\but^2$, which is done \rev{by solving (\ref{eq:spalding})} using Newton's method.

To enforce the computed value of $\btw$ at the wall, the following approach is used.
The no-slip condition for  the filtered velocity is retained, although due to the coarseness of the employed mesh it is to be considered as part of the wall modeling methodology and not a physical condition, see \cite{sagaut,bose:14}.
To \rev{enforce the predicted} $\btw$, the subgrid viscosity is modified at the wall.
Since in the discretization procedure the wall stress is evaluated using a simple first-order accurate finite difference stencil, the appropriate value of $\nu_{\sgs}$ is obtained as,
\begin{equation}\label{eq:wmBC}
\nu_{\sgs} = \frac{\btw \Delta y_w}{\rho} \left[ \bu_{x,c}^2 +\bu_{z,c}^2 \right]^{-1/2} - \nu,
\end{equation}
where $\Delta y_w$ is the distance from the wall to the center of the wall-adjacent cell,  $\bu_{x,c}$ and $\bu_{z,c}$ are the wall-parallel components of the instantaneous filtered velocity in the center of the wall-adjacent cell.
A more detailed discussion on how the described wall modeling fits into the framework of finite volume discretization can be found in~\cite{wmReport:17}.
For the details of the open-source implementation of this and also several other wall models within the framework of \of, see~\cite{mukha:18}.

\subsection{Uncertainty quantification techniques}\label{sec:uq}
As pointed out in \sect~\ref{sec:intro}, there are different factors affecting the quality of WMLES.
\rev{Because of the wall modeling, there are more such factors to consider than for WRLES.} 
Consequently, evaluating the sensitivity of the WMLES results to the variations in the combination of influential modeling factors requires a large number of simulations.
\tim{However, this number can be reduced by}  making use of the relevant uncertainty propagation and sensitivity analysis techniques developed in the UQ framework, see~\cite{smith}.
In what follows, the specific techniques used in the present study are reviewed\revCom{, after a few~definitions}.

\tim{The responses, $\cR$, are generally defined as a set of quantities that can be computed based on the results of a simulation.
Hence, every time a simulation is performed a realization of $\cR$ is obtained.
In particular, in line with the goal of the current study, here $\cR$ will be defined as the set of obtained errors in the QoIs, see \sect~\ref{sec:simDetails} below.
In a simulation campaign, those parameters\rev{, denoted by~$\fq$,} whose effect on $\cR$ is aimed to be quantified are assumed to be uncertain, meaning that they are allowed to adopt different values over some defined range referred to as their admissible space $\BQ$. 
The rest of the influential factors in the particular simulation set are taken to be fixed parameters and are denoted by $\chi$.
In UQ terms, the goal of a study is thus to construct an approximate surface of $\cR$ over the admissible space of $\fq$.
To that end, different techniques exist, see \eg~\cite{smith,lematre:10}.
Here, non-intrusive~gPCE \cite{xiu_gPCE} with collocation method \cite{xiu:07} is employed. 
This approach has been successfully used in WRLES, for instance refer to~\cite{lucor:07,meldi:12,salehPoF:18}.
Hence, for sake of brevity, only a general overview is provided here. 
}

Consider $p\geq 1$ mutually-independent uncertain parameters $\fq$ varying over the~admissible~space~$\BQ\subset \BR^{p}$.
Note that by tensor-product, $\BQ=\bigotimes_{i=1}^p \BQ_i$\revCom{, where $\BQ_i$ is the admissible range of the $i$-th parameter.}
Let $\BQ_i$ be mapped one-to-one to $\Gamma_i=[-1,1]$.
Accordingly, a one-to-one correspondence between any sample $\fq\in \BQ$ and $\fQ\in \Gamma =\bigotimes_{i=1}^p \Gamma_i$ is established.
Ideally, for a set of $\chi$, a parameterized functional form between~$\cR$ and the parameters~$\fq$ can be constructed as~$\cR= f(\chi,\fq)$ based on \rev{an infinite number} of realizations. 
However, to avoid the excessive computational cost, a surrogate or metamodel~$\tilde{f}(\chi,\fq)$ can be constructed instead.
Adopting~gPCE for this purpose results in, 
\begin{equation}\label{eq:pce}
\cR\approx\tilde{f}(\chi,\fQ)=\sum_{\rk=0}^N \hat{f}_{\rk}(\chi)\Psi_{\rk}(\fQ) \, ,
\end{equation}
where, $\Psi_{\rk}(\fQ)$ represent the orthogonal bases that are known by construction, \revCom{see below}.  
\revCom{The orthogonality is with respect to $\rho (\fQ)=\prod_{i=1}^p \rho_{i} (Q_i)$ that is the joint probability density of the mapped parameters.} 
The deterministic coefficients~$\hat{f}_{\rk}(\chi)$ are determined from
\begin{equation}\label{eq:fHat}
\hat{f}_\rk(\chi)=\frac{1}{\gamma_\rk}\BE[\tilde{f}(\chi,\fQ)\Psi_\rk(\fQ)] \,,
\end{equation}
where, $\gamma_\rk = \BE[\Psi_\rk(\fQ) \Psi_\rk(\fQ)]$, and \rev{the expected value of any arbitrary $g(\fQ)$ is} \revCom{defined as,}
$$
\BE[g(\fQ) \Psi_\rk(\fQ)]=\int_\Gamma g(\fQ) \Psi_\rk(\fQ) \rho (\fQ) \dd \fQ \,.
$$

\rev{Each parameter space $\BQ_i$ is sampled by $n_i$ collocation points, see~\cite{xiu:07}.}
Using the tensor product method,~$N$ in \eq~(\ref{eq:pce}) becomes equal to $\prod_{i=1}^p n_i -1$. 
For this construction, $\Psi_{\rk}(\fQ)=\prod_{i=1}^p \psi_{k_i}(Q_i)$, where~$\rk$ is a unique re-index associated with any combination of $\{ k_i\}_{i=1}^p$.
The coefficients $\hat{f}_\rk(\chi)$ in~\eq~(\ref{eq:pce}) are computed from \revCom{numerical integration of (\ref{eq:fHat}), see \cite{xiu:07}}.
All the parameters studied in \revCom{\sects~\ref{sec:wmParams} and \ref{sec:gridAnisot} and also~\ref{app:LSA}} are considered to have either uniform or Gaussian distributions.
Choosing respectively Legendre and Hermite polynomial bases for these two types of distributions \rev{is required}, see~\cite{xiu_gPCE}.
\rev{Consequently, the collocation samples become the roots of these polynomials and hence are the Gauss quadrature points.}
\revCom{
The convergence of the gPCE metamodels is briefly discussed in~\ref{app:gPCEConvergence}.
}

As a complement to the propagation of the uncertainties in the model response, global sensitivity analysis (GSA) can be performed. 
In GSA, the sensitivity of the response $\cR$ is determined with respect to a single parameter, while all parameters are allowed to simultaneously vary over their admissible ranges. 
This is in contrast to the local sensitivity analysis (LSA) in which the sensitivity of $\cR$ to the small variation of each parameter is obtained keeping all other parameters fixed at their nominal values. 
The results of the GSA can be reported in terms of the Sobol indices~\cite{sobol} which are here directly computable from the coefficients in the expansion~(\ref{eq:pce}), see~\cite{sudret:08}.
According to~\cite{sobol}, the total sensitivity index of response $\cR$ with respect to the $i$-th parameter is obtained from,
\begin{equation}\label{eq:sobol}
S_{T_i}=1-\frac{\var(\BE(\cR|q_{\sim i}))}{\var(\cR)} \,,
\end{equation}
where, $q_{\sim i}=\{q_1,\ldots,q_{i-1},q_{i+1},\ldots,q_p\}$, and $\var(\cdot)$ denotes the variance.
Both LSA and GSA techniques are employed in \sects~\ref{sec:LSAGSA}, \ref{sec:wmParams}, and \ref{sec:gridAnisot}. 
By post-processing the results of the flow simulated by \of~for any combination of the sampled parameters, the associated response~$\cR$ is evaluated. 
Once the responses corresponding to all sampled parameters are evaluated, the metamodel (\ref{eq:pce}) is constructed by the open-source library \dakota, \cite{dakotaMan}, and the Sobol indices are computed.

\section{Simulations Setup and Error Measures}\label{sec:simDetails}
To investigate the effect of different factors on the predictive accuracy of WMLES, 
\rev{the, perhaps,  most canonical wall-bounded turbulent flow}, fully developed turbulent channel flow, is considered. 
This is a standard case for evaluation and development of different aspects of WMLES, see \eg~\cite{cabotMoin:00,nicoud:01,piomelli:03,templeton:05,lee:13,park:14,yang:17}. 
The flow is periodic in streamwise and spanwise directions with associated coordinates~$x$ and~$z$, respectively. 
All simulations in \sect~\ref{sec:results} are carried out over a computational domain $l_x\times l_y \times l_z = 9\delta \times 2\,\delta \times 4\delta$, where $\delta$ denotes the channel half-height. 
Insensitivity of the results with respect to different combinations of $l_x$ and $l_z$ was confirmed via \rev{additional simulations varying these spatial dimensions} (results are not shown here).
\rev{A structured grid with orthogonal hexahedral cells is used.}
\rev{In the majority of the simulations, the grid consists of cubical cells with side length $\delta/n$.}
However, the effect of grid anisotropy is also studied, separately.

The nominal friction-based Reynolds number $\reyt=\ut \delta/\nu$, for all simulations except those in \sect~\ref{sec:ReEffect}, is $5200$, with the DNS data of Lee and Moser~\cite{lee-moser:15} used as reference. 
To set up the simulations, the bulk velocity $U_b$, the kinematic viscosity $\nu$ and $\delta$ are chosen according to the DNS bulk Reynolds number, $\rey_b=U_b\delta/\nu$. 
The time step size is $\Delta t = 0.01 \delta/U_b$. 
All simulations started from the same initial field obtained from a separate WMLES of channel flow at the same $\rey$-number.
After simulating for $500 \delta/U_b$, when the initial transients have disappeared, time-averaging was started. 
The statistics were gathered over the time interval $1500 \delta/U_b$, corresponding to~$\approx 167$ domain flow-throughs.

\tim{The QoIs consist of} averaged values of channel flow quantities.
The averaging is performed in both time and spatially homogeneous directions, $x$~and~$z$, and is denoted by $\langle \cdot \rangle$.
In particular, the considered QoIs include $\lbut$, and the profiles of mean streamwise velocity $\U$, resolved \tim{Reynolds shear} stress $\buv$, and \tim{the} resolved turbulent kinetic energy, $\bk$.
\rev{The latter is~defined as,~\mbox{$\bk = \frac{1}{2} \langle \bu'_k \bu'_k\rangle$} with summation over $k$.}

The measures for assessing the accuracy of the flow QoIs are defined to be the relative errors between the QoIs predicted by WMLES and the corresponding values given by reference DNS.
\tim{In line with the discussion in \sect~\ref{sec:uq},} the errors in different QoIs are referred to as the simulation responses, $\cR$. 
For $\lbut$, the error measure is defined as,
$$
\epsilon[\lbut] = \left(\lbut - \lut^\circ \right)/\lut^\circ \,,
$$
where the superscript circle specifies DNS values. 
Consider a profile of some QoI, $g$, across the wall-normal~interval~$y\in[0,\delta]$.
The associated error can be measured by,
\begin{equation}\label{eq:einf}
\einf[g] = \frac{\max_{y\in[\delta_1,\delta]} |g(y)-g^\circ(y)|} {\| g^\circ\|_\infty} \,,
\end{equation}
where $\|g^\circ\|_\infty = \max_{y\in[0,\delta]} |g^\circ(y)|$. 
Measuring the errors in the $L_\infty$ norm rather than the $L_p$ ones with $p\geq 1$, is advantageous since it removes the need for numerical integration across the channel. 
Note that reducing the error in the $L_\infty$-sense automatically guarantees the error reduction in the~$L_p$-sense. 

As thoroughly discussed below, the near-wall flow field of WMLES can be contaminated by spurious overshoots. 
To bypass them and instead assess the responses in the outer layer, $\delta_1$ is taken to be $0.2\delta$. 
In wall units, $\delta_1^+=\delta_1/\delta_\nu$ is $1040$, based on the target $\reyt=5200$.
Therefore, the chosen~$\delta_1$ in (\ref{eq:einf}) falls in the overlap region of TBL. 
It is noteworthy that according to Zanoun \et~\cite{zanoun:03}, the outer part of  channel flow's mean velocity profile starts from $y^+\approx 150$ almost independently of the $\rey$-number.
The upper limit of the overlap region described by the logarithmic law increases with the $\rey$-number and extends up to $\approx 0.75\delta$ for $\reyt=5200$. 
This also corresponds to the upper limit of the region over which the Spalding law (\ref{eq:spalding}) is~valid.

\section{Preliminary Discussions}\label{sec:prelimDisc}
\subsection{A-priori sensitivity analysis of the wall model}\label{sec:LSAGSA}
\begin{figure}
\centering
    \begin{tabular}{cc}
    \includegraphics[scale=0.45]{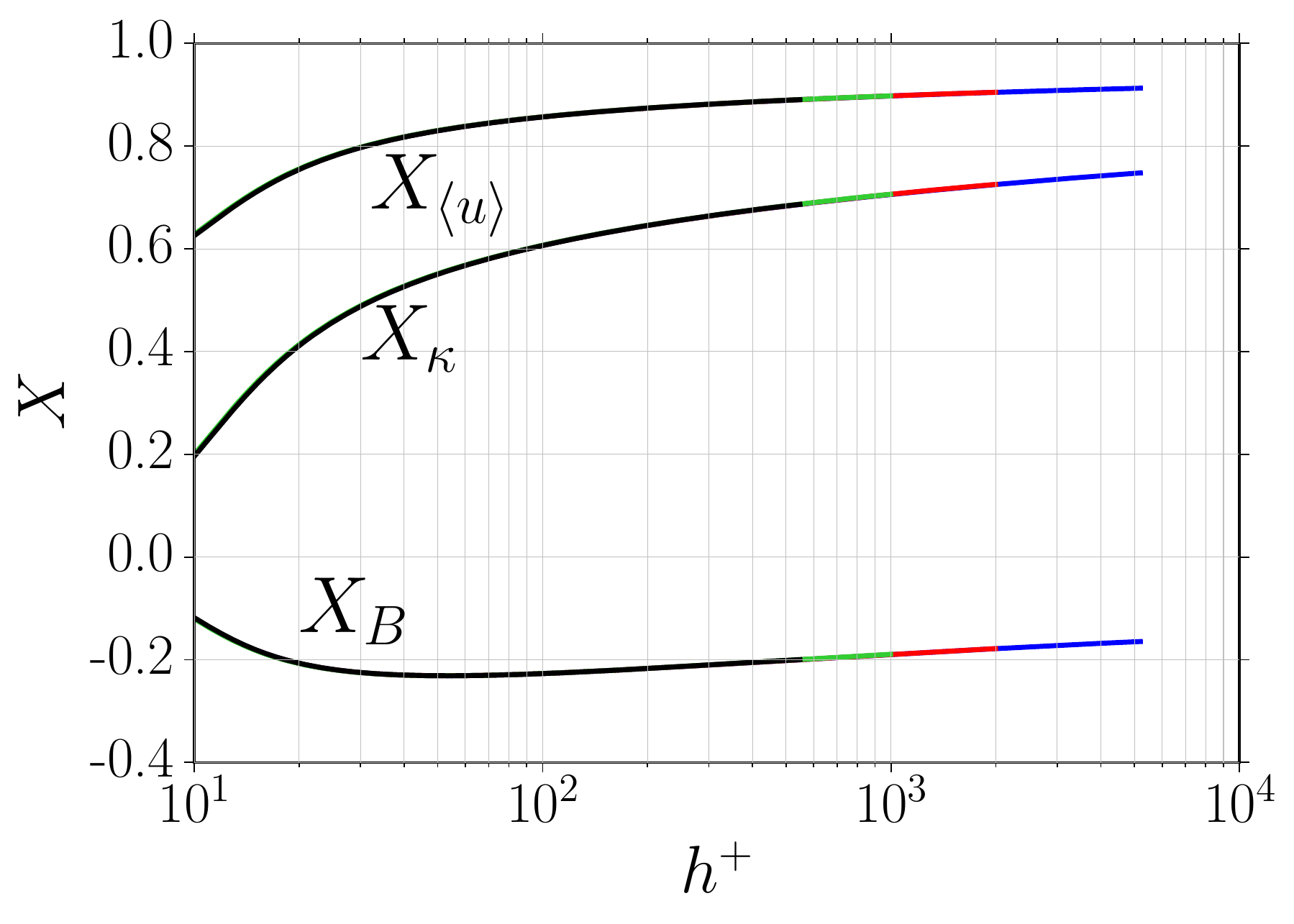} &
    \includegraphics[scale=0.45]{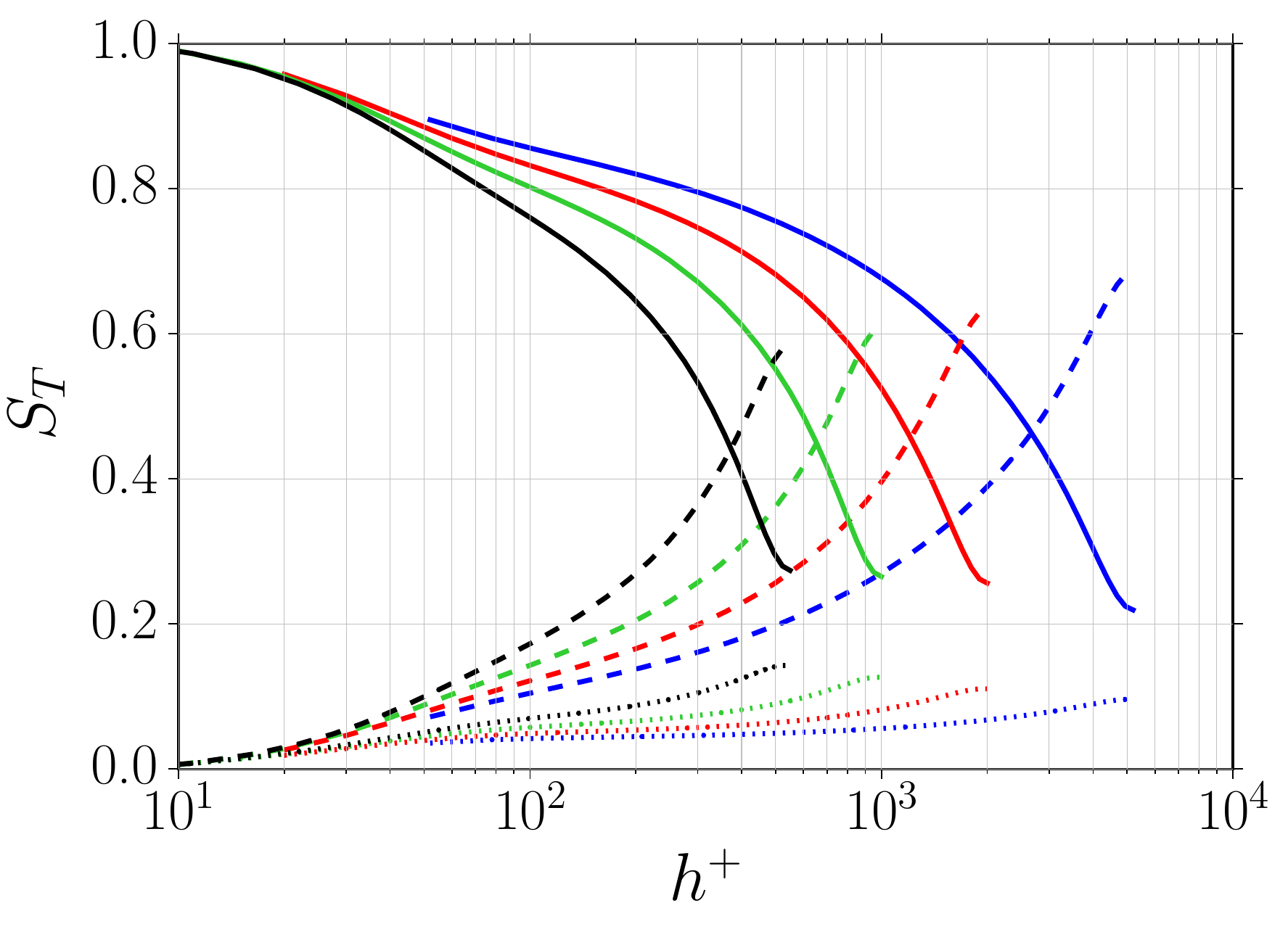} \\
    {\small (a)} &    {\small (b)} \\
    \end{tabular}
    \caption{The local (a) and global (total Sobol) (b) sensitivity indices of \revCom{the wall shear stress predicted by} the Spalding law, \revCom{plotted} against~$h^+$. DNS data of \cite{lee-moser:15} for velocity samples are used at $\reyt=550$ (black), $1000$ (green), $2000$ (red), and $5200$ (blue) along with the nominal parameter values $\kappa_0=0.395$ and $B_0=4.8$. In (b), the Sobol indices with respect to $\bu$, $\kappa$, and $B$ are represented by solid, dashed, and dotted lines, respectively.}\label{fig:SAIndices}
\end{figure}

The wall shear stress predicted by algebraic wall models, such as the one used in this study, depends on the functional form of the law of the wall, \eg~the Spalding law (\ref{eq:spalding}), the associated parameters \revCom{ (here~$\kappa$ and~$B$)}, and, finally, the input velocity sampled from the LES domain at a distance $h$ from the wall.
In order to rank the influence of these contributing factors on the predicted wall shear stress, both local and global sensitivity analyses reviewed in \sect~\ref{sec:uq} can be employed \revCom{in an a-priori way}. 
To this end, no WMLES simulation is performed and instead the available DNS data for velocity are used\revCom{, for details see\app~\ref{app:LSA}.} 
\revCom{The resulting sensitivity indices at different $\reyt$ are plotted in \fig~\ref{fig:SAIndices} against the inner-scaled sampling height $h^+=h\langle u_{\tau_0}\rangle/\nu$.
The nominal values of mean wall friction velocity,~$\langle u_{\tau_0} \rangle$, are taken from DNS \cite{lee-moser:15}.
}

\revCom{Based on the LSA, see \fig~\ref{fig:SAIndices}(a), the predicted mean wall shear stress is most sensitive to the variations in the sampled velocity, independent of the sampling height and $\reyt$.
In contrast, in GSA (\fig~\ref{fig:SAIndices}(b)) which considers the fluctuations of the velocity samples in addition to the mean values, 
the Sobol indices at different $\reyt$ do not match.} 
At all $\reyt$, the highest sensitivity is observed with respect to velocity and the lowest one to parameter $B$. 
Sensitivity to the sampled velocity decreases as the sampling height moves farther from the wall and it eventually becomes less than the sensitivity to $\kappa$.
\revCom{This is due to the fact} that as the sampling height increases, the ratio of the fluctuating to the mean velocity decreases, hence reducing the influence of the velocity on the predicted wall shear stress.

Although the analyses conducted here are based on using perfect DNS data fed in the wall models, the crucial influence of the velocity samples on the predicted wall shear stress\rev{, as compared to potential inaccuracies in the law of the wall parameters,} is well clarified.  
\rev{This stresses the importance of highly accurate velocity values in WMLES in order for the wall modeling to be~precise.}
Besides this, \revCom{it is concluded that} if varying the wall model parameters is going to be used as a controller of the error in the wall shear stress, changing the \vk~coeffcient $\kappa$ can be more effective than \revCom{$B$}.

\subsection{Relation between the QoIs of channel flow}\label{sec:qoiRelation}
An analysis of the basic connections between different quantities of channel flow is provided here to help motivate the observations in the upcoming sections. 
\rev{For this flow, averaging the streamwise component of the LES momentum equation (\ref{eq:LESmomentum}) in time and the wall-parallel directions results in,}
\begin{eqnarray}\label{eq:xMomChan}
\lbut^2 (1-\eta)= \nu\frac{\dd \U}{\dd y} - \buv - \langle B_{xy} \rangle  \,,
\end{eqnarray}
where, $B_{xy}=\overline{uv} -\bu \bv$ and $\eta=y/\delta$.
According to this equation, the total shear stress linearly varies across the channel and is balanced with the summation of the viscous shear stress and the total Reynolds shear stress.
Dividing all terms by~$\lbut^2$, the inner-scaled version of the equation~is~derived,
\begin{equation}\label{eq:xMomChan_inScal}
(1-\eta) = \frac{\dd \U^+}{\dd y^+} - \buv^+ - \langle B_{xy}\rangle^+ \,,
\end{equation}
where, $\U^+= \U/\lbut$ and $y^+=y\lbut/\nu$. 
It is noticed that the left-hand-side of this equation is independent of the flow quantities. 
\revCom{Later, in \fig~\ref{fig:uvBalance_lust}, different terms in \eq~(\ref{eq:xMomChan_inScal}) are evaluated for a set of WMLES.}

In the overlap region of the TBL where the logarithmic law describes the mean velocity profile, the viscous stress ${\dd \U^+}/{\dd y^+}$ is inversely proportional to $y^+$. 
Therefore, as $y^+$ increases, the contribution of the viscous stress continuously becomes smaller than the effect of the turbulent counterpart.
This immediately leads to the conclusion that whether or not the predicted mean wall shear stress is accurate,
the inner-scaled \tim{total} turbulent \tim{shear} stress $-(\buv^+ + \langle B_{xy}\rangle^+)$ \tim{will} match well with the benchmark data in the outer layer of the TBL.
Therefore, using this quantity as a measure of the simulation quality can be misleading.
Motivated by this, separate evaluation of the errors in the computed wall shear stress, \rev{and outer-scaled} Reynolds shear stresses and \tim{mean} velocity, is essential, as shown in the next sections.

Based on the same discussion, over- and under-prediction of $\lbut$ (associated with negative and positive LLM, respectively) are followed by over- and under-prediction of the Reynolds shear stress in the outer layer.
Consequently, upon accurate prediction of $\lbut$, accurate Reynolds shear stress in the outer layer of the TBL will be obtained. 
If the contribution of the SGS component, $-\langle B_{xy} \rangle$, becomes negligible --due to the particular combination of the numerical scheme, SGS model and grid resolution-- the accurate $\lbut$ is accompanied by an accurate resolved Reynolds shear stress~$-\buv$.

\section{Simulation Results and Discussions}\label{sec:results}
The influence of different factors on WMLES of channel flow is thoroughly discussed in the following sections. 
\rev{Each section is focused on one particular aspect.}
\rev{Inner-scaled quantities are denoted by superscript~$^+$.} 
\reva{The other reported QoIs, $\U$, $-\buv$, and $\bk$ are outer-scaled using bulk velocity $U_b$ and are considered as profiles of $\eta=y/\delta$.}
\rev{It is emphasized that}, when discussing the results, the phrase ``error in \tim{the} profile~$\ldots$ far from the wall \emph{or} in the outer layer" may be used by which the error in a cross-channel profile measured by~(\ref{eq:einf}) is meant. 
\rev{It is recalled that in all sections, except \sect~\ref{sec:ReEffect}, $\reyt=5200$.}

\subsection{Effect of numerics, SGS modeling, and wall modeling}\label{sec:numSGSwm}
The three main contributors to the results of WMLES are the numerical scheme for solving~(\ref{eq:LESmomentum}), the SGS model required to close those equations, and finally, the wall model for predicting and imposing the boundary condition at the wall. 
\rev{As quantified} below by systematic simulations, even with accurate wall shear stress, other flow quantities still \rev{contain} numerical and SGS errors. 
\rev{See also, \cite{cabot:96,yang:17}.}

The discussion \tim{below is separated into} three subsections.
In the first one, the influence of numerics and SGS modeling without wall modeling is discussed. 
Then, it is investigated if and how those effects might be modified when wall modeling is added.
Based on \rev{the discussions}, a mechanism explaining how different QoIs of WMLES of channel flow are connected is developed. 
The final subsection is devoted to demonstrate the role of SGS modeling in removing spurious overshoots in the profiles of velocity's second-order moments.

\subsubsection{Footprint of numerics and SGS modeling}\label{sec:footprint}
At the first step, \rev{even before considering wall modeling,} the aim is to characterize the footprint of the \tim{considered} numerical scheme\tim{s} and SGS model\tim{s}.
To this end, a set of channel flow simulations with \tim{an} isotropic grid of resolution~$\nd=28$ is carried out with no wall modeling involved.
Two interpolation schemes, Linear and LUST, are used for the convective term in \eq~(\ref{eq:LESmomentum}), see \sect~\ref{sec:cfd}. 
To make the impact of the numerical scheme distinct, simulations with no explicit SGS model \tim{are} also included.

\begin{table}[!b]
\centering
\caption{The error in $\lbut$ and $\U$ for LES without any wall modeling with resolution $\nd=28$.}\label{tab:numSGSDuTau}
\begin{small}
   \begin{tabular}{c|cccc}
   \toprule\toprule
   {} & \multicolumn{2}{c}{$\epsilon[\lbut]\, \%$} & \multicolumn{2}{c}{$\einf[\U]\, \%$}\\
   SGS model & Linear  & LUST   & Linear  & LUST \\
   \hline
   No SGS & -53.36 & -56.73 & 2.38 & 5.16\\
   WALE & -5.56 & -11.75 & 2.98 & 1.19 \\
   Smagorinsky & -5.13 & -6.06 & 0.83 & 0.95\\
   Dynamic $k_{\sgs}$-eqn & -6.77 & -12.62 & 0.58 & 1.56\\
   \bottomrule
   \end{tabular}
\end{small}
\end{table}

\begin{figure}[!h]
\centering
   \begin{tabular}{ccc}
   \includegraphics[scale=0.43]{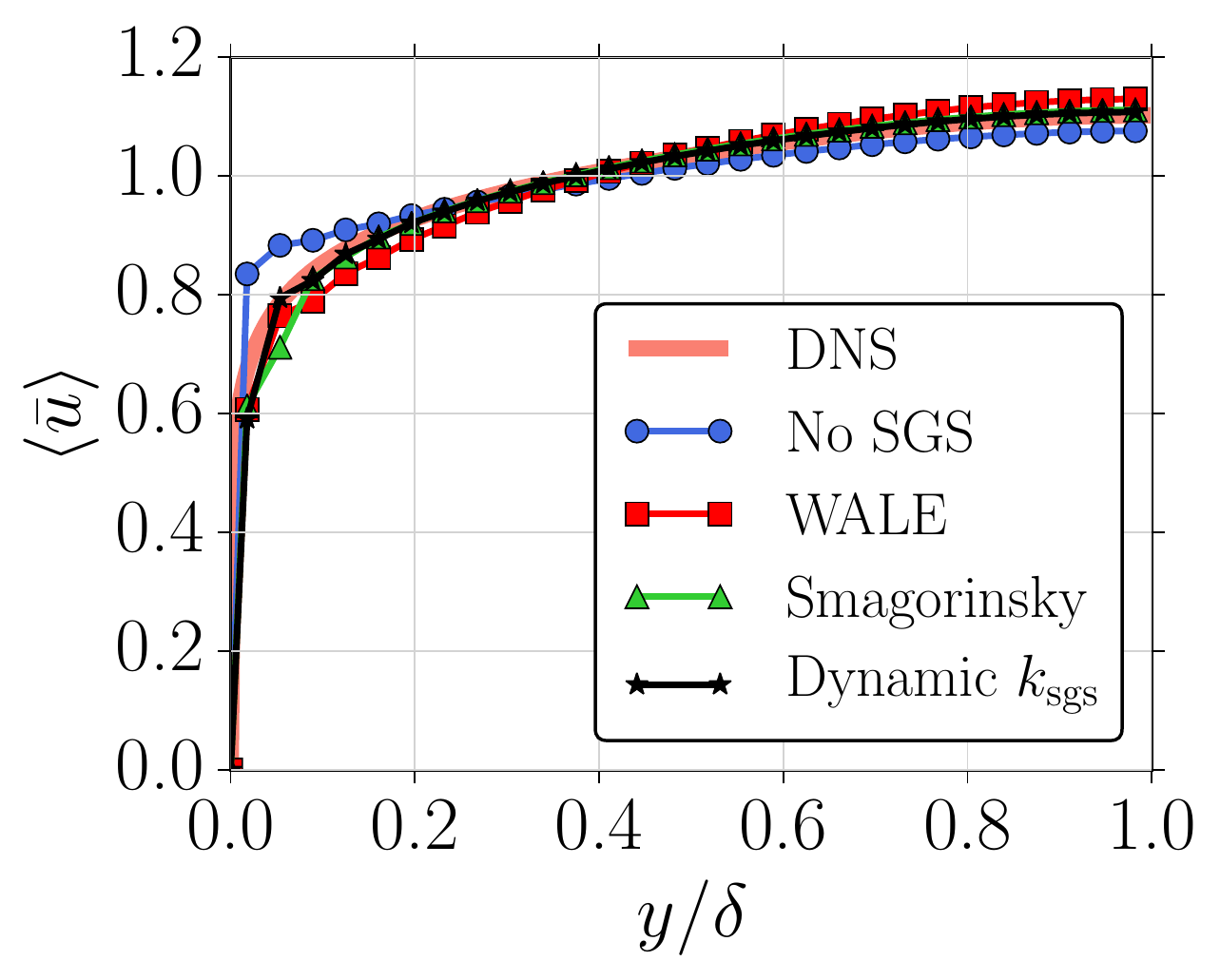} & \hspace{-0.6cm} 
   \includegraphics[scale=0.43]{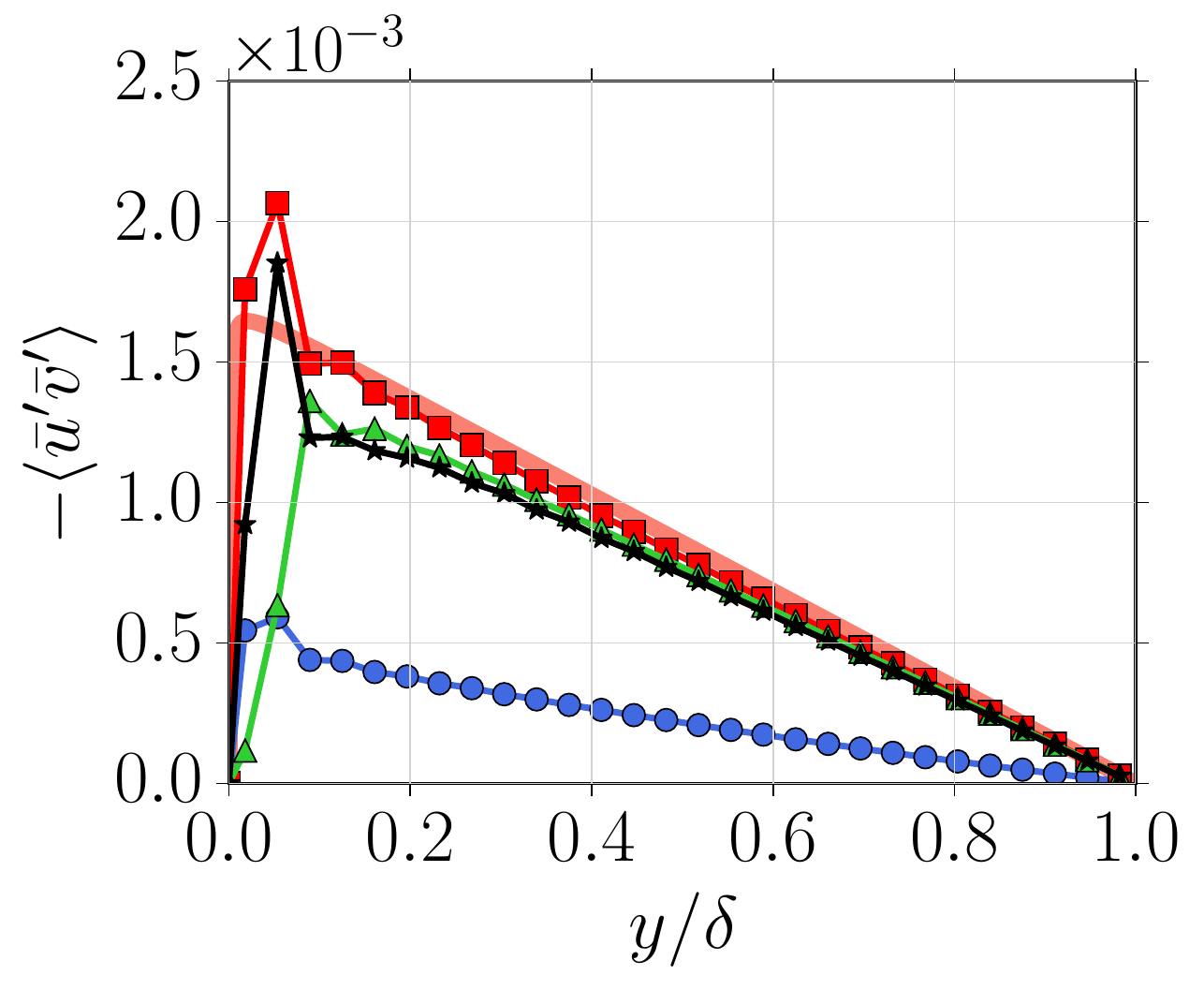} &   \hspace{-0.6cm}
   \includegraphics[scale=0.43]{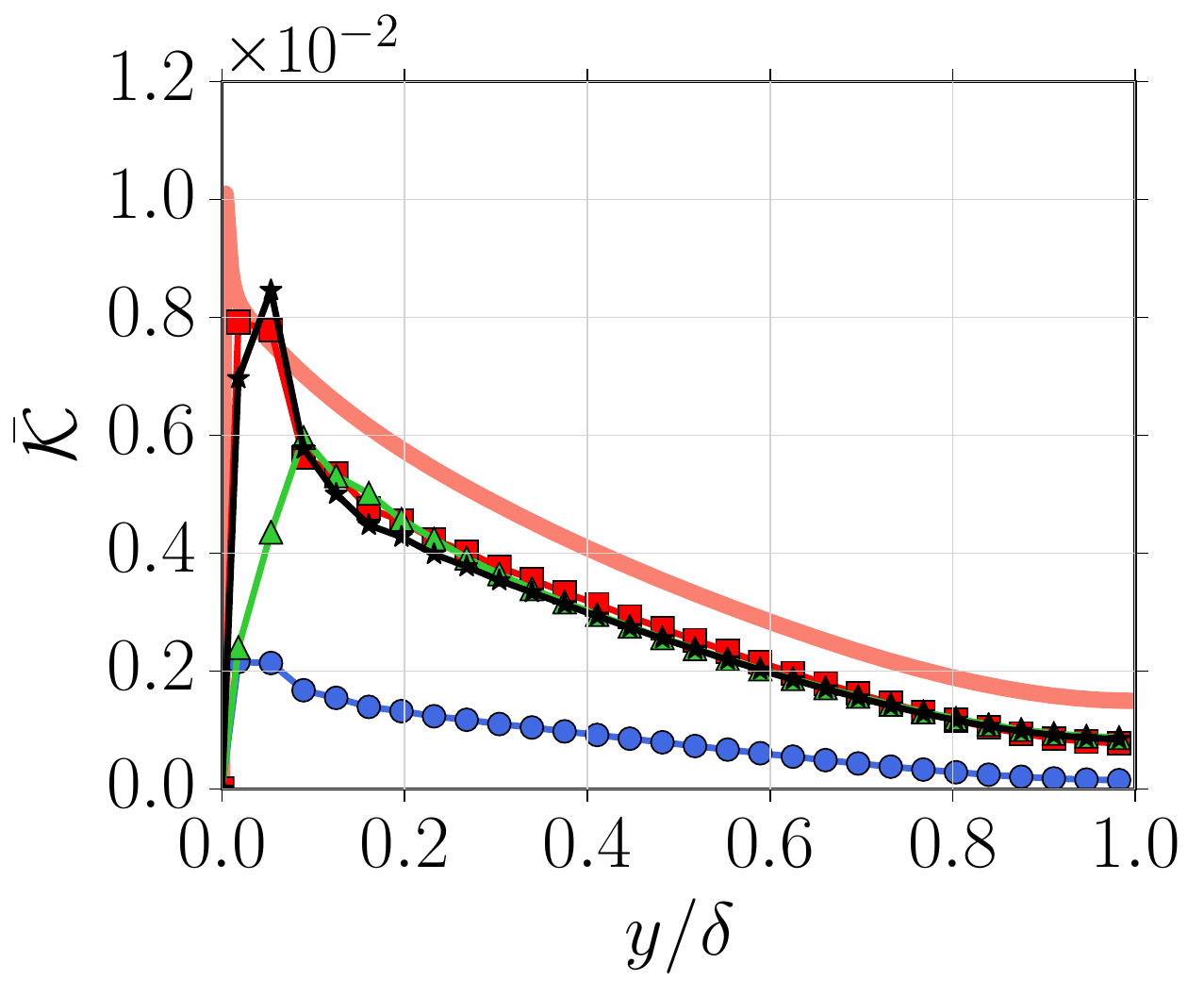} \\
   {\small{(a)}} &    {\small{(b)}} &    {\small{(c)}}\\      
   \includegraphics[scale=0.43]{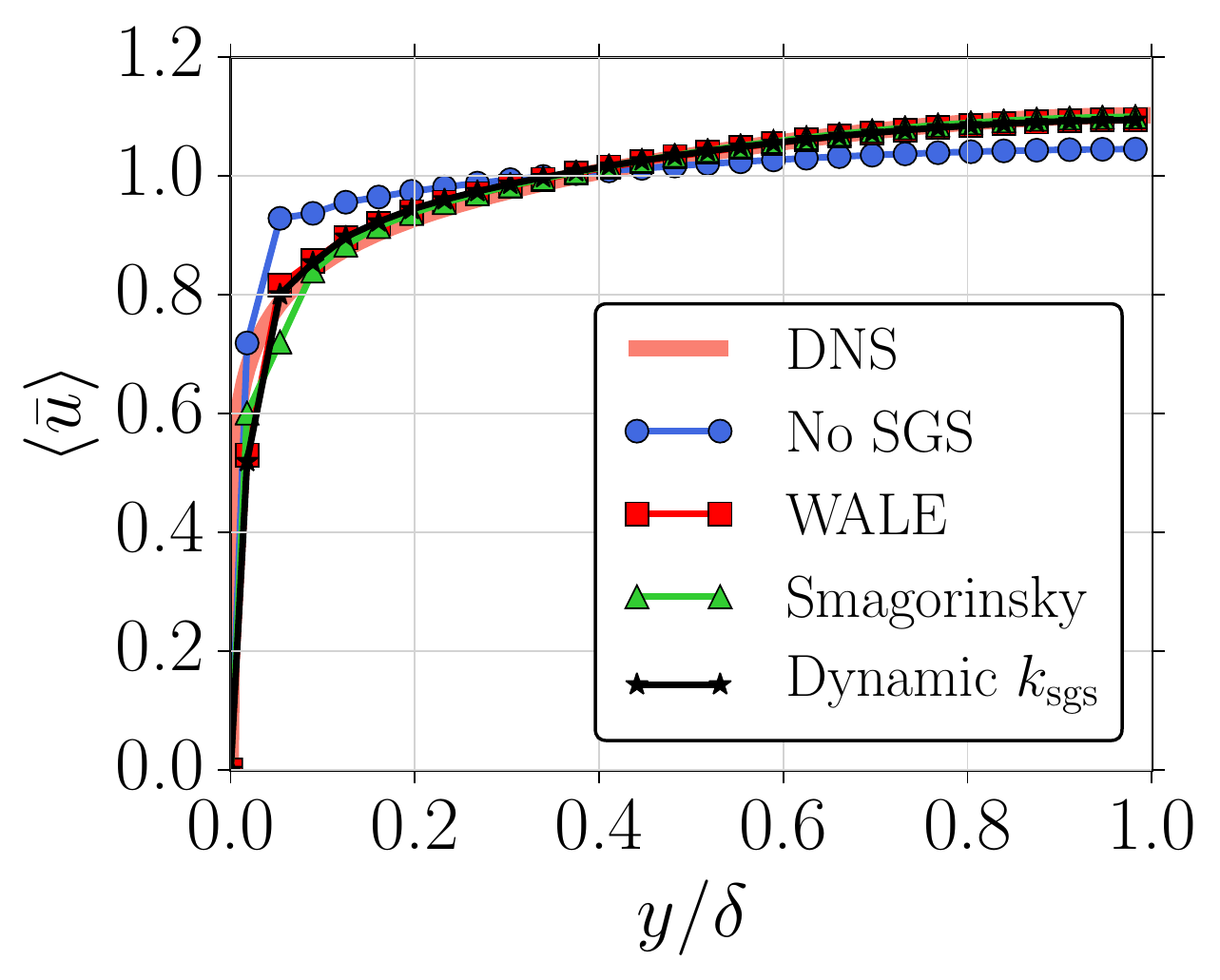} &\hspace{-0.6cm}
   \includegraphics[scale=0.43]{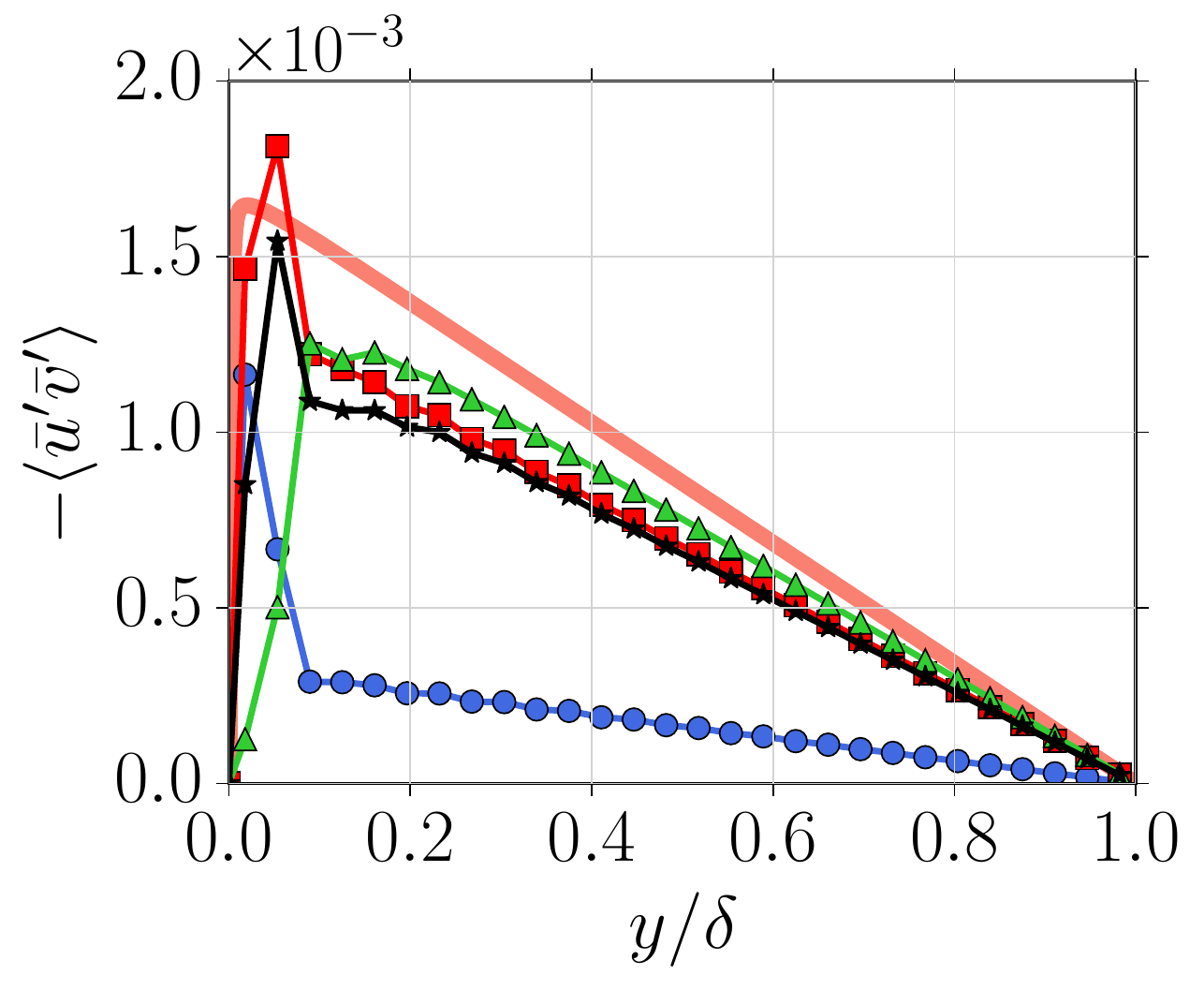} &   \hspace{-0.6cm}
   \includegraphics[scale=0.43]{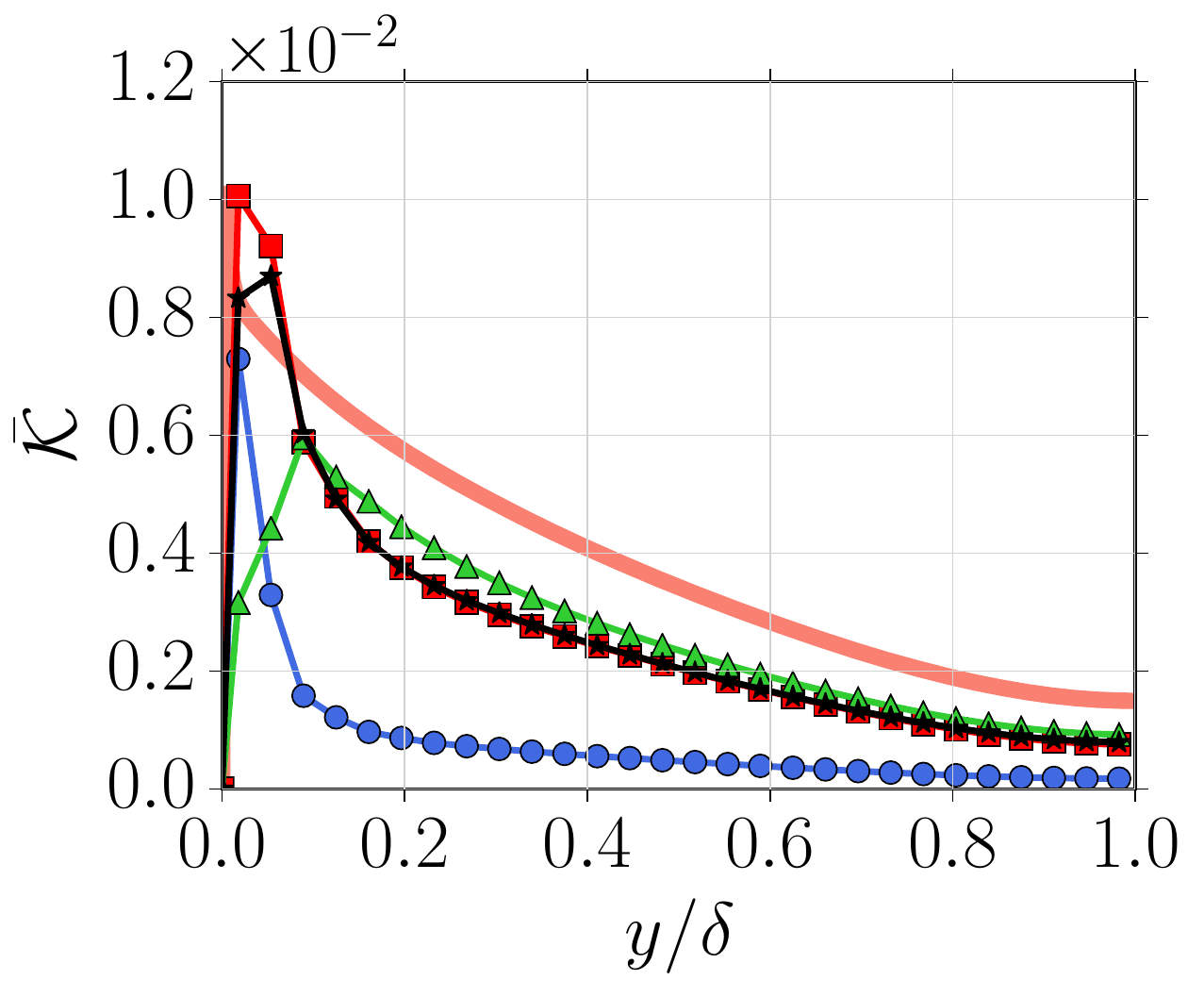} \\   
   {\small{(d)}} &    {\small{(e)}} &    {\small{(f)}}\\       
   \end{tabular}
   \caption{Profiles resulted from simulations with no wall modeling using the Linear scheme (top) and the LUST scheme~(bottom) associated with \tab~\ref{tab:numSGSDuTau}.}\label{fig:numSGSfoot}
\end{figure}

The resulting errors in $\lbut$ and \tim{the} $\U$ profile far from the wall are listed in \tab~\ref{tab:numSGSDuTau}.
For both interpolation schemes, the least accurate $\lbut$ is obtained when excluding explicit SGS model\tim{ing}. 
The improving effect of including \tim{an} SGS model for the Linear scheme is more than the corresponding gain when using the LUST scheme.
The Smagorinsky model leads to the lowest magnitude of~$\epsilon[\lbut]$ compared to the other options for SGS modeling. 
Further investigation reveals that the Smagorinsky model has comparatively higher values of $\langle \nu_{\sgs}\rangle$ across the channel, particularly in the first few off-wall cells (not shown here).

For the same set of simulations, \fig~\ref{fig:numSGSfoot} shows the resulting profiles of the mean velocity $\U$, resolved Reynolds shear stress $-\buv$, and resolved TKE,~$\bk$.
Similar to $\lbut$, the largest deviation between the LES profiles and the reference DNS \cite{lee-moser:15} is observed when SGS model\tim{ing} is excluded, independent \tim{of} the numerical \rev{interpolation} scheme used.
\tim{In particular, it is clear that without the additional turbulent mixing introduced by the subgrid viscosity, the wall friction has little impact on the mean velocity profile, which is almost flat.
Thus, SGS modeling plays a critical role in compensating for the under-resolution of the \rev{near-wall region} of the TBL.
} 
\rev{This is in contrast to WRLES for which it has been shown that, see~\cite{GMR:1}, numerical dissipation alone can be sufficient to account for the unresolved scales.}

For the LUST scheme, the resulting $\U$ profile \tim{above} $\yd\gtrsim 0.1$, is predicted with similar high accuracy for different SGS models, see also \tab~\ref{tab:numSGSDuTau}. 
The low sensitivity of these simulations to SGS modeling makes the LUST scheme well suited for WMLES.
This will be further supported in the rest of this article. 
In contrast, the choice for the SGS model can significantly influence the simulations with the Linear scheme, see how \tim{the} WALE model fails in capturing $\U$ compared to the Smagorinsky and Dynamic $k_{\sgs}$-eqn models.  

\begin{figure}[!h]
\centering
   \begin{tabular}{cc}
   \includegraphics[scale=0.4]{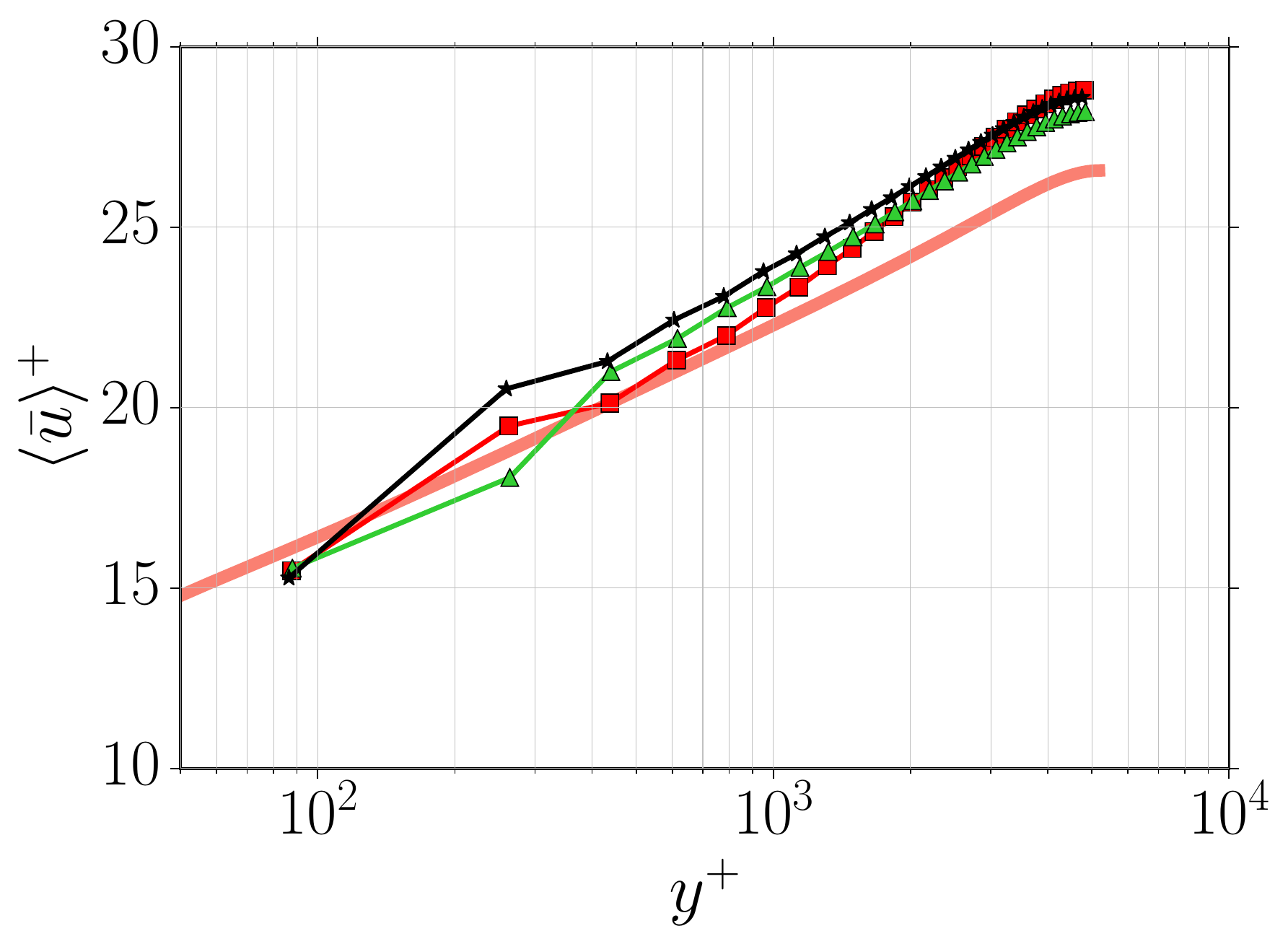}&
   \includegraphics[scale=0.4]{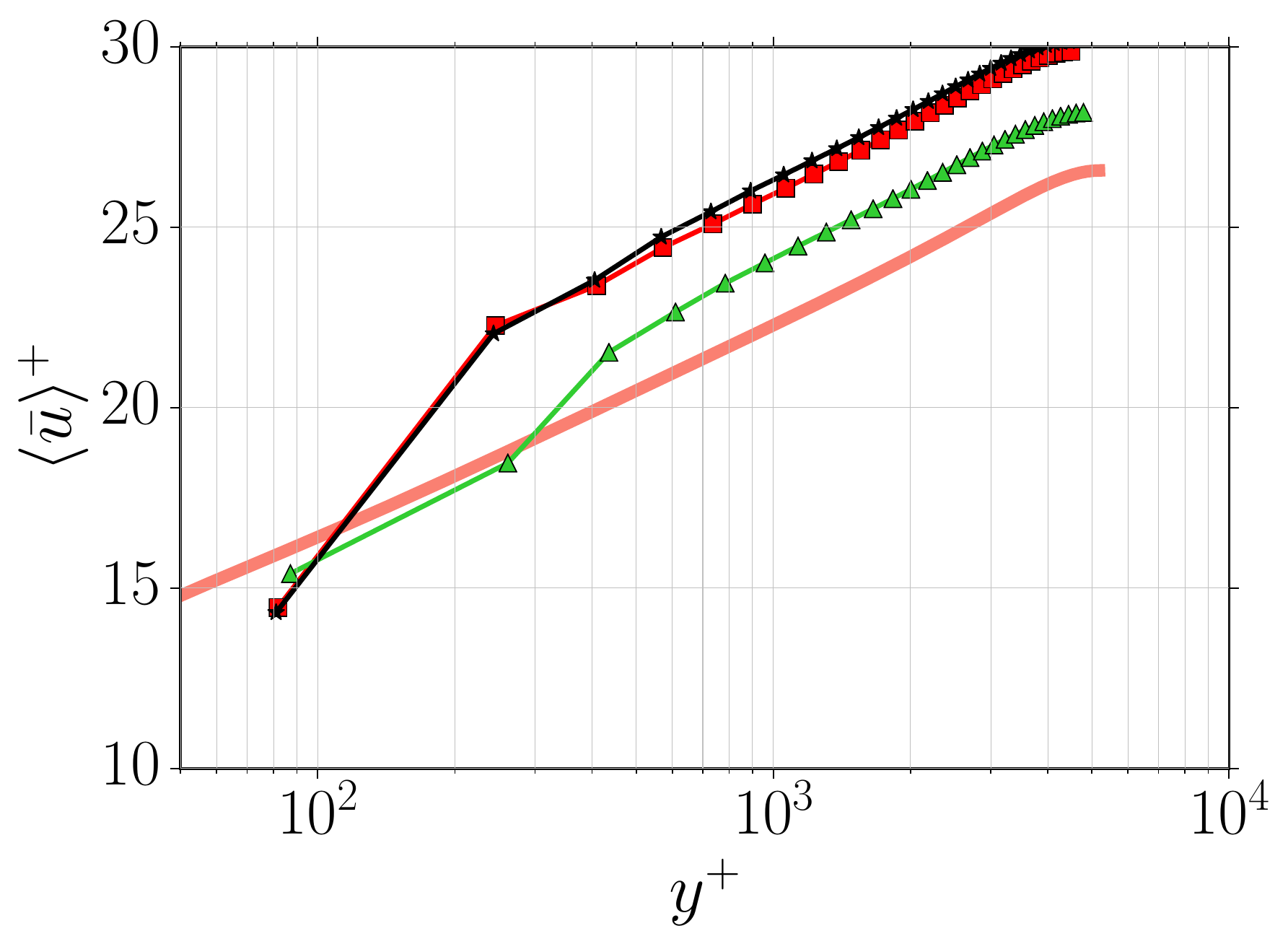}\\
   {\small{(a)}} &    {\small{(b)}} \\
   \end{tabular}
   \caption{Inner-scaled mean velocity profiles of the simulations in \fig~\ref{fig:numSGSfoot} obtained using the Linear (a) and LUST~(b) schemes.}
   \label{fig:numSGSfoot_uPls}
\end{figure}

\rev{
It is also observed that a relatively accurate $\U$ in the outer layer can be obtained despite having a significant error in $\lbut$, compare \fig~\ref{fig:numSGSfoot}(a,d) and \tab~\ref{tab:numSGSDuTau}.
This is even more clear from \fig~\ref{fig:numSGSfoot_uPls}, where except for the combination of the Linear scheme and the WALE model, other profiles seem to be approximately similar to the DNS data in the outer layer, but shifted upward (positive LLM). 
This observation is further explained in the next section. 
}

Shown in \fig~\ref{fig:numSGSfoot}\tim{(b,e)} are the profiles of the resolved Reynolds shear stress $-\buv$. 
Recall that according to \sect~\ref{sec:qoiRelation}, the under-prediction of $\lbut$ listed in \tab~\ref{tab:numSGSDuTau} should be accompanied by the under-prediction of the \revCom{total Reynolds shear stress $-\uv:=-(\buv+\langle B_{xy}\rangle)$} profile in the outer layer when compared to the reference DNS data. 
This seems to hold also for the resolved $-\buv$ profiles in \fig~\ref{fig:numSGSfoot}\tim{(b,e)}, except for the combination of Linear-WALE.
By this, the contribution of the SGS Reynolds shear stress is indicated to be negligible. 
\rev{The variation of the resolved TKE profiles with the explicit SGS model is shown in \fig~\ref{fig:numSGSfoot}(c,f).}
The favorable effect of including explicit SGS modeling is~clear.

\begin{figure}[!h]
\centering
   \begin{tabular}{cc}
   \includegraphics[scale=0.4]{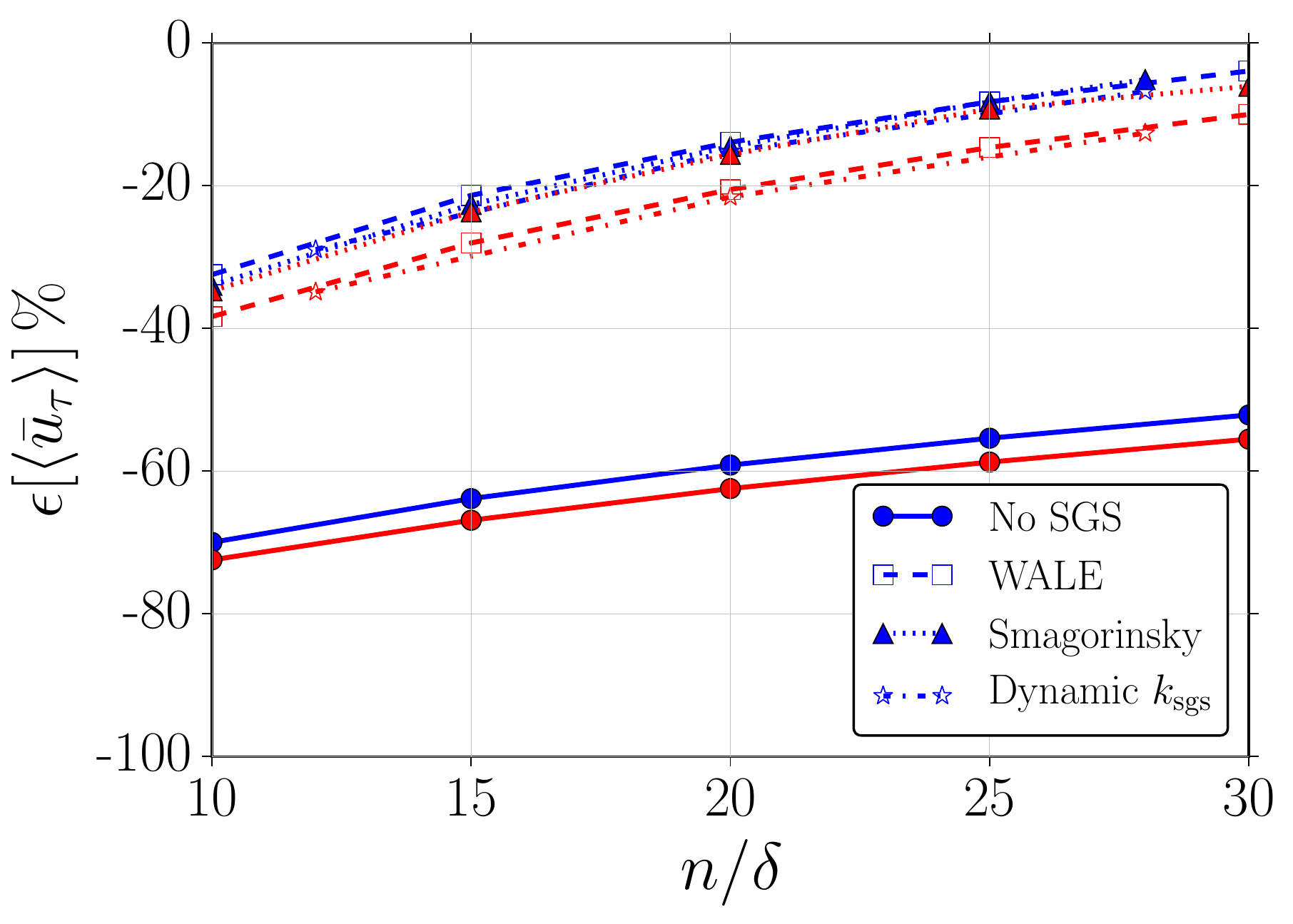}   &
   \includegraphics[scale=0.4]{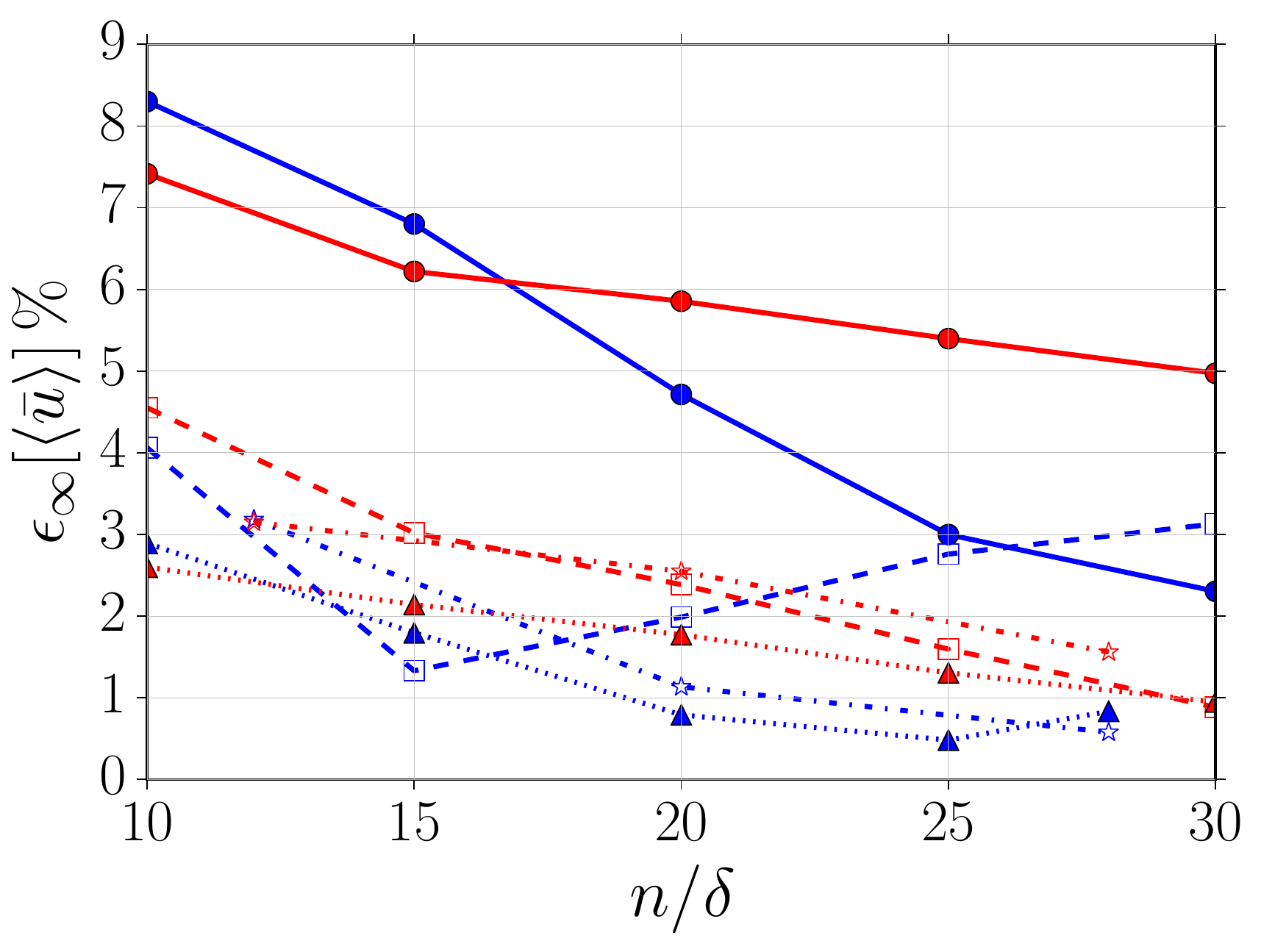}   \\   
   {\small{(a)}} &    {\small{(b)}} \\   
   \includegraphics[scale=0.4]{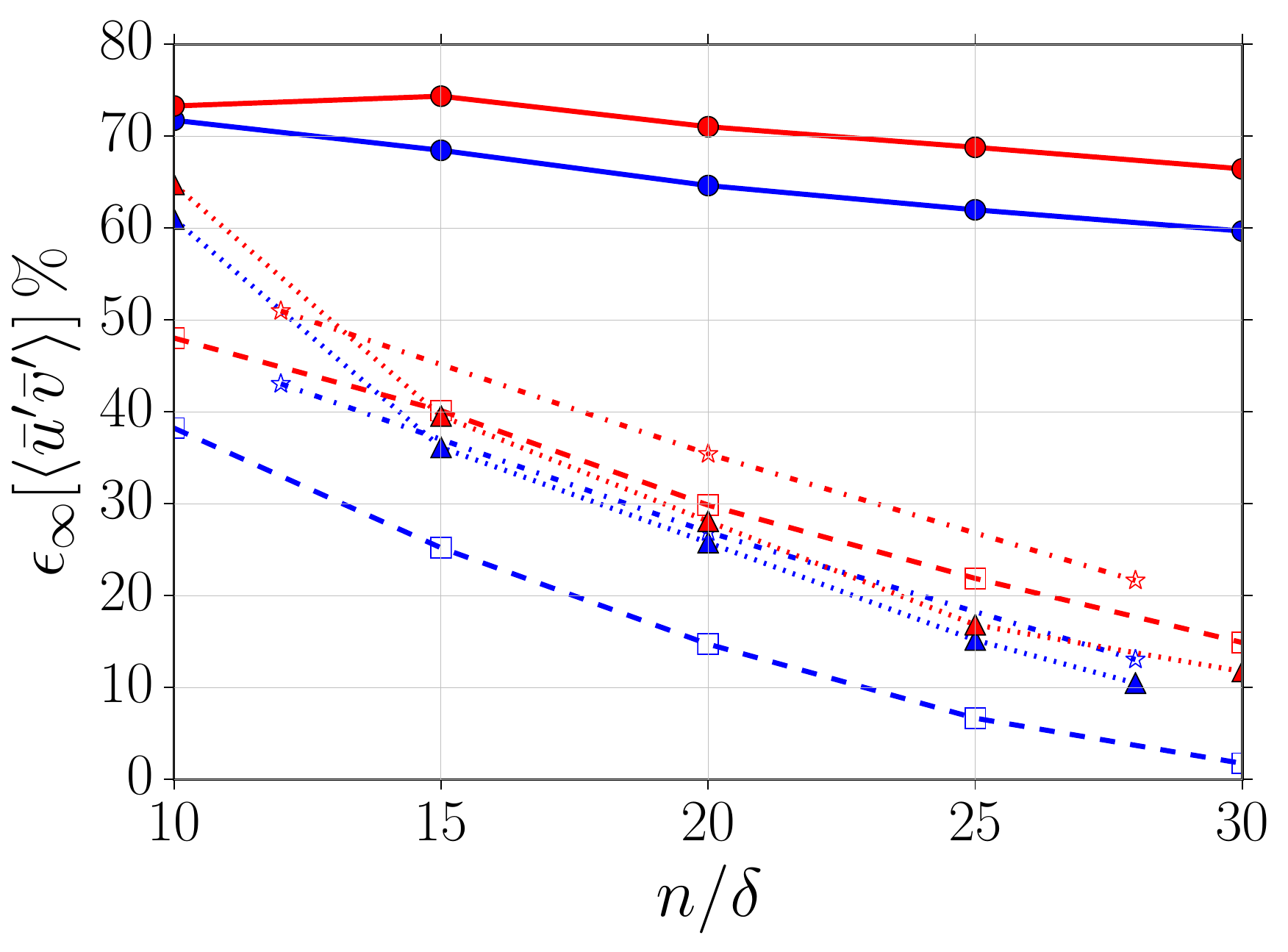}   &
   \includegraphics[scale=0.4]{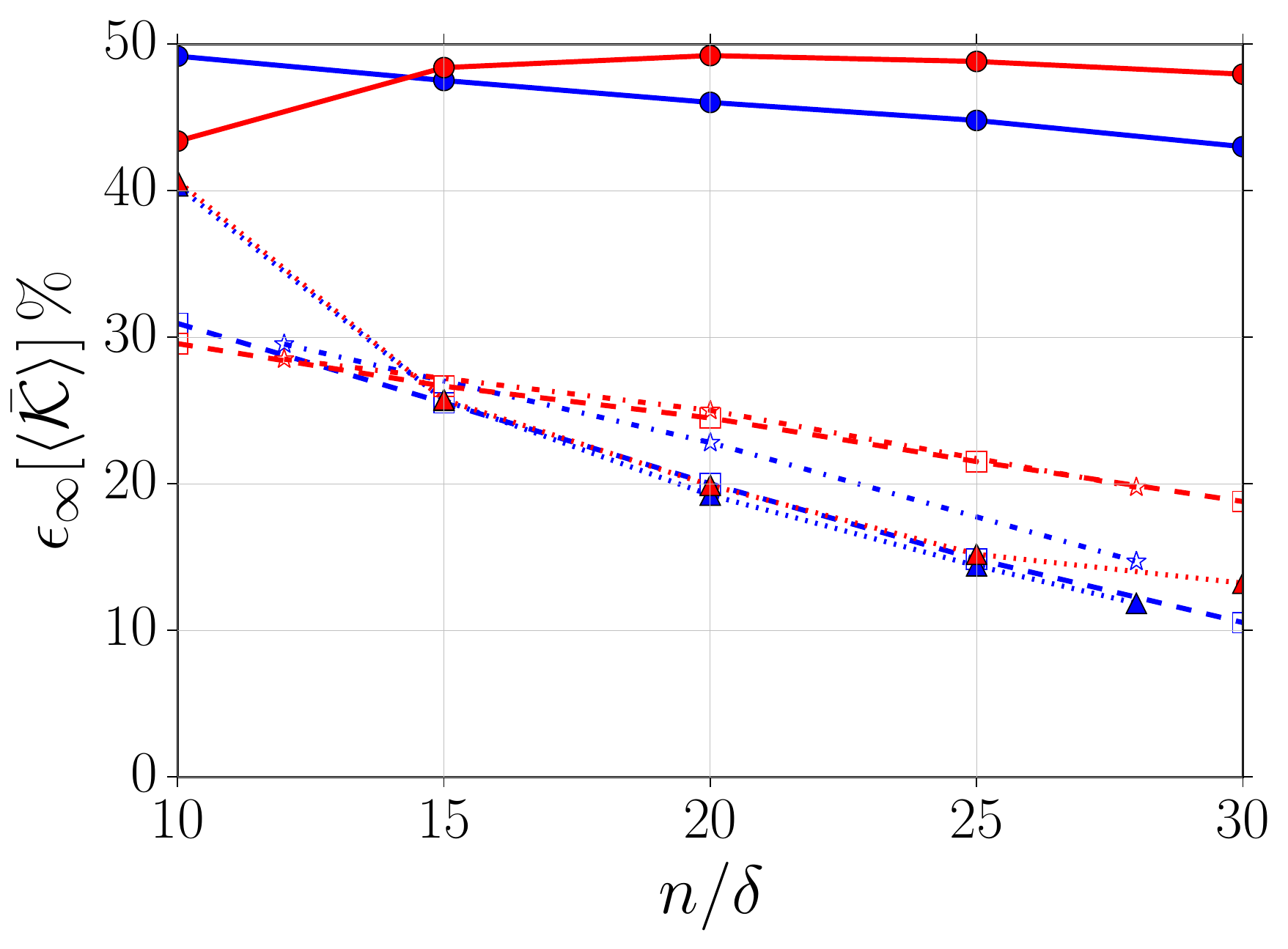}   \\      
   {\small{(c)}} &    {\small{(d)}} \\   
   \end{tabular}
\caption{Variation of the errors in the LES QoIs with grid resolution $\nd$ for different combinations of SGS models and numerical interpolation schemes: Linear (blue) and LUST (red). No wall model is used.}\label{fig:noWM_conv}
\end{figure}

The above observations were made for a fixed resolution $\nd=28$. 
To infer how the error in different QoIs of channel flow in the absence of wall modeling vary with the grid resolution $\nd$, \fig~\ref{fig:noWM_conv} is provided. 
As expected, by increasing the grid resolution, the magnitude of $\epsilon[\lbut]$ as well as the errors in the velocity statistical moments in the outer layer monotonically decrease, almost for all cases.
The exceptions are the $\einf[\U]$ for low-dissipative combinations of the Linear scheme with \tim{the} WALE and no-SGS models. 
In \tim{the} case of using \tim{an explicit} SGS model, the rate of reduction of the errors in $\lbut$ and $-\buv$ is almost independent of the numerical scheme.
In contrast to this,~$\einf[\bk]$ for the Linear scheme decreases with increasing $\nd$ faster than that for the LUST scheme. 
It is also interesting to note that for the simulations with no explicit SGS model, little improvement \revCom{in QoIs} is achieved by increasing the resolution. 

\subsubsection{Gain by wall modeling}\label{sec:WMgain} 
The same set of simulations as in the previous section are repeated here, but with wall modeling.
The Spalding law~(\ref{eq:spalding}) with $\kappa=0.395$ and $B=4.8$ is utilized with instantaneous streamwise velocity sampled from the~$\frth$ \tim{consecutive off-wall} cell center located at $\hd=0.125$.
\tim{This particular choice of the wall model parameters is not of primary importance here, and} the sensitivity of the wall model to different factors will be extensively discussed \tim{in the next sections}.

\begin{table}[!b]
\centering
\caption{The error in $\lbut$ and $\U$ for WMLES with $\nd=28$. The Spalding law (\ref{eq:spalding}) with $\kappa=0.395$ and $B=4.8$ is the wall model with the sampled velocity taken from $\hd=0.125$ ($\frth$~cell center).}\label{tab:WMfootDuTau}
\begin{small}
   \begin{tabular}{c|cccc}
   \toprule\toprule
   {} & \multicolumn{2}{c}{$\epsilon[\lbut]\, \%$} & \multicolumn{2}{c}{$\einf[\U]\, \%$}\\
   SGS model & Linear  & LUST  & Linear  & LUST  \\
   \hline
   No SGS &  -6.15  & 0.20 & 6.03 & 0.70\\
   WALE & -4.28 &  0.38 & 3.08 & 0.58 \\
   Smagorinsky &  -1.80 & -0.00 & 1.16 & 0.70\\
   Dynamic $k_{\sgs}$-eqn  & -1.60 &  0.43 & 1.01 & 0.76\\
   \bottomrule
   \end{tabular}
\end{small}   
\end{table}

As listed in \tab~\ref{tab:WMfootDuTau}, the resulting $\lbut$ are significantly improved by the inclusion of the wall model. 
The highest gain is obtained for the simulations with no explicit SGS model. 
\tim{The increased wall shear imposed by the wall model is, therefore, capable of compensating for the lack in wall-normal momentum diffusion coming from the absence of SGS viscosity.
Employing a dissipative scheme, such as LUST, the numerical viscosity, coupled with the resolved turbulent motions, is sufficient to properly distribute the momentum across the channel, making the results independent of SGS modeling.
Conversely, when using the Linear scheme, results are improved when more dissipative SGS models are employed.
}

\begin{figure}[!h]
\centering
   \begin{tabular}{ccc}
   \includegraphics[scale=0.43]{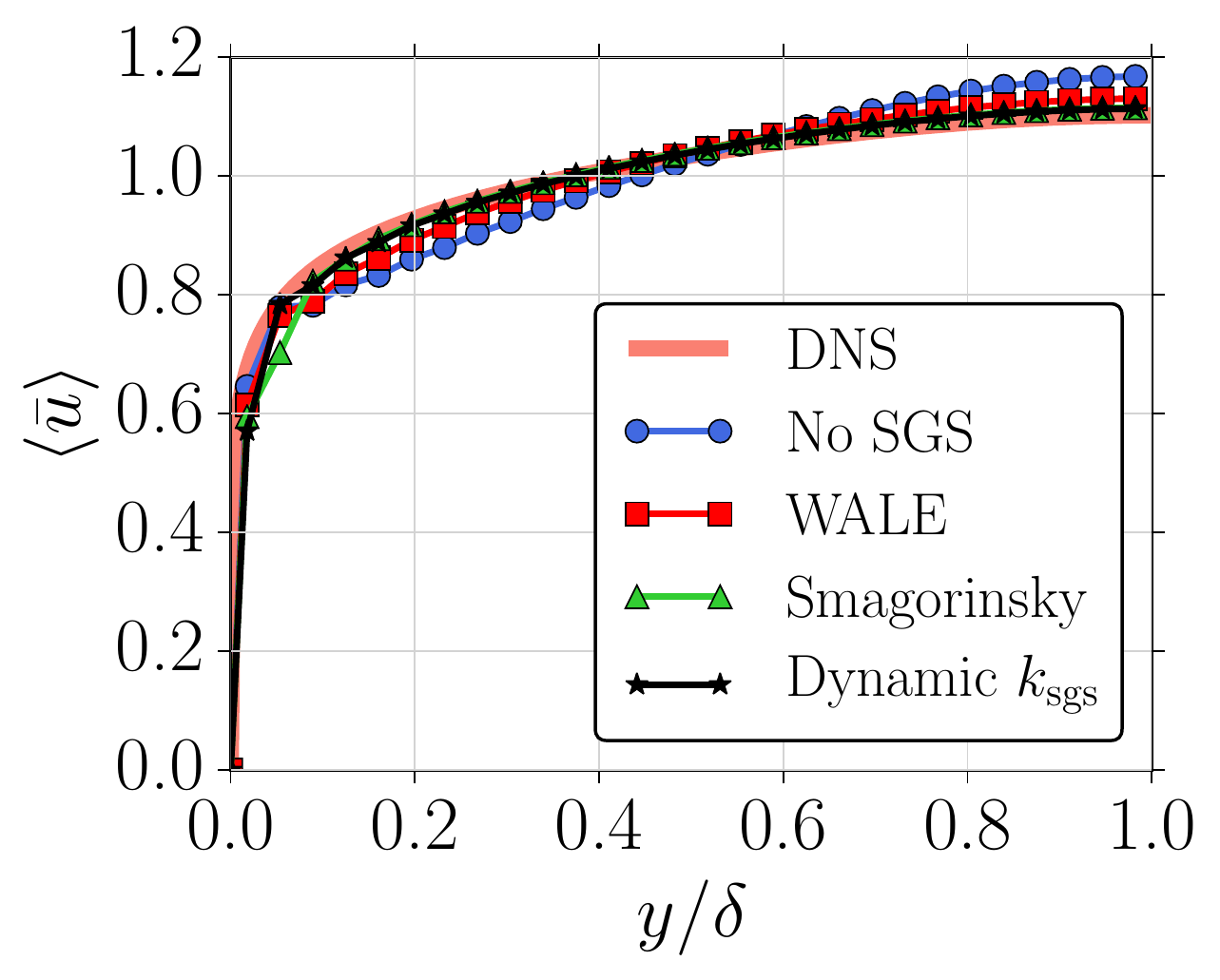}& \hspace{-0.52cm}
   \includegraphics[scale=0.43]{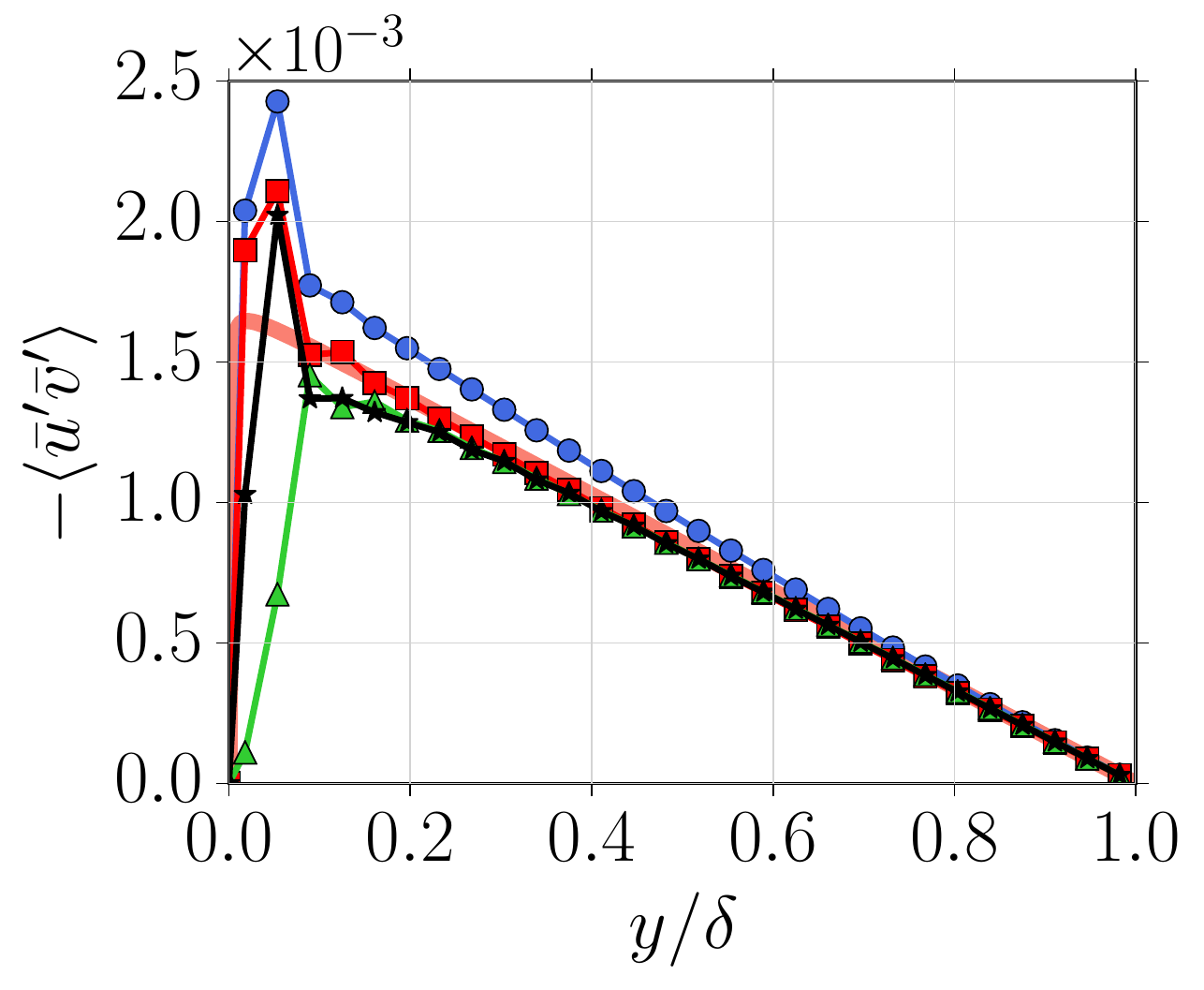}& \hspace{-0.54cm}
   \includegraphics[scale=0.43]{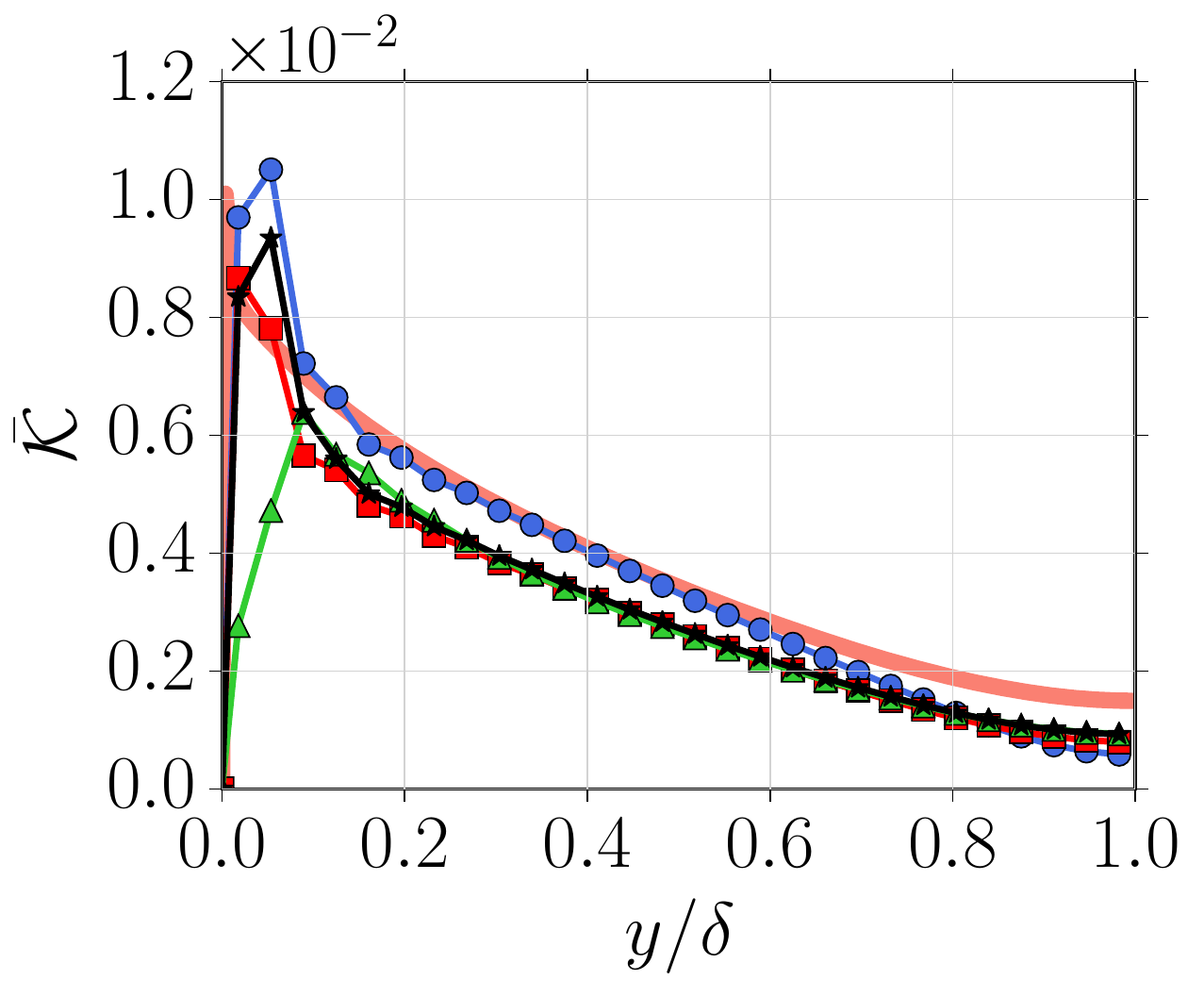} \\   
   {\small{(a)}} &    {\small{(b)}} &    {\small{(c)}}\\
   \includegraphics[scale=0.43]{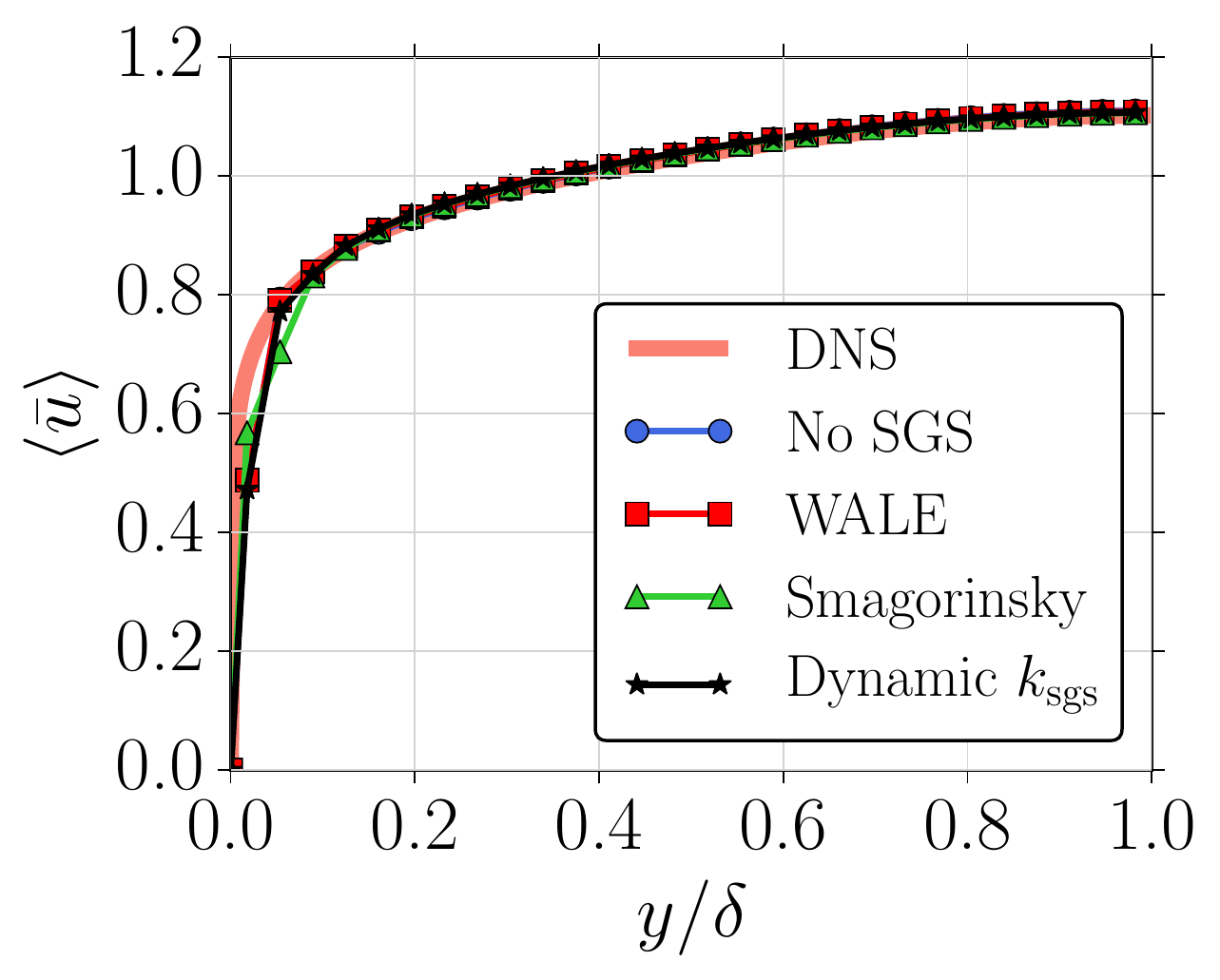}& \hspace{-0.52cm}
   \includegraphics[scale=0.43]{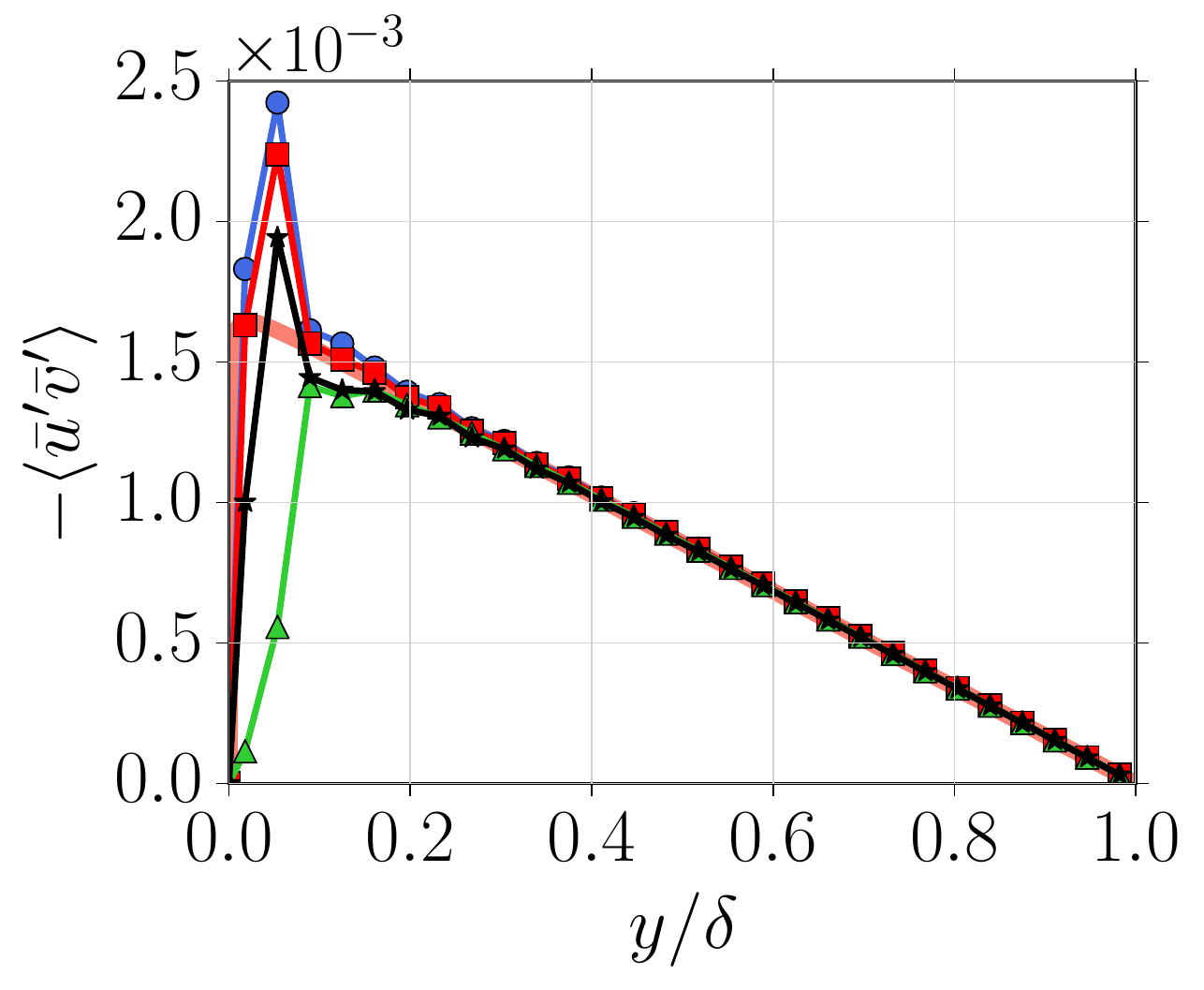}& \hspace{-0.54cm}
   \includegraphics[scale=0.43]{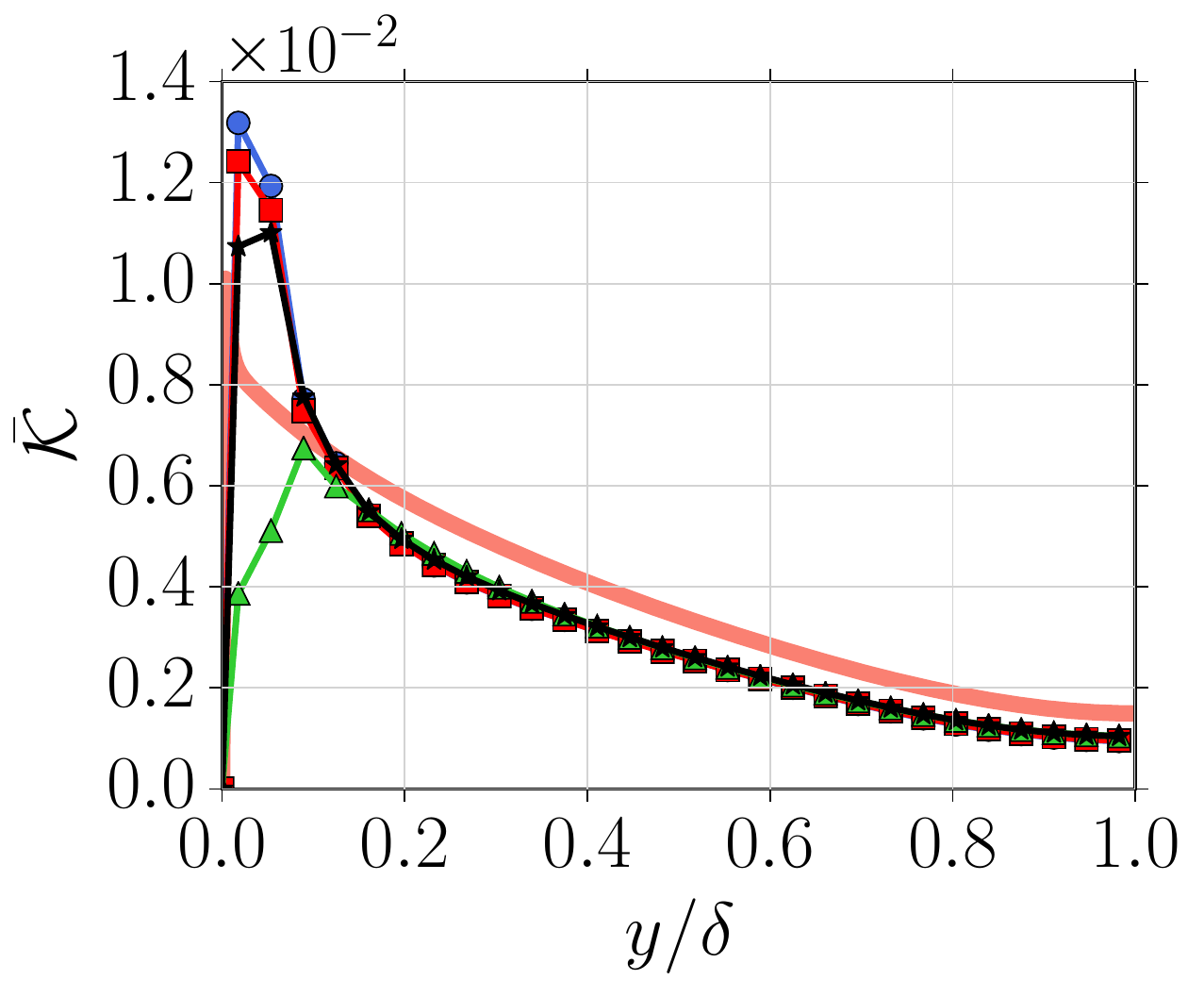} \\   
   {\small{(d)}} &    {\small{(e)}} &    {\small{(f)}}\\
   \end{tabular}
   \caption{Profiles of WMLES \revCom{obtained using} different SGS models and the Linear (top) and LUST (bottom) schemes. \revCom{For the details of simulations, see the caption of} \tab~\ref{tab:WMfootDuTau}.}
   \label{fig:WMfoot1}
\end{figure}

\begin{figure}[!h]
\centering
   \begin{tabular}{cc}
   \includegraphics[scale=0.4]{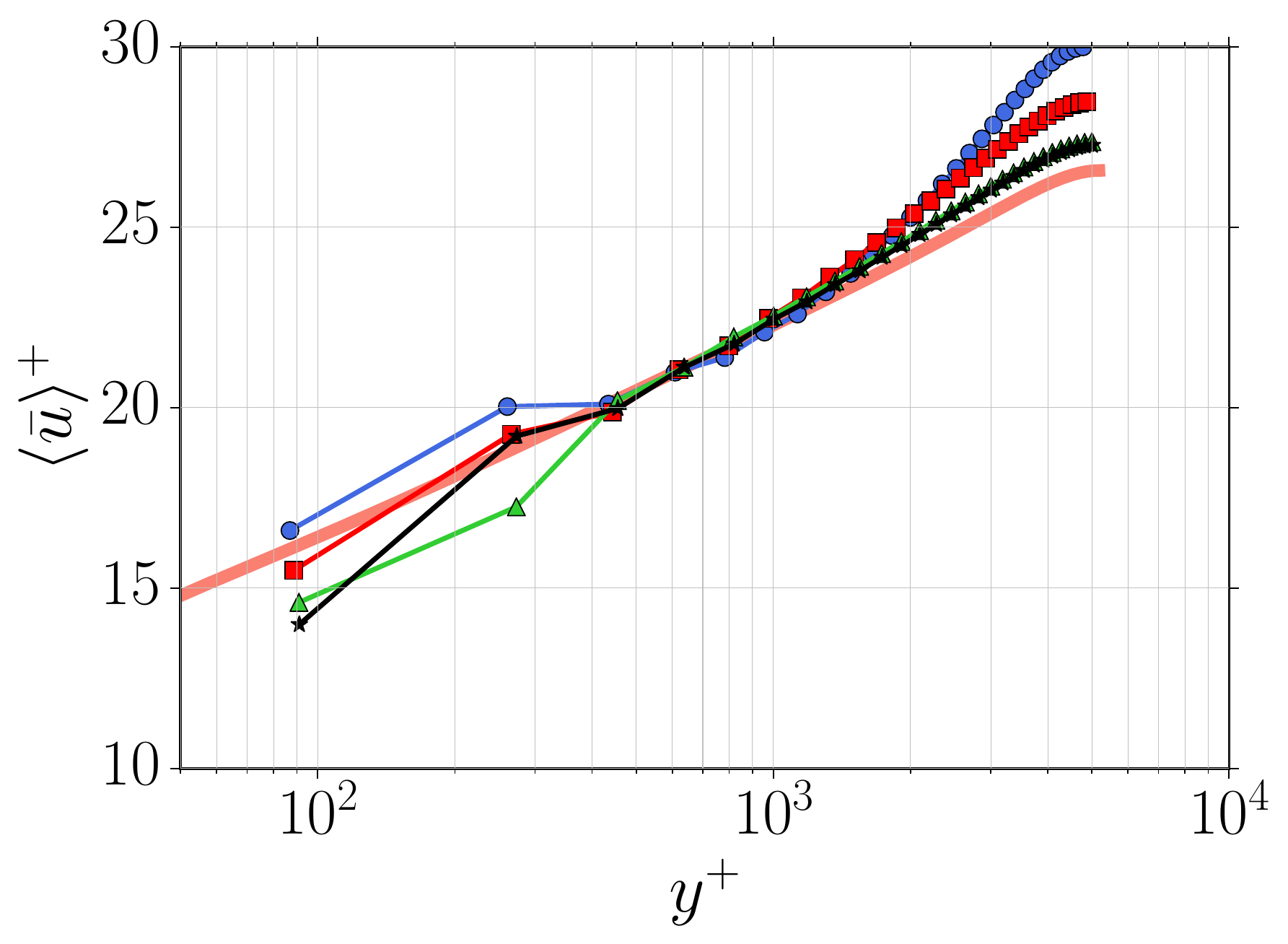}&
   \includegraphics[scale=0.4]{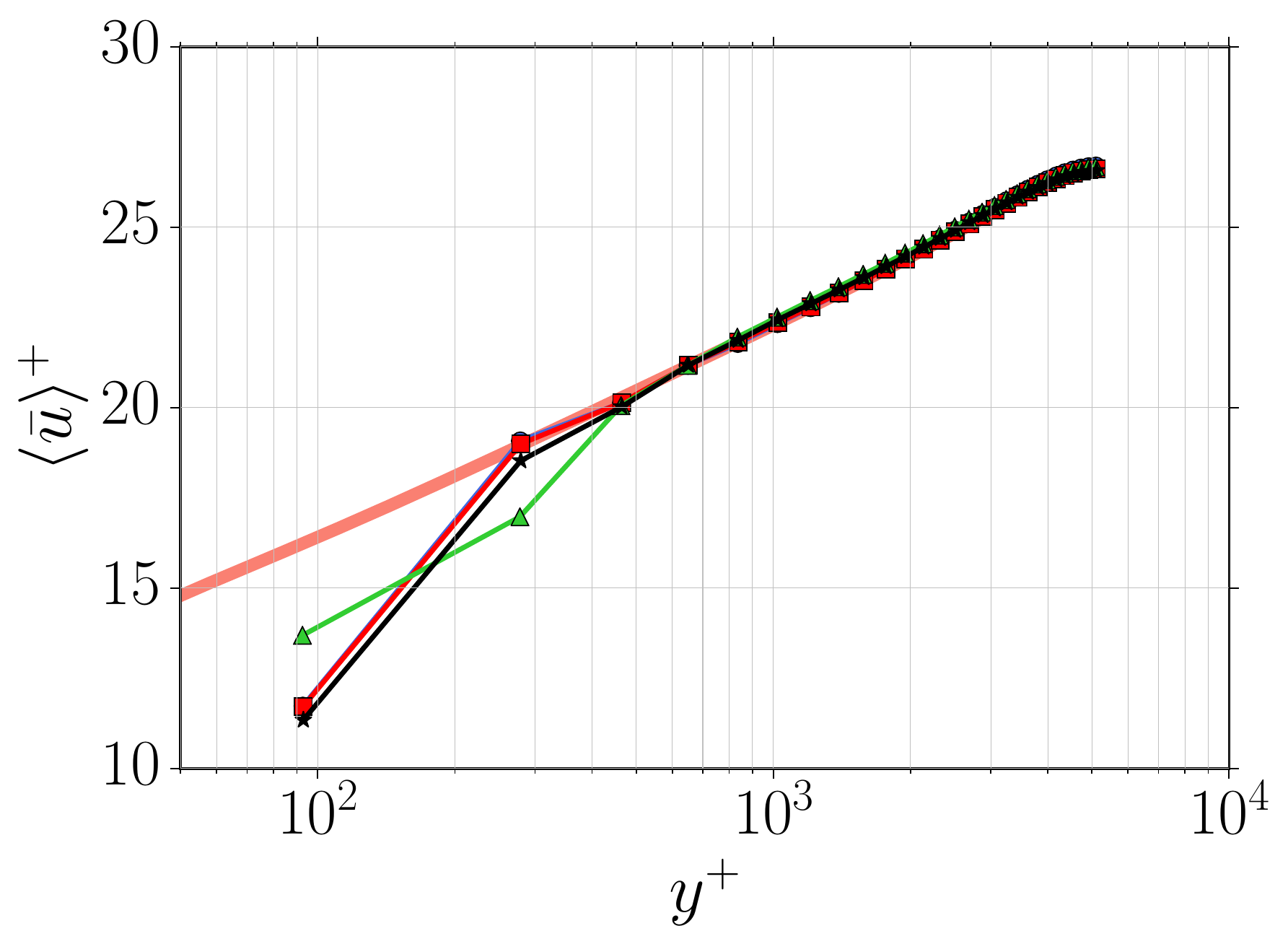}\\
   {\small{(a)}} &    {\small{(b)}} \\
   \end{tabular}
   \caption{Inner-scaled mean velocity profiles of the simulations in \fig~\ref{fig:WMfoot1}, obtained using the Linear (a) and LUST~(b) schemes. }
   \label{fig:WMfoot_uPls}
\end{figure}

The resulting profiles of $\U$, $-\buv$, and $\bk$ are shown in \fig~\ref{fig:WMfoot1}, along with the corresponding inner-scaled mean velocity profiles in \fig~\ref{fig:WMfoot_uPls}. 
The profiles in the outer layer, obtained in simulations using the LUST scheme, are insensitive to the choice of the SGS model. 
Moreover, corresponding profiles of~$\U^+$ are in excellent agreement with the DNS data of \cite{lee-moser:15}.
In contrast, it is clearly understood that, similar to~$\lbut$, the profiles of the Linear scheme are strongly affected by the SGS modeling. 
\tim{This is in line with the discussion above.}
In particular, the Smagorinsky and Dynamic $k_{\sgs}$-eqn models give the best predictions for~$\U$ in the outer layer, whether or not the wall model is used, compare \fig~\ref{fig:WMfoot1}(d) and \fig~\ref{fig:numSGSfoot}(d).
As the~$\lbut$ obtained when using these two \revCom{SGS} models are improved by wall modeling, the associated profiles of~$\U^+$ also become closer to the reference data, see \fig~\ref{fig:WMfoot_uPls}.

\tab~\ref{tab:WMfootDuTau} quantitatively reveals the connection between the error in the predicted~$\lbut$ and the error in~$\U$ far from the wall: \tim{the} higher the $\einf[\U]$, \tim{the higher the corresponding error in}~$\lbut$.
\tim{This in contrast to the weak connection between these two quantities \revCom{observed} when no wall modeling was employed.}
\tim{The introduced coupling is explained by the} high sensitivity of the wall model predictions to the velocity in the outer layer, as pointed out in \sect~\ref{sec:LSAGSA}.

Motivated by \tim{the above observations}, a mechanism explaining the connection between QoIs and numerical scheme, SGS model and wall modeling can be hypothesized.

First, compare the profiles of $\U$ across the channel in the presence and absence of wall modeling, respectively shown in \fig~\ref{fig:WMfoot1}(a,d) and \fig~\ref{fig:numSGSfoot}(a,d).
If the combination of the numerical scheme and SGS model provides enough dissipation, then upon refining the grid \rev{to $\nd\gtrsim 25$}, see \fig~\ref{fig:noWM_conv}, even without an accurate prediction of~$\lbut$, the mean velocity profile far from the near-wall region can be computed reasonably well. 
\tim{This is also evident from} the values of $\einf[\U]$ in \tab~\ref{tab:numSGSDuTau} for the Linear scheme combined with Smagorinsky and Dynamic $k_{\sgs}$-eqn models, and also for the LUST with all explicit SGS models. 
Once these appropriate numerical-SGS-modeling combinations are employed in WMLES, reasonably accurate velocity samples can be imported to the wall model, if the sampling point resides in the \rev{overlap region} of the TBL. 
Consequently, a wall model is able to predict the mean value of wall shear stress accurately.

\reva{Based on these observations, it is therefore possible to obtain an accurate outer-scaled profile of $\U$ \revCom{(\ie~normalized by $U_b$)} versus $y/\delta$ in the outer layer of the TBL, independent of the accuracy of the wall shear stress.
This can be an indication of the large independency of the outer layer dynamics from the near-wall region, as shown by \cite{flores:06,mizuno:13,chung:14}.
However, it must be noted that for the flows where the thickness of the TBL, $\delta$, is not defined by the geometry, the predictions for wall shear stress strongly affect this quantity. 
}

When the \rev{predicted} wall shear stress is imposed as the wall boundary condition via \eq~(\ref{eq:wmBC}), the velocity in the first cell center off the wall is accordingly modified. 
\tim{Since the total volumetric flux is enforced, this leads to a redistribution of the streamwise momentum across the whole channel.}
However, such influence turns out to be negligible \reva{when enough dissipation is added by the numerical scheme and SGS model, compare the values of $\einf[\U]$ in \tab~\ref{tab:numSGSDuTau} and~\tab~\ref{tab:WMfootDuTau}.} 
\rev{In contrast, for low-dissipation simulations}, the effect of the imposed wall shear stress \rev{on other LES QoIs} cannot be damped, and consequently, the shape of the $\U$ profile is significantly affected across the channel.
\rev{Due to this}, for the choice of the Linear scheme without SGS modeling, the inclusion of wall modeling \rev{even} increases $\einf[\U]$, see \tabs~\ref{tab:numSGSDuTau} and~\ref{tab:WMfootDuTau}.

\tim{In summary, on meshes typical of WMLES, the accuracy of the mean velocity profile is heavily reliant on extra turbulent mixing introduced either through numerical dissipation, explicit SGS modeling, or a combination of the two.
An accurate $\U$ profile leads to good performance of the wall model, but the backward coupling \rev{between these quantities} is negligible unless no SGS modeling is used, in which case the increased wall shear stress plays a key role in the redistribution of momentum across the channel.}

The described mechanism can be further extended to include the connection between $\lbut$ and the profiles of Reynolds shear stresses far from the wall.
To this end, the discussion in \sect~\ref{sec:qoiRelation} is followed.  
When accurate wall shear stress is predicted by the wall model and imposed at the wall, accurate total Reynolds shear stress is expected in the outer part of the TBL, see \eq~(\ref{eq:xMomChan}). 
\rev{What is important to realize is the contribution by the resolved stress $-\buv$. 
For all simulations either by the Linear or LUST schemes, the magnitude of $-\langle B_{xy}\rangle$ in the outer layer is observed to be much smaller than $-\buv$. This will be shown for the use of the LUST scheme further below, in \fig~\ref{fig:uvBalance_lust}.}

What drives the accuracy in development of fluctuations $\bu'$ and $\bv'$ and hence $-\buv$, is the combination of wall modeling, numerical method and SGS modeling.
\rev{For the LUST scheme, upon accurate predictions of~$\lbut$, accurate profiles of $-\buv$ in the outer layer are obtained independent of the SGS model, see \tab~\ref{tab:WMfootDuTau} and \fig~\ref{fig:WMfoot1}(e). }
\rev{In contrast,} the $-\buv$ profiles of the Linear scheme depend on the SGS model, as clearly discernible from \fig~\ref{fig:WMfoot1}(b).
Without an SGS model,~$\lbut$ is under-predicted, therefore an under-prediction of \revCom{the total Reynolds shear stress} $-\uv$ in the outer layer is expected. 
However, the resolved stress~$-\buv$ is found to be over-predicted \rev{due to spurious fluctuations of $\bu$ and $\bv$.}
Therefore, the extra level of fluctuations should be taken away by SGS model.
\rev{All the employed SGS models are effective for this purpose, however}, the performance of the Smagorinsky and Dynamic $k_{\sgs}$-eqn models is \rev{superior because the profiles of $\U$ and $-\buv$ far from the wall are simultaneously predicted accurately}.

Similar to the simulations with no wall modeling, the resolved TKE, $\bk$, is less sensitive to different types of explicit SGS models, for both Linear and LUST schemes. 
Due to wall modeling, the prediction of $\bk$ in the outer layer can be improved, specifically when the LUST scheme is used, compare \fig~\ref{fig:WMfoot1}(c,f) with \fig~\ref{fig:numSGSfoot}(c,f).
As shown in \fig~\ref{fig:WMfoot1}(c), up to the wake region, the Linear scheme without an explicit SGS model seems to return accurate resolved TKE. 
However, as already discussed, this improvement is due to the spurious excessive level of the resolved fluctuations.

Combining the observations and discussions in this section and the previous one, the following remarks can be made.
Depending on the combinations of the numerical scheme and SGS model, specific pattern in the mean streamwise velocity profile is achieved that is preserved if wall modeling is used. 
In particular, for the LUST scheme with either no explicit SGS model or the WALE and Dynamic $k_{\sgs}$-eqn models, the velocity in the first cell center is significantly under-predicted. 
Upon using Smagorinsky with either the LUST or Linear scheme, the velocity in the first two off-wall cell centers is under-predicted.
For combinations of the Linear scheme with less dissipative SGS models, the values in the first few cell centers are erroneous.
\rev{These observations will be referred to in the next sections.}

\rev{Another important conclusion can be made for simulations with the LUST scheme on meshes typical of WMLES. 
Without wall modeling, the choice of SGS model affects the QoIs predicted by LES.
However, when wall modeling is involved, the role of SGS models on the accuracy of the QoIs becomes less significant.
}

\subsubsection{Spurious overshoots in the profiles of velocity fluctuations}\label{sec:spuriousOvershoot}
A common issue when a coarse grid resolution is used for simulation of wall-bounded turbulence is that spurious overshoots may appear in the second-order statistical moments of the velocity. 
These can be observed whether or not wall modeling is used, see \figs~\ref{fig:numSGSfoot} and \ref{fig:WMfoot1}.
The issue is known and discussed in several researches, see \eg~Refs.~\cite{mason:92,baggett:98,cabotMoin:00,nicoud:01,piomelli:03,brasseur:10} and the references therein, and is shortly \tim{analyzed} here in order to \tim{illustrate how SGS modeling can be employed to rectify it}.

\tim{The origin of the spurious fluctuations lies in the multi-scale nature of TBLs.}
According to the streamwise momentum balance \eq~(\ref{eq:xMomChan}), in the near wall region the mean shear $\dd\U/\/\dd y$ is the main contributor to the shear stress \tim{and balances out} the mean pressure gradient. 
As discussed in \sect~\ref{sec:qoiRelation}, in the outer layer, the mean shear is totally dominated by the turbulent stress. 
\rev{At the wall-distance where the near-wall region and outer layer meet, }the shear stresses need to match. 
However, \tim{when the computational grid is coarse} close to the wall, here $\yd\lesssim 0.1$,  overshoots in the plots of $\bk$ and $-\buv$ profiles may appear.

Through computational experiments, Baggett \cite{baggett:98} showed that for coarse grid resolutions (in the wall-parallel directions), artificial horizontal streaks are generated that are self-sustained and lead to a spurious increase in the simulated velocity fluctuations. 
As a consequence, instantaneous~$\bar{u}'\bar{v}'$ and hence averaged~$\buv$ increase unphysically. 
The latter leads to extra production of the resolved TKE. 
To prevent the creation of such streaks, adding stochastic fluctuations to the SGS model (stochastic backscatter) has been successfully tested, see Refs. \cite{mason:92,baggett:98,piomelli:03}.
Using a highly dissipative SGS model can be another remedy.
The improving effect of this method is confirmed observing the overshoots in $-\buv$ and $\bk$ profiles are removed when \tim{the} Smagorinsky model is used with or without wall modeling, see \figs~\ref{fig:numSGSfoot} and \ref{fig:WMfoot1}. 
It is also noteworthy that the Dynamic $k_{\sgs}$-eqn model results in a slightly smaller overshoot as compared to WALE, showing the improving effect of dynamic adjustment of the SGS stress on the velocity fluctuations.

\begin{figure}[!h]
\centering
   \begin{tabular}{cc}
   \includegraphics[scale=0.35]{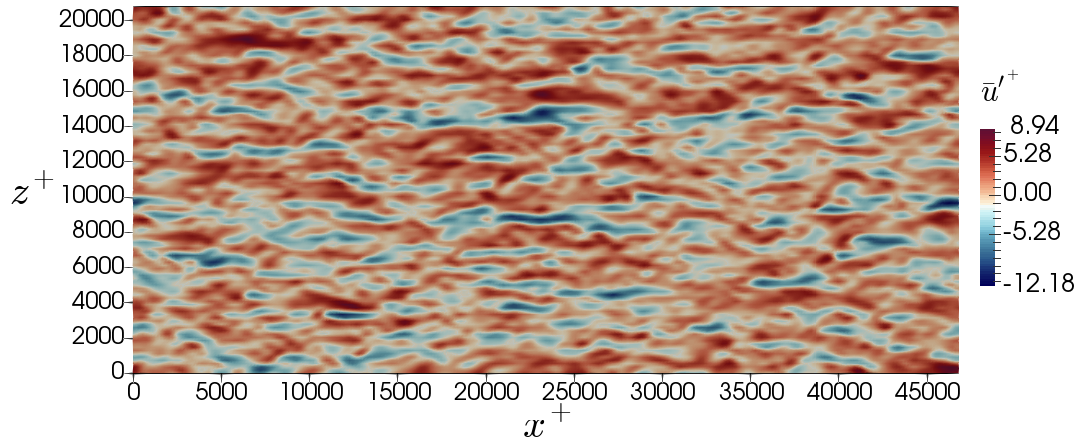}&
   \includegraphics[scale=0.35]{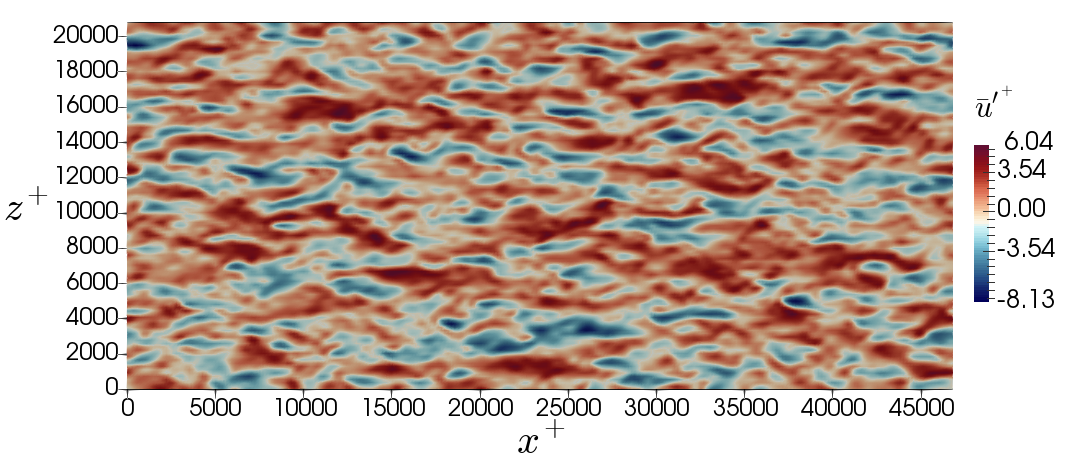}\\      
   {\small{(a)}} &    {\small{(b)}} \\
   \includegraphics[scale=0.35]{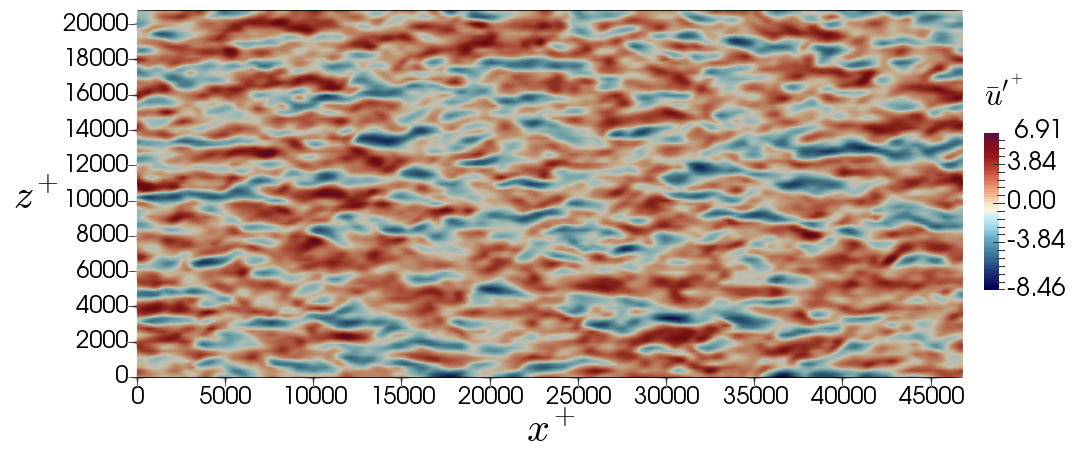}&
   \includegraphics[scale=0.35]{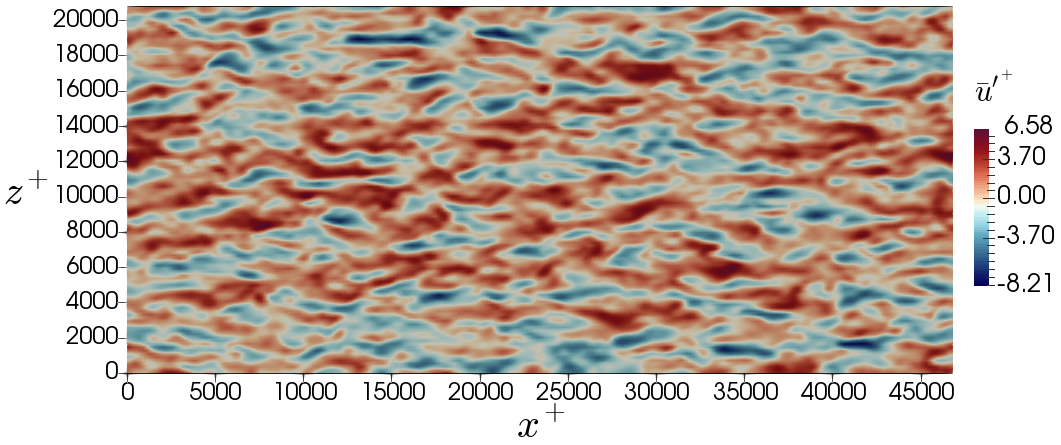}\\      
   {\small{(c)}} &    {\small{(d)}} \\   
   \includegraphics[scale=0.35]{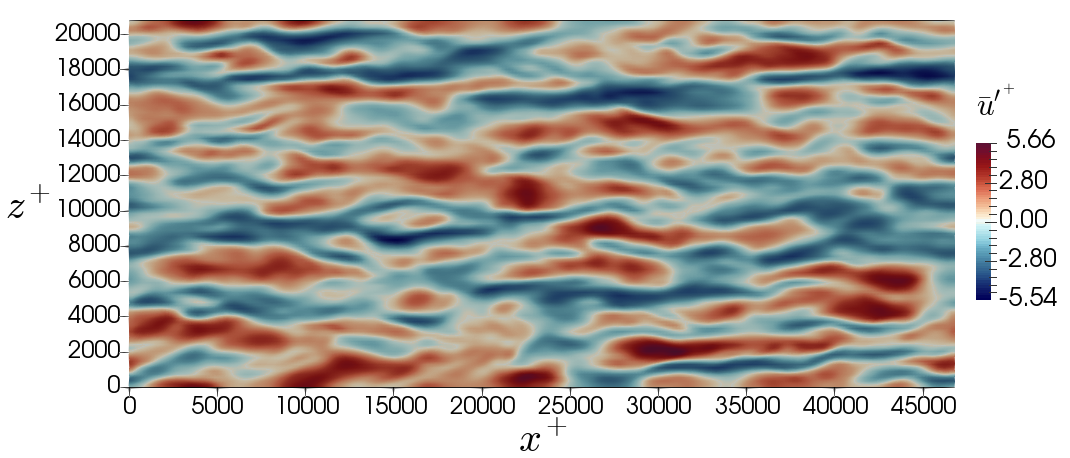}&
   \includegraphics[scale=0.35]{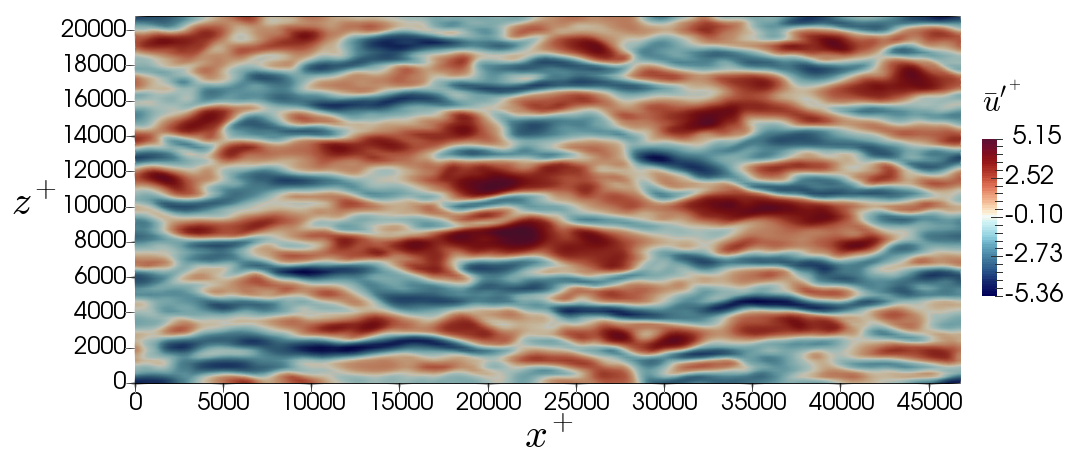}\\      
   {\small{(e)}} &    {\small{(f)}} \\      
   \end{tabular}
   \caption{Snapshots of normalized instantaneous streamwise velocity, $\bu'^+ = \bu'/\lbut$,  at $y/\delta = 0.05$ ($y^+=260$) obtained by the LUST scheme used with no wall model (left: a,c,e), and with the Spalding wall model~(right:~b,d,f). Simulations are performed with no SGS (top: a,b), WALE (middle: c,d), and  Smagorinsky (bottom: e,f) models, and correspond to those in the bottom row of \figs~\ref{fig:numSGSfoot} and~\ref{fig:WMfoot1}.}
   \label{fig:uContour_LUST_y05}
\end{figure}

\begin{figure}[!h]
\centering
\includegraphics[scale=0.6]{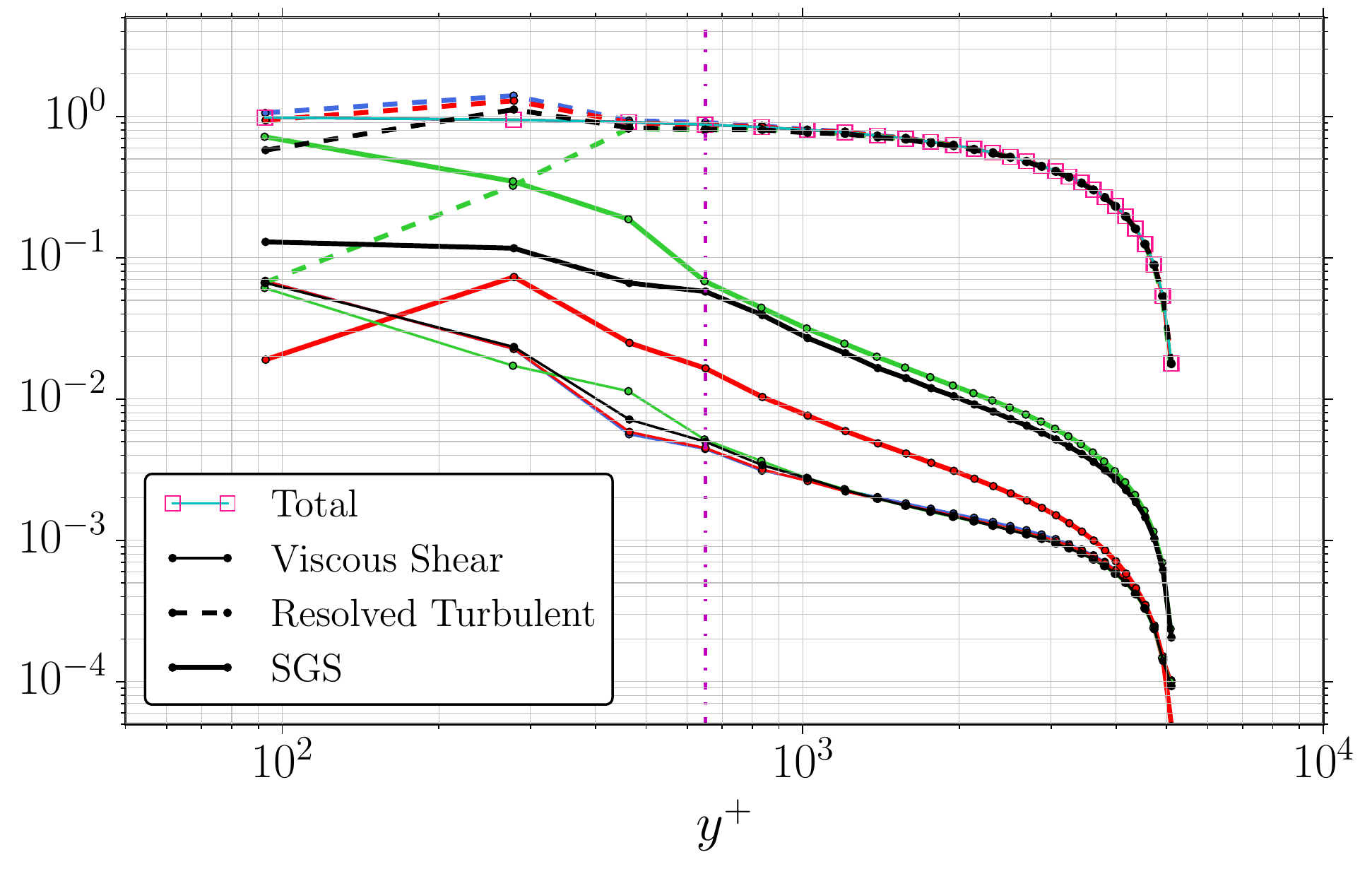}
\caption{Terms in \eq~(\ref{eq:xMomChan_inScal}) plotted across the channel half-height for the WMLES with the LUST scheme associated with the bottom row in \fig~\ref{fig:WMfoot1}. 
The terms are total stress $(1-\eta)$, viscous shear stress $\dd\U^+/\dd y^+$, resolved Reynolds shear stress~$-\buv^+$, and SGS shear stress $-\langle B_{xy}\rangle ^+$. The vertical dashed-dotted line specifies $y^+=650$ ($\frth$~cell center) at which LES velocity is sampled for wall modeling. Colors specify SGS models:
no SGS (blue), WALE (red), Smagorinsky (green), and dynamic $k_{\sgs}$-eqn model (black).}
\label{fig:uvBalance_lust}
\end{figure}

The impact of \tim{the} SGS model can be also noted \tim{in} the snapshots of streamwise velocity \tim{when the LUST scheme is used, see} \fig~\ref{fig:uContour_LUST_y05}.
Comparing the figures in each row, it is evident that the wall modeling has negligible effect on the patterns of the velocity fluctuations. 
Moreover, using \tim{the} WALE model does not make any major difference in the structures compared to the case without any SGS model. 
But, \tim{the} Smagorinsky model \rev{damps the structures} so substantially that the streaks become thick in both streamwise and spanwise directions. 
Consequently, for a given grid resolution, these structures are resolved better and the spurious overshoots in the velocity fluctuations diminish.

\tim{The damping mechanism of excessive SGS viscosity} \rev{can be better understood from}~\fig~\ref{fig:uvBalance_lust}.
The \tim{exhibited results are from} WMLES performed with the LUST scheme\tim{, see also} \fig~\ref{fig:WMfoot1} \tim{and} \tab~\ref{tab:WMfootDuTau}. 
\rev{For the $\thrd$ cell center and above}, the resolved Reynolds shear stress by far has the most dominant contribution to the total wall shear stress, independent of the SGS model. 
The contribution of the viscous shear stress is small across the channel and is approximately the same for all SGS models.   
Between the wall and the sampling height, when no SGS model is used and also for \tim{the} WALE model, excessive resolved stress is produced.
Contrary to these, for the Smagorinsky model, the resolved Reynolds shear stress is reduced, which is compensated by the excessive amount of SGS stress to keep the balance (\ref{eq:xMomChan_inScal}) valid.
\revCom{In \ref{app:xMomBalance_Converge}, the validity of \eq~(\ref{eq:xMomChan_inScal}) for the WMLES of \fig~\ref{fig:uvBalance_lust} is briefly discussed.}

\subsection{Sensitivity of WMLES to grid resolution}\label{sec:nTests}
So far, the \rev{important} role of the numerical scheme and SGS model in determining the accuracy of WMLES~velocity profile $\U$, \rev{as well as, the pivotal role of wall modeling in correctly predicting~$\lbut$} has been demonstrated. 
The aim of this section is to evaluate how the flow QoIs \tim{react} to the variation of the grid resolution, \tim{which} is the main controller of the numerical and modeling errors in implicitly-filtered LES.

The grid is assumed to be isotropic in all spatial directions. 
However, the influence of the grid anisotropy on the errors in the QoIs will be discussed later in \sect~\ref{sec:gridAnisot}.
The performance of different combinations of the Linear and LUST schemes with the  WALE, Smagorinsky and Dynamic $k_{\sgs}$-eqn models is examined.
The Spalding law (\ref{eq:spalding}) with parameters $\kappa=0.395$ and $B=4.8$ is used as the wall model with the velocity sampled from $\hd=0.125$ and $0.25$.  
\tim{Note that} for different resolutions these heights are associated with different cell centers. 
Relying on the tests in \sect~\ref{sec:hTests}, the conclusions below are also valid for $\hd>0.25$.

\begin{figure}[!t]
\centering
   \begin{tabular}{cc}
   \includegraphics[scale=0.43]{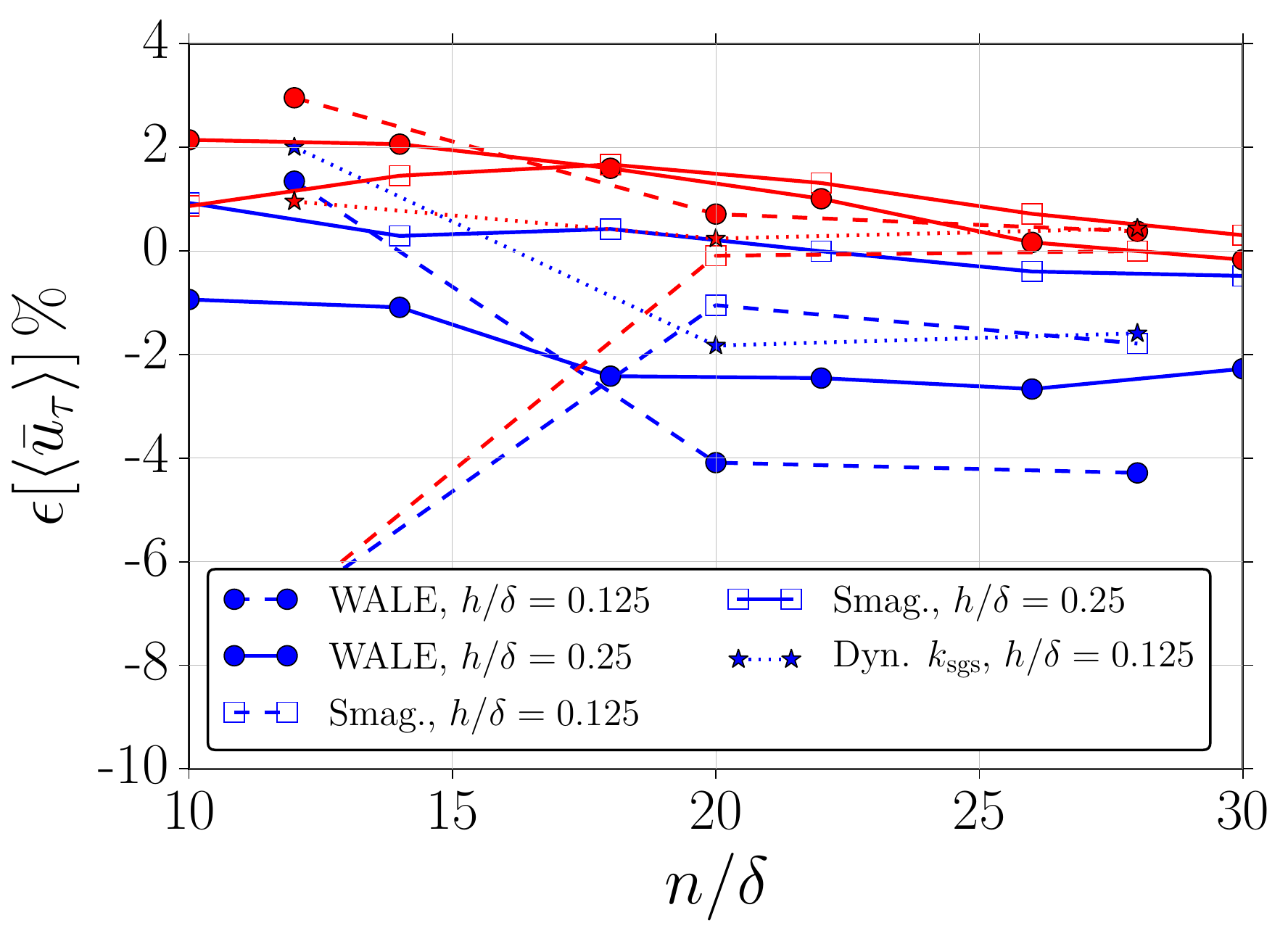}   &
   \includegraphics[scale=0.43]{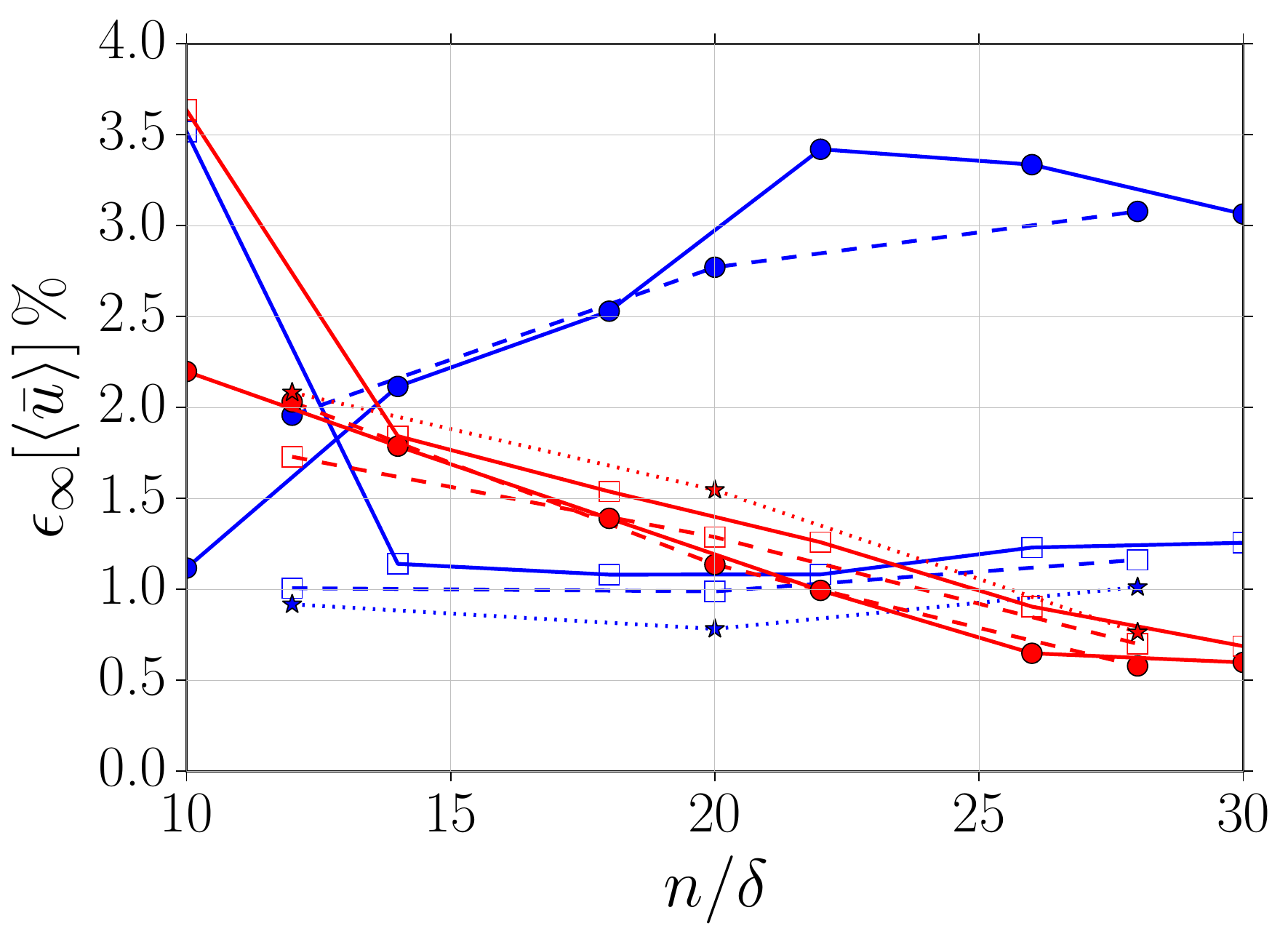}   \\   
   {\small{(a)}} &    {\small{(b)}} \\   
   \includegraphics[scale=0.43]{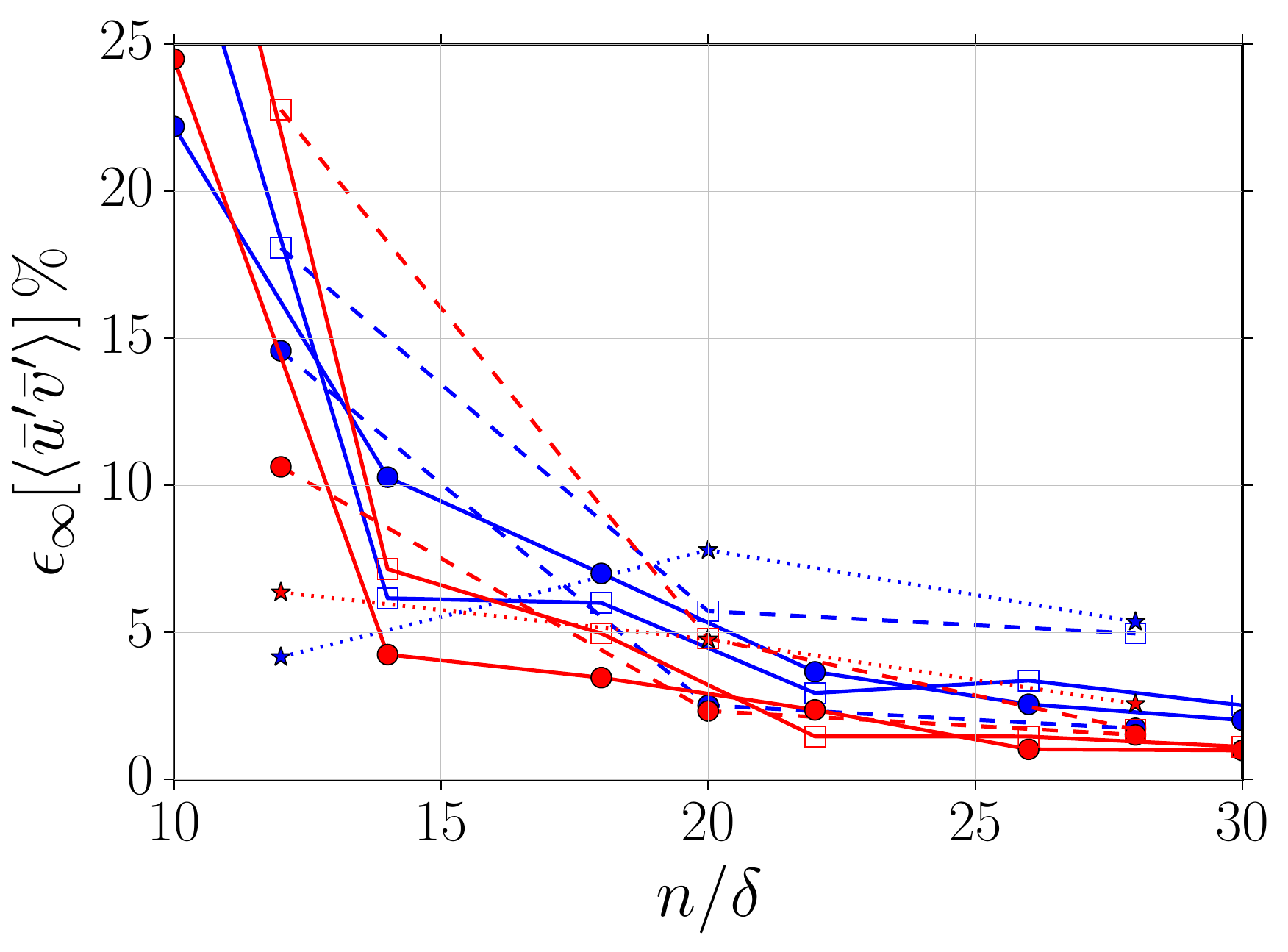}   &
   \includegraphics[scale=0.43]{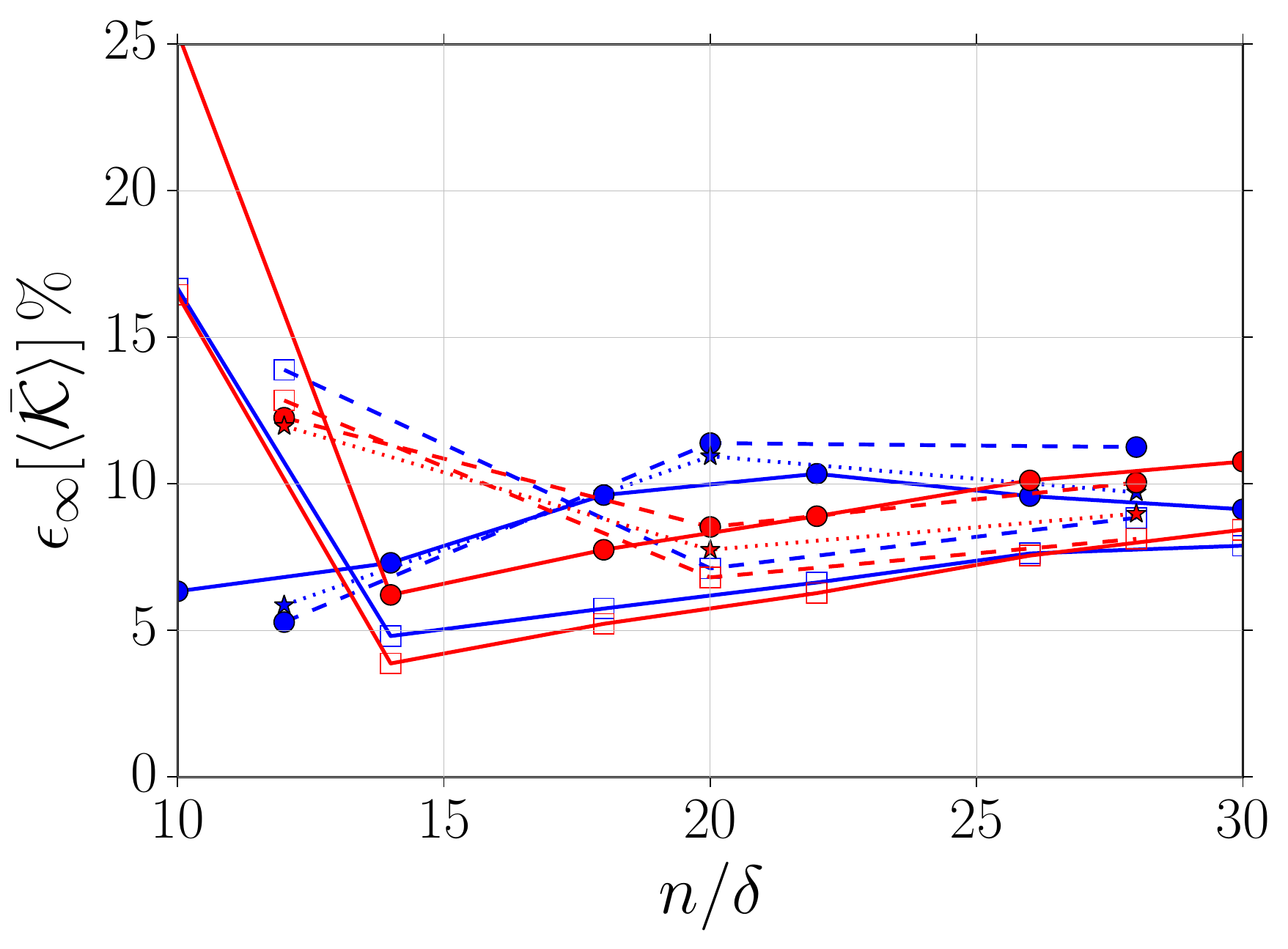}   \\      
   {\small{(c)}} &    {\small{(d)}} \\   
   \end{tabular}
\caption{Variation of \revCom{the error in the WMLES QoIs} with grid resolution $\nd$. The Spalding law is used with $\kappa=0.395$ and $B=4.8$. Results of the Linear and LUST schemes are plotted in blue and red, respectively.}
\label{fig:convtest_n} 
\end{figure}

Plotted in \fig~\ref{fig:convtest_n} are the errors in $\lbut$, $\U$, $\buv$, and $\bk$ obtained by adopting different grid resolutions~$\nd$.
The mean velocity in the outer layer is expected to remain almost insensitive to the near-wall treatment.
This is confirmed by comparing the graphs in \fig~\ref{fig:convtest_n}(b) and \fig~\ref{fig:noWM_conv}(b).
In particular, the value of $\einf[\U]$ increases with $\nd$ for the combination of Linear scheme with WALE model, whether or not \tim{a} wall model is used.
This unexpected behavior can be due to the numerical artefacts of the scheme, as formerly experienced in WRLES, see \cite{salehPoF:18}. 
Although using \tim{the} Smagorinsky \tim{and} \tim{the} Dynamic $k_{\sgs}$-eqn models improves the $\einf[\U]$ of the Linear scheme, this error becomes nearly constant and does not reduce further with increasing $\nd$.
Contrary to the Linear scheme, for the WMLES with the LUST scheme, $\einf[\U]$ monotonically decreases with $\nd$. 
This is in agreement with the trend in \fig~\ref{fig:noWM_conv} in the absence of wall modeling.

According to \fig~\ref{fig:convtest_n}(a), by refining the grid \rev{$\epsilon[\lbut]$ seem to tend to fixed values}.
For the simulations \tim{using} the LUST scheme, the \rev{$\lbut$ converges to} the reference DNS value, but for the Linear scheme, the value of $\lbut$ \rev{at high $\nd$} depends on the SGS model and the sampling height. 
As a consequence of the sampling from a less accurate velocity field, the use of \tim{the} WALE SGS model with the Linear scheme results in the most inaccurate $\lbut$.
The defect of the Linear scheme is minimized by adding Smagorinsky model and sampling from $\hd=0.25$.

Almost for all simulations, $\einf[\buv]$ decreases with $\nd$ and reaches a plateau for $\nd\gtrsim 25$. 
These constant values are lower for the LUST scheme which have more accurate $\lbut$ compared to the Linear~scheme. 

\begin{figure}
\centering
   \begin{tabular}{cc}
   \includegraphics[scale=0.65]{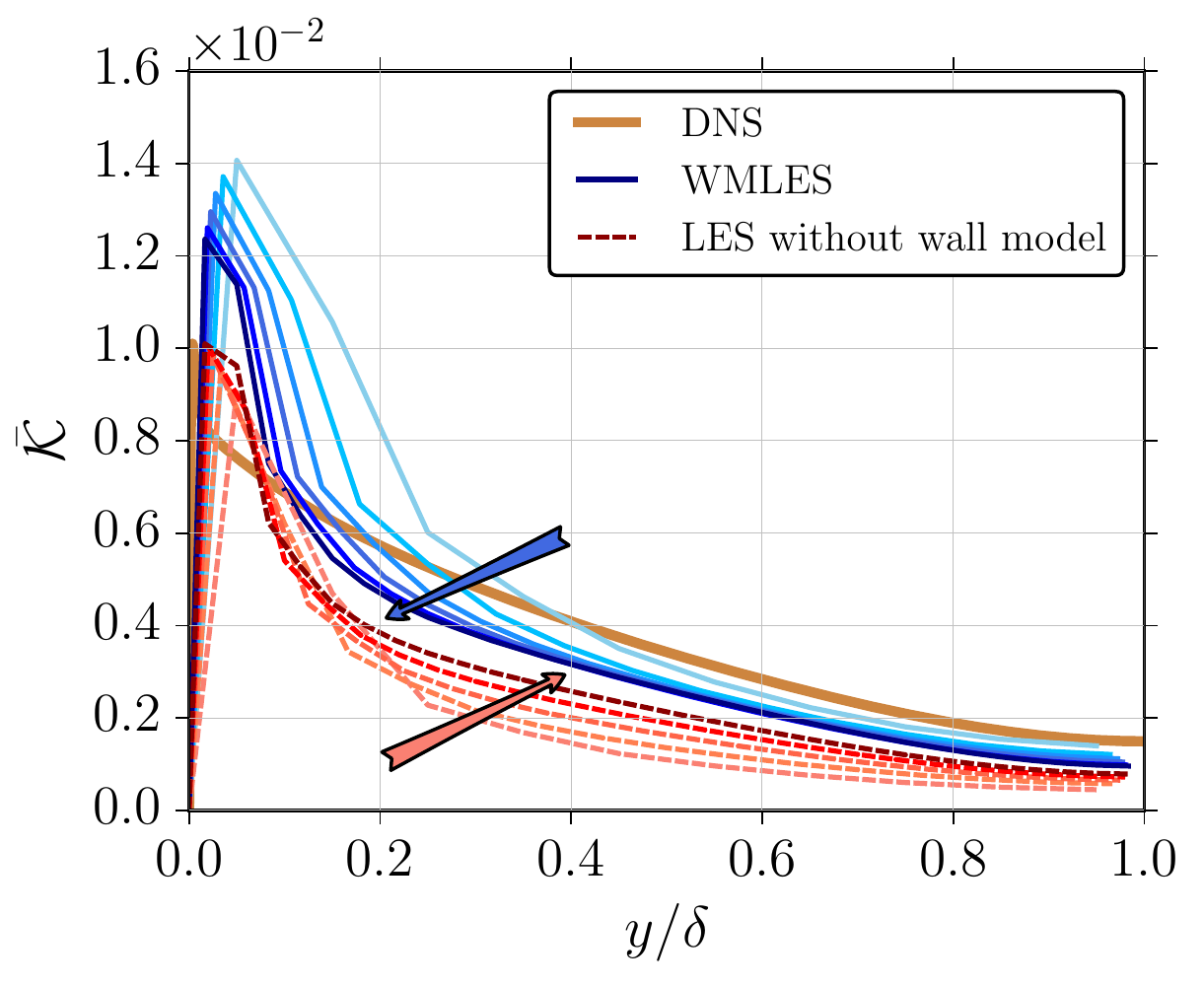} &
   \includegraphics[scale=0.65]{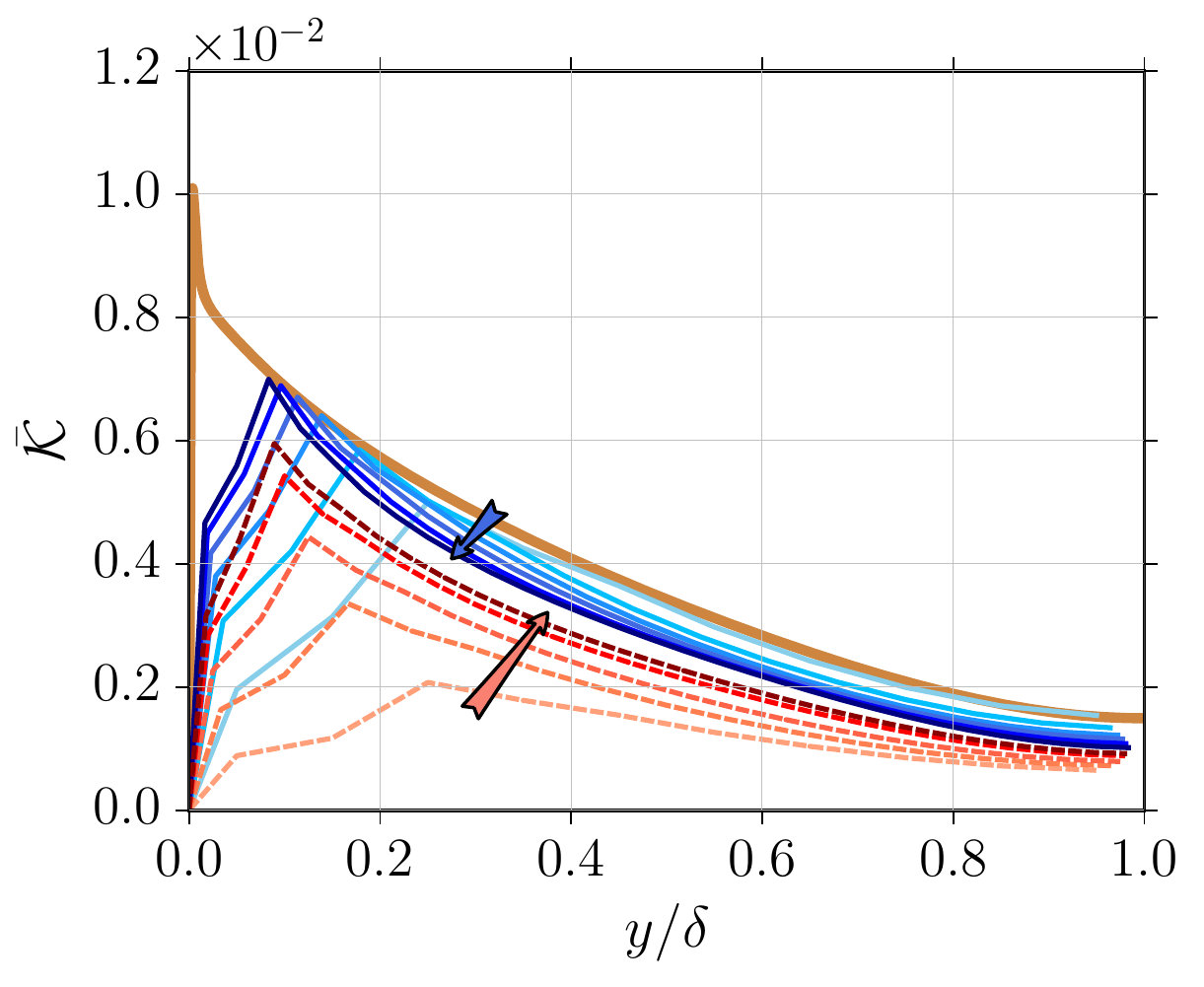} \\
   {\small{(a)}} &    {\small{(b)}} \\       
   \end{tabular}
   \caption{Profiles of resolved TKE across the channel resulted from WMLES (solid lines) and LES with no wall modeling (dashed lines). Simulations are carried out using the LUST scheme with WALE (a) and Smagorinsky (b) models. The grid resolution $\nd$ increases \revCom{by~5} from $10$ (lightest color) to $30$ (darkest color) in direction of the arrows.}\label{fig:kProfs_nTest}
\end{figure}

Compared to other QoIs, the error in the resolved TKE far from the wall has a different trend.
In fact,~$\einf[\bk]$ may increase by a few percent in spite of refining the grid, see \fig~\ref{fig:convtest_n}(d).    
This is opposite to the monotonous reduction of $\einf[\bk]$ for the LES without wall modeling previously shown in \fig~\ref{fig:noWM_conv}(d).
Despite this, for any \tim{given} combination of numerical scheme and SGS model and at any grid resolution, using wall modeling lowers the error in $\bk$ in the outer layer. 
In particular, \tim{consider} \fig~\ref{fig:kProfs_nTest} representing the profiles of $\bk$ for the LUST scheme employed along with \tim{the} WALE and Smagorinsky SGS models. 
Without wall modeling, increasing $\nd$ results in resolving more fluctuations over $\yd\ \gtrsim 0.2$, although the near-wall resolution is still coarse. 
Even for $\nd=30$ the resulting $\bk$ profile over $\yd\gtrsim 0.2$ is less accurate than that by WMLES at the same resolution. 

Interestingly, at the coarse resolution $\nd=10$ and over $\yd\ \gtrsim 0.2$, the WMLES profile of $\bk$ for both SGS models is the closest one to the DNS data.
By increasing $\nd$, the $\bk$ profiles in the outer layer become more deviated from the DNS data.
A possible explanation for this unexpected behavior can be stated as follows. 
For WMLES with the WALE model, the excessive generation of the velocity fluctuations, specially in the streamwise direction, results in the spurious overshoots near the wall.
Subsequently, the excessive fluctuations may propagate to the outer layer. 
When the grid resolution increases, the overshoots decrease and consequently the spurious fluctuations in the outer layer are lowered.  
In the case of using Smagorinsky model, the damping of fluctuations is so high that no overshoot appears across the channel. 
When increasing the resolution, the peak of $\bk$ moves toward the wall and the values in the outer layer are adjusted \revCom{(lowered)} accordingly.

What is also noteworthy regarding \fig~\ref{fig:kProfs_nTest} is that, neither in the presence nor the absence of wall modeling, the profiles of the resolved TKE  become entirely over-predicted across the channel. 
The same behavior is also observed when using the Linear scheme (not shown here).
This is despite applying the no-slip boundary condition for the velocity at the wall, that was formerly reported to lead to over-prediction of the streamwise velocity fluctuations in coarse-grid LES,~see~\eg~\cite{bae:18}.

Summarizing this section, the influence of the grid resolution on WMLES is found to be tightly connected to the numerical scheme and SGS model.
Also, the influence is \rev{more complex than what is the case for} LES without wall modeling. 
The error in the resolved TKE profile far from the wall slightly increases with grid resolution. 
For inappropriate combinations such as the Linear scheme and WALE model, $\einf[\U]$ does not decrease with $\nd$. 
Consequently, the errors in the velocity propagate to the predictions of \tim{the} wall model for the wall shear velocity. 
In contrast, the errors in the velocity profiles and hence $\epsilon[\lbut]$ associated with the LUST scheme are reduced with increasing the grid resolution.
Eventually for $\nd\gtrsim 25$, these errors approximately tend to constant values. 
Based on these observations, the grid resolution $\nd\approx 25$-$30$ is recommended as a best practice guideline for WMLES in the settings of the present study.
It is emphasized that the optimal choice for grid resolution depends on the particular solver used for WMLES, for instance see the values given in \cite{kawai:12,lee:13}.

\subsection{Sensitivity of WMLES to the velocity sampling height}\label{sec:hTests}
In this section, the emphasis is on the influence of the sampling height $h$, from which the velocity samples are fed into the wall model. 
\rev{
The aim is to possibly derive guidelines for \tim{the} appropriate sampling height, given the numerical scheme and SGS model.}

\begin{figure}[!h]
\centering
   \begin{tabular}{cc}
   \includegraphics[scale=0.44]{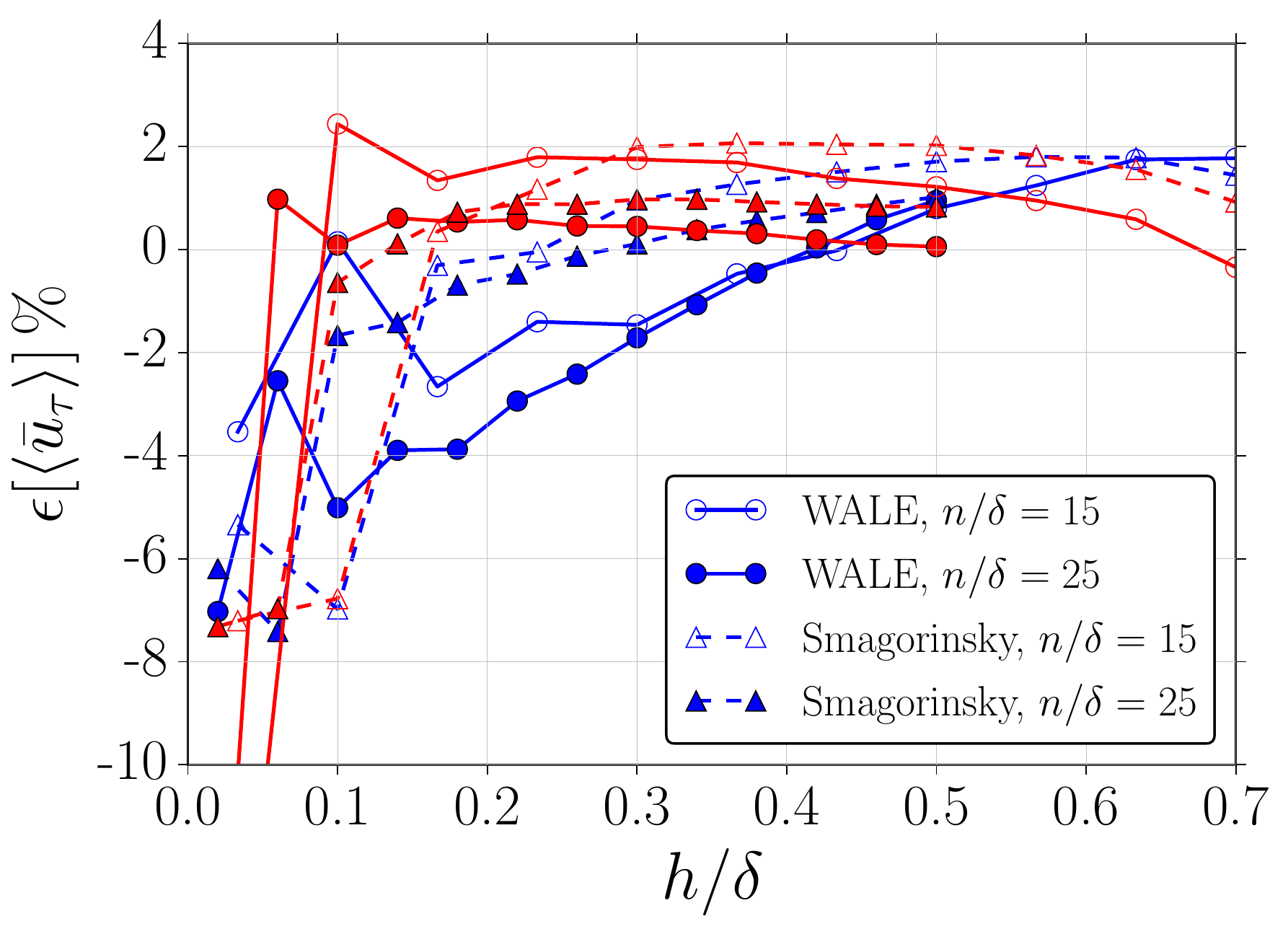}   &
   \includegraphics[scale=0.44]{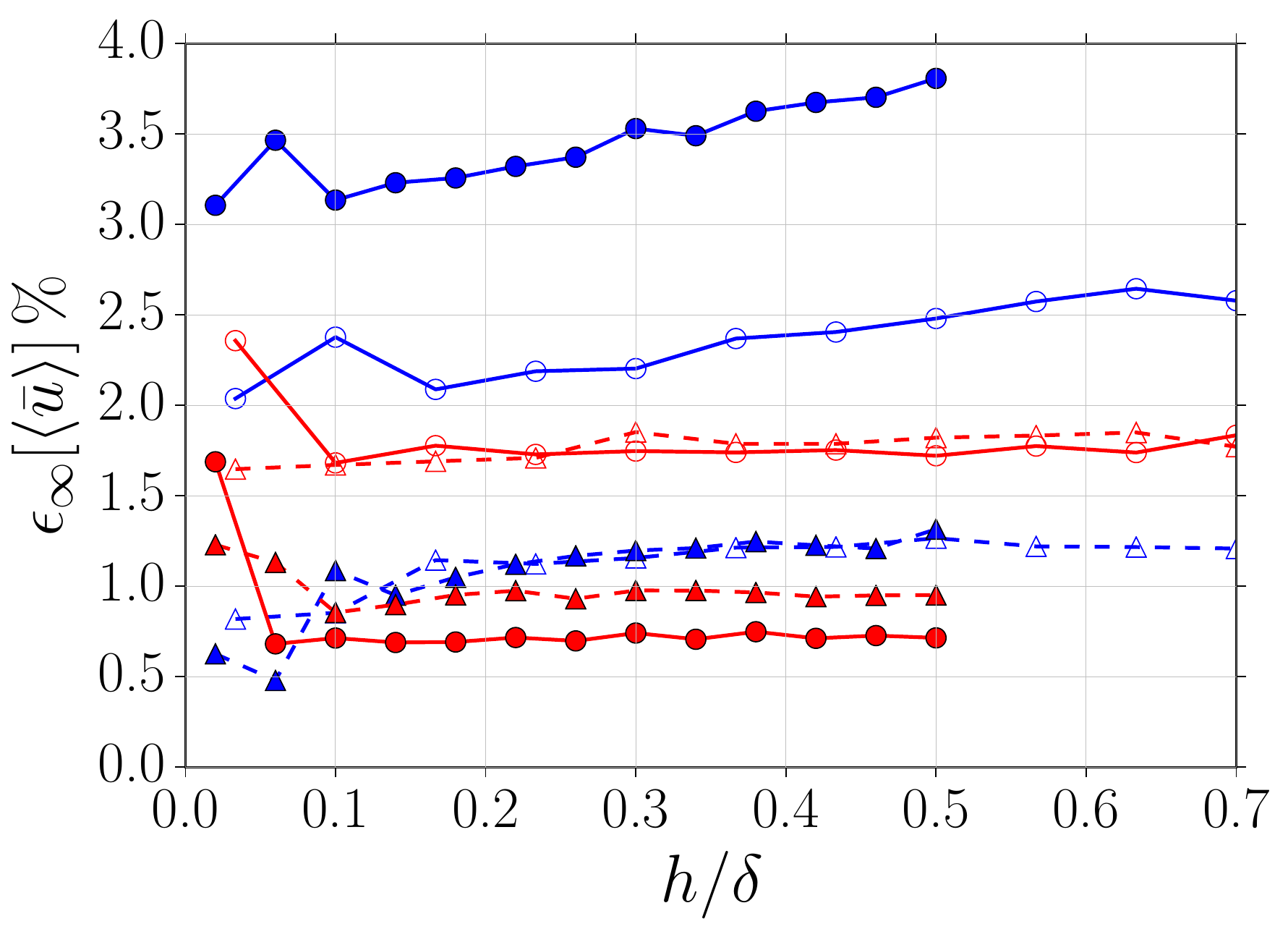}   \\   
   {\small{(a)}} &    {\small{(b)}} \\   
   \includegraphics[scale=0.44]{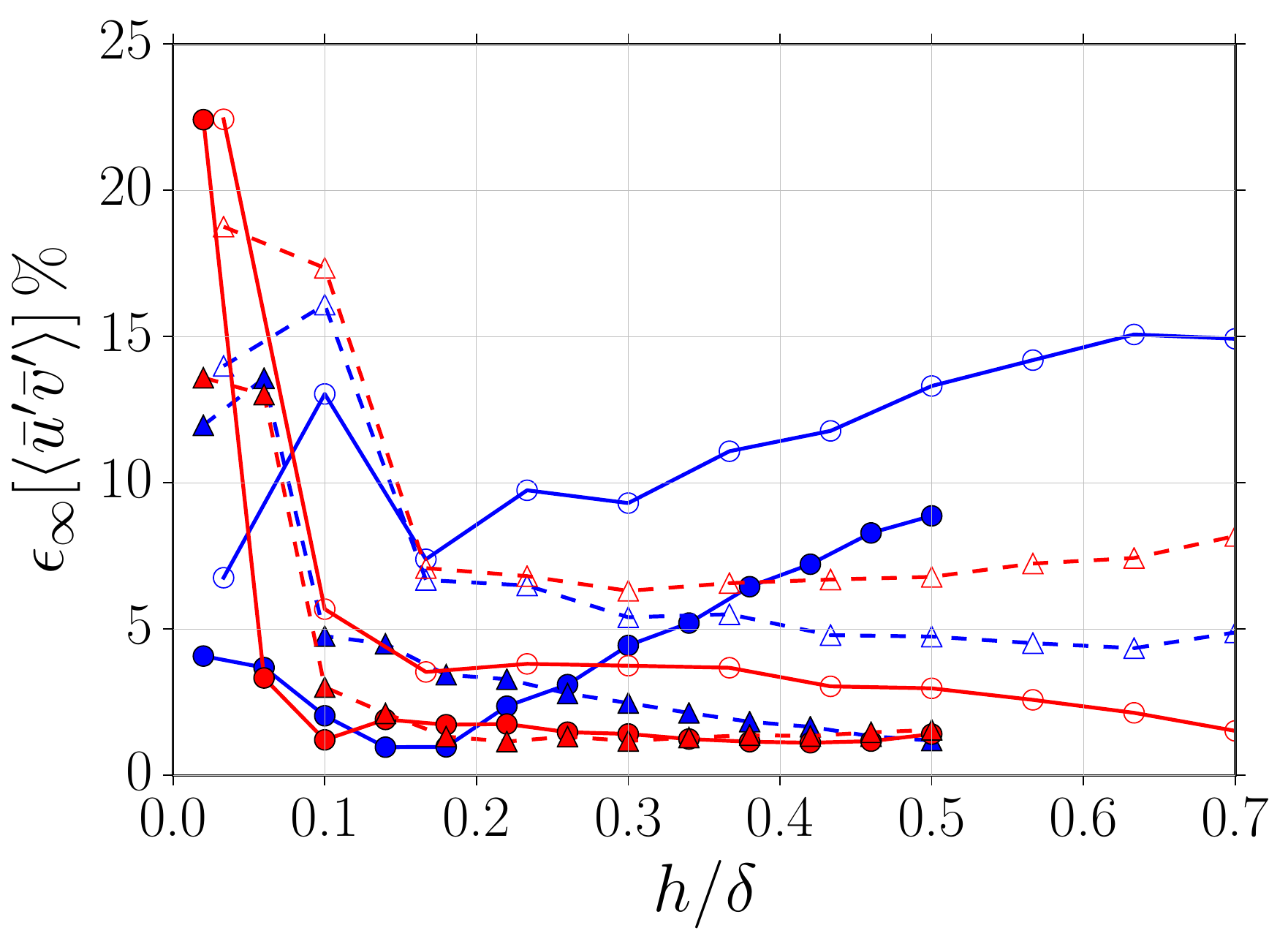}   &
   \includegraphics[scale=0.44]{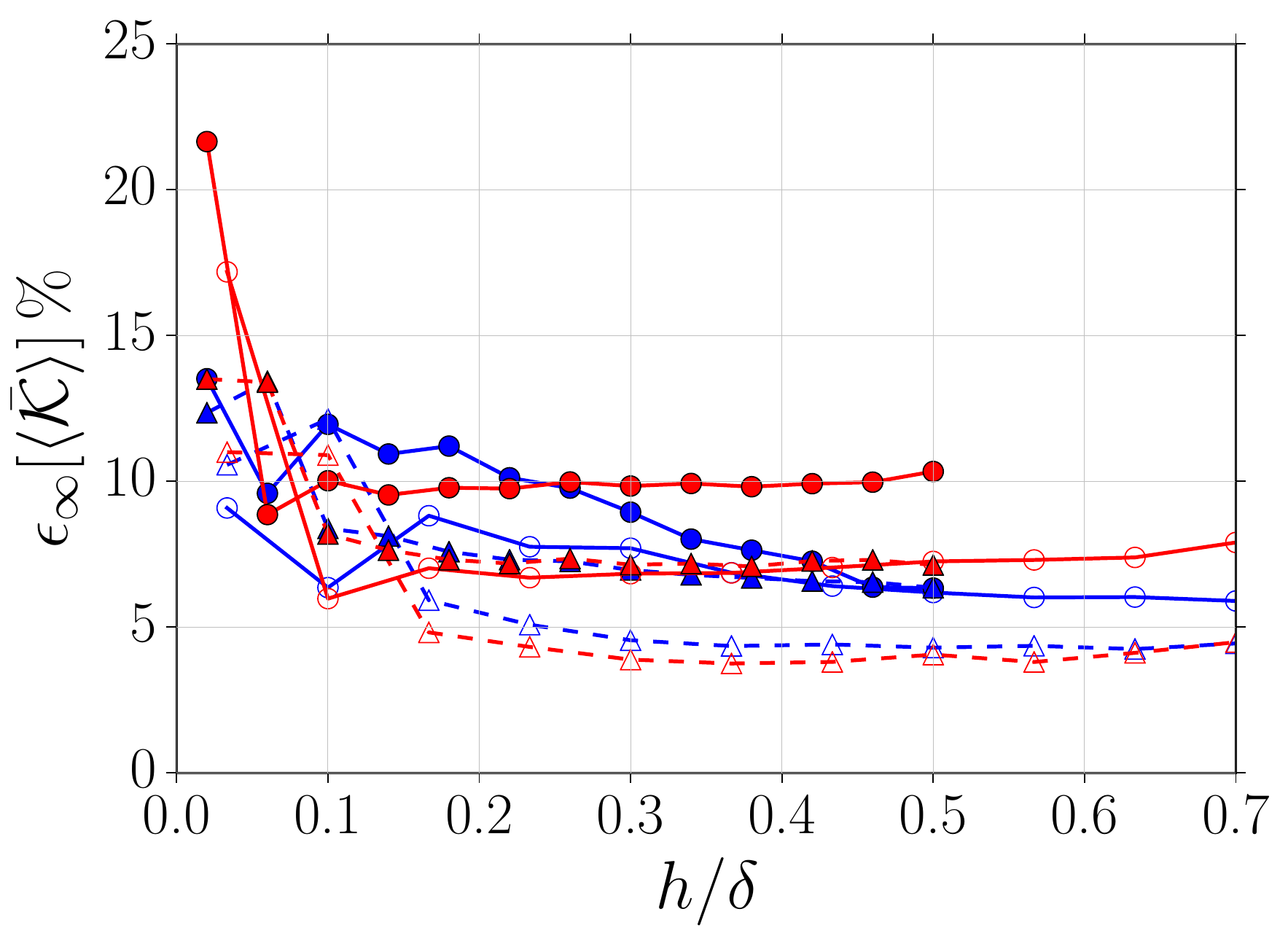}   \\      
   {\small{(c)}} &    {\small{(d)}} \\   
   \end{tabular}
\caption{Variation of the error in the WMLES QoIs with $\hd$ for different combinations of numerical scheme, SGS model, and grid resolution. The Spalding law is used with $\kappa=0.395$ and $B=4.8$. Results of the Linear and LUST schemes are plotted in blue and red, respectively. }\label{fig:convtest_h}
\end{figure}

A set of WMLES is conducted using both Linear and LUST schemes along with the WALE and Smagorinsky SGS models. 
Isotropic cells with resolutions $\nd=15$ and $25$ are considered. 
The velocity sampling points are chosen to be the cell centers located at different distances from the wall. 
Similar to the previous sections, \tim{the} Spalding law~(\ref{eq:spalding}) with $\kappa=3.95$ and $B=4.8$ is the wall model. 
The resulting errors in the QoIs of channel flow are plotted in \fig~\ref{fig:convtest_h}.
Sampling \rev{from cell centers with associated} heights up to $\hd\approx0.6 \dash 0.7$ \tim{are} investigated, taking into account the validity range of the Spalding law~(\ref{eq:spalding}).

\rev{As explained earlier, the velocity at the first few cell centers off the wall can be significantly deviated from the} \revCom{reference} DNS data.
\rev{Therefore, sampling from these cell centers for wall modeling can lead to significant error (here, under-prediction) in $\lbut$, see \fig~\ref{fig:convtest_h}(a)}.
The footprint of the velocity \tim{on} $\lbut$, can be understood \tim{by} comparing the shape of mean velocity profiles in \fig~\ref{fig:WMfoot_uPls} and $\epsilon[\lbut]$ in \fig~\ref{fig:convtest_h}(a). 
For instance, sampling from the first off-wall cell for the simulations with LUST, or from the first two cells in simulations using the Smagorinsky model will result in a large under-prediction of $\lbut$\tim{, independent of the distance from the corresponding cell centers to the wall.}
Therefore, more important is the consecutive cell number associated with the sampling height.
Recall that according to \sect~\ref{sec:numSGSwm}, the specific pattern of the error in the mean velocity profile is the footprint of numerical and modeling errors, and hence can\tim{, in principle,} be known prior to the course of wall modeling.
Note that although instantaneous velocity samples are used for the purpose of wall modeling, the reasoning provided above to connect the under-predications of mean velocity $\U$ and $\lbut$ remains valid.

The accuracy of the wall shear stress imposed at the wall has \revCom{generally} a low influence on $\einf[\U]$, see the plots in \fig~\ref{fig:convtest_h}(a,b). 
\revCom{However,} consider the case of sampling from the first off-wall cell when using the LUST scheme.
The under-prediction of $\lbut$ is so large that the profile of $\U$ in $\yd\geq 0.2$ is also affected. 
However, the resulting~$\einf[\U]$ is only~$\approx 1\%$ higher than the simulations with sampling from other cells.
Recall that the lack of dissipation allows for the error in $\lbut$ to slightly propagate to $\U$.
Therefore, the connection between the velocity and wall shear stress is generally mutual, although it is only strong in one direction.

Next, the errors in the velocity\tim{'s} second-order moments are considered in \fig~\ref{fig:convtest_h}(c,d).
The trend of changing~$\einf[\buv]$ with~$\hd$ is directly related to that of~$\epsilon[\lbut]$, confirming the discussion in \sect~\ref{sec:qoiRelation}. 
Independent of \tim{the} SGS model,~$\einf[\buv]$ is minimized for~$\nd=25$.
\tim{On the other hand,} similar to the previous section, \tim{the} most accurate profiles of~$\bk$ in the outer layer are obtained by using Smagorinsky model with \tim{a} coarse resolution,~$\nd=15$. 
\rev{
To understand these behaviors further studies are required, in which, the interaction between the fluctuations of the sampled velocity and predicted wall shear stress is investigated.}

Similar to~$\U$, for the simulations with enough dissipation, \ie~LUST and also \tim{the} combination of \tim{the} Linear and Smagorinsky model, both $\einf[\buv]$ and $\einf[\bk]$ remain almost unchanged far from the wall.
This is an important observation that the fluctuations in the predicted wall shear stress do not impact the velocity fluctuations in the outer layer over $\yd \geq 0.2$, when appropriate dissipation is provided.

As a main conclusion of the discussion, it is vital to avoid sampling the velocity for wall modeling from the first few cells near the wall. 
Except for the low-dissipative combination of the Linear scheme and the WALE model, the variation of the location of the sampling point for $\hd\gtrsim 0.2$ does not make any difference in the mean velocity and resolved TKE profiles away from the wall. 
Moreover, the associated variations in~$\epsilon[\lbut]$ and~$\einf[\buv]$ are negligible.

\subsection{Influence of the wall model parameters}\label{sec:wmParams}
In this section, the focus is on the influence of the wall model parameters, i.e. $\kappa$ and $B$ in the Spalding law (\ref{eq:spalding}).
As pointed out in \sect~\ref{sec:LSAGSA} and also discussed in \cite{salehECCOMAS:18}, these parameters can potentially have considerable influence on the wall model predictions for $\lbut$. 
Motivated by this, the possibility of using those parameters to control the accuracy of the WMLES QoIs is discussed here.
To the authors' best of knowledge, there is not any previous detailed study on the sensitivity of the WMLES to the variation of the parameters appearing in the laws of the wall which form the algebraic equilibrium type of wall models. 
\tim{By contrast}, dynamic adjustment of \tim{the} parameter $\kappa$, the von K\'arm\'an coefficient, appearing in wall models other than the laws of the wall, has been previously addressed, see \cite{cabotMoin:00,wangMoin:02,kawai:13,park:14}.
Similar to these studies, the parameters in the laws of the wall are here treated as freely-adjustable parameters that can be modified to enhance the accuracy of the modeling.

In the following subsections, first a motivational example is given to show the importance of the wall model parameters. 
\rev{Then, it is investigated how adjustment of the parameters affects different WMLES QoIs.
This is followed by showing that through the adjustments, the dependency of the WMLES accuracy on the velocity sampling height can be removed.}

\subsubsection{Motivational example} \label{sec:motivational}
As a basic requirement, any law of the wall that is aimed to be used as wall model should be capable of accurately representing the true reference data of the inner-scaled mean velocity over some specific range of off-wall distance, hereafter called $\Omega^+_h$. 
Clearly, in the course of WMLES, the velocity samples imported to the wall model must be taken from an $h$ falling in this range.
Therefore, before being used as wall model, a law of the wall should be tuned up by adjusting its model parameters such that the best match with the reference data over $\Omega^+_h$ is obtained. 
This becomes more important noting that many of the laws of the wall, such as the Spalding law (\ref{eq:spalding}), were calibrated in accordance with the low $\rey$-number data available at the time. 
Following this, the parameter values $\kappa=0.395$ and $B=4.8$, giving the best fit for the Spalding law against the DNS data of channel flow at $\reyt=5200$ \cite{lee-moser:15} over $\Omega^+_h = [0,0.6\reyt]$ (in wall unit\tim{s}) have been used throughout the present study, instead of the original values suggested by Spalding \cite{spalding}.

\begin{figure}
\centering
    \begin{tabular}{cc}
    \includegraphics[scale=0.47]{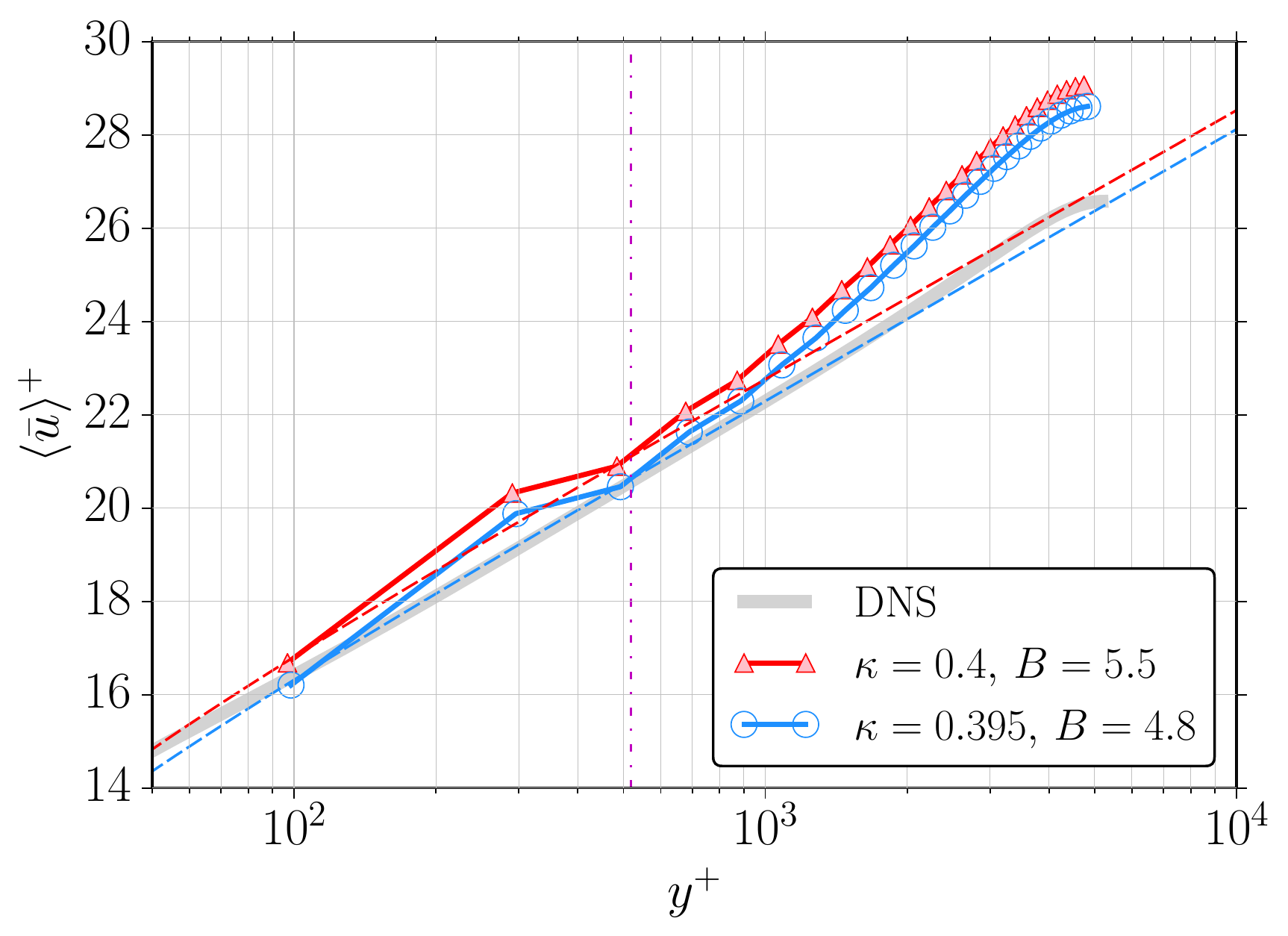} &
    \includegraphics[scale=0.47]{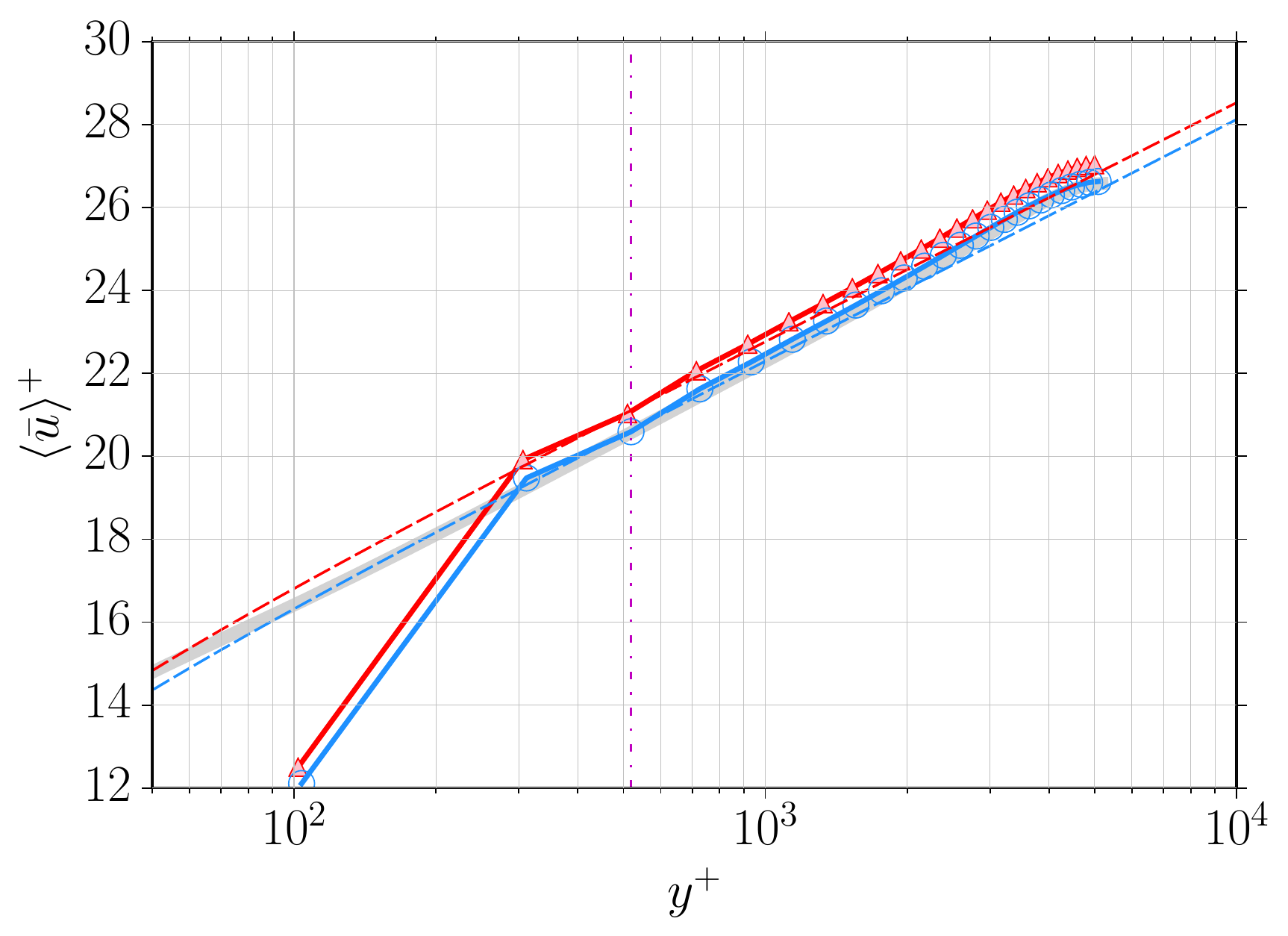} \\    
    {\small{(a)}} &    {\small{(b)}} \\
    \end{tabular}
    \caption{Influence of the wall model parameters on the inner-scaled mean velocity profiles obtained by using the Linear (a) and LUST (b) schemes with \tim{the} WALE model and $\nd=25$. The vertical dashed-dotted line specifies the sampling height associated with $\hd=0.1$. The dashed lines show the Spalding law (\ref{eq:spalding}) for the two sets of parameter values.
     \revCom{Plots of the same color correspond to the same values of $\kappa$ and $B$.}}
     \label{fig:kapBUpls}
\end{figure}

As thoroughly discussed in the previous sections, what determines the accuracy of the WMLES QoIs outside the near-wall region of the TBL is mainly the overall interaction of numerics, SGS modeling, and grid resolution.
Therefore, the a-priori estimated parameters do not necessarily result in accurate predictions of mean wall shear stress and other QoIs in WMLES. 
This is rooted in the fact that by the wall modeling described in \sect~\ref{sec:wallModels}, the sampling point is the only point across the TBL at which the resulting instantaneous inner-scaled velocity $\bu_h/\but$ and off-wall distance, $h \but/\nu$, are forced to match those of the law of the wall.
If the sampling point belongs the calibration range $\Omega^+_h$ and is not among those cells at which the velocity field is highly erroneous, \eg~the first off-wall cell for LUST scheme or the first two off-wall cells when Smagorinsky is used, then the a-priori estimated parameters may possibly \tim{lead to increased accuracy}. 
This argument is supported by the plots in~\fig~\ref{fig:kapBUpls}. 
Using parameters $\kappa=0.395$ and $B=4.8$ instead of the original values, the inner-scaled velocity profiles are shifted downward such that the $(\U^+,y^+)$ at $\hd=0.1$ becomes very close to the reference DNS data.
This is in contradiction with \cite{lee:13}, p.4, where the influence of changing $\kappa$ and $B$ in the wall model was stated to be negligible. 
Hereafter, the wall model parameter values that lead to accurate WMLES results are called optimal parameters, $\fq_\opt$\rev{,~see~below.}

\begin{figure}[!htbp]
\centering
    \begin{tabular}{ccc}
    \includegraphics[scale=0.26]{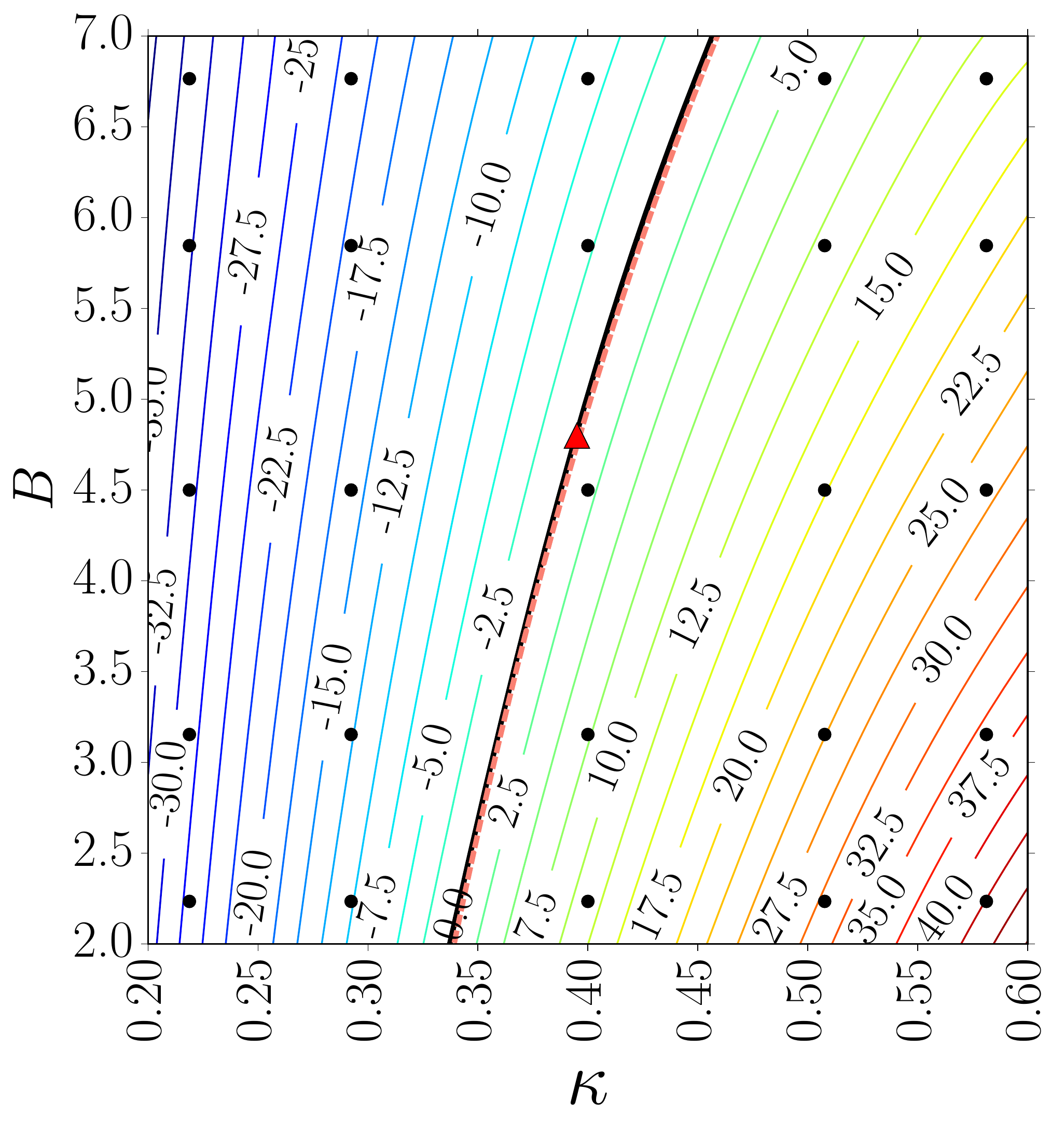} &   \hspace{-0.45cm}  
    \includegraphics[scale=0.26]{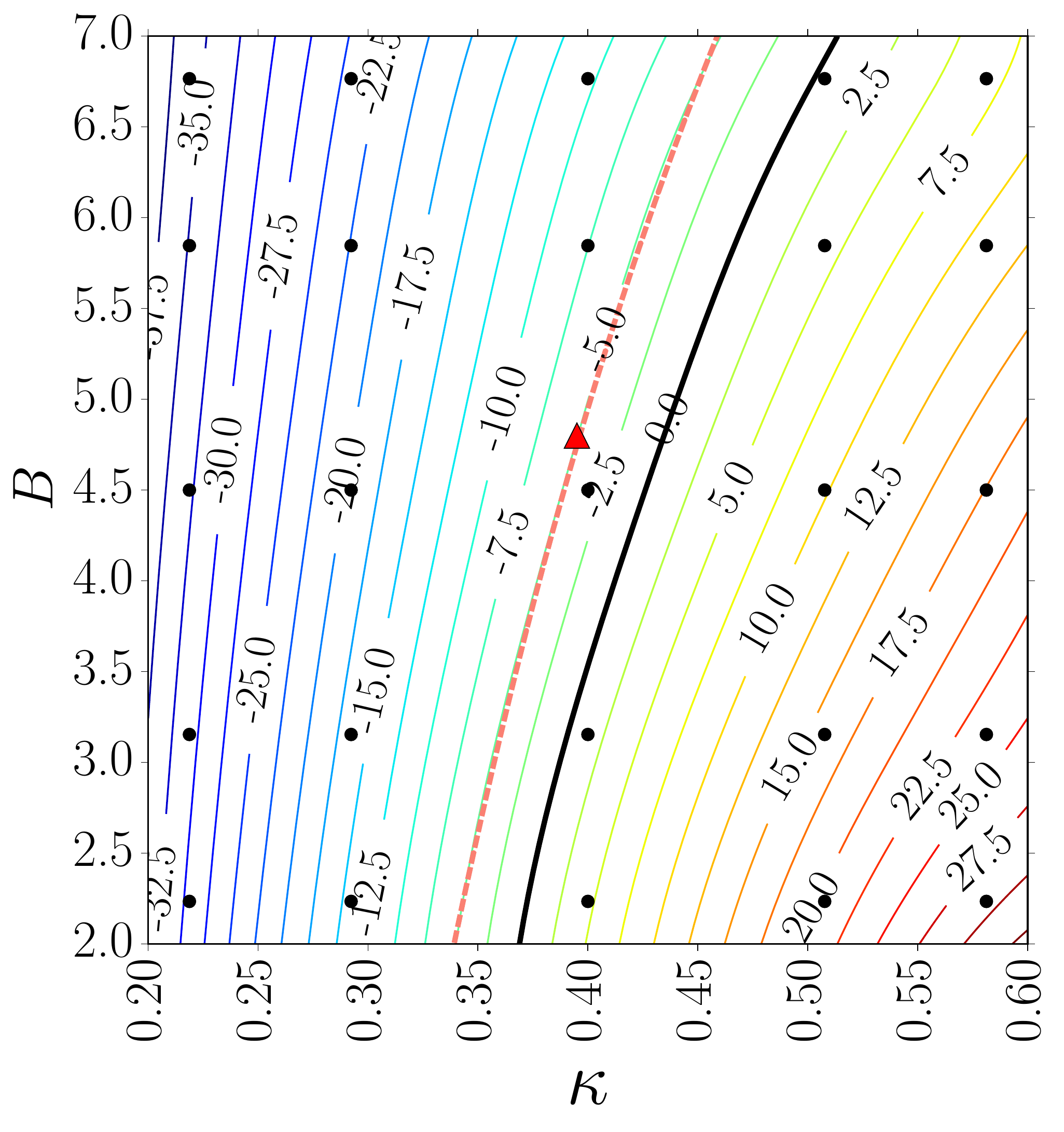} &   \hspace{-0.45cm}
    \includegraphics[scale=0.26]{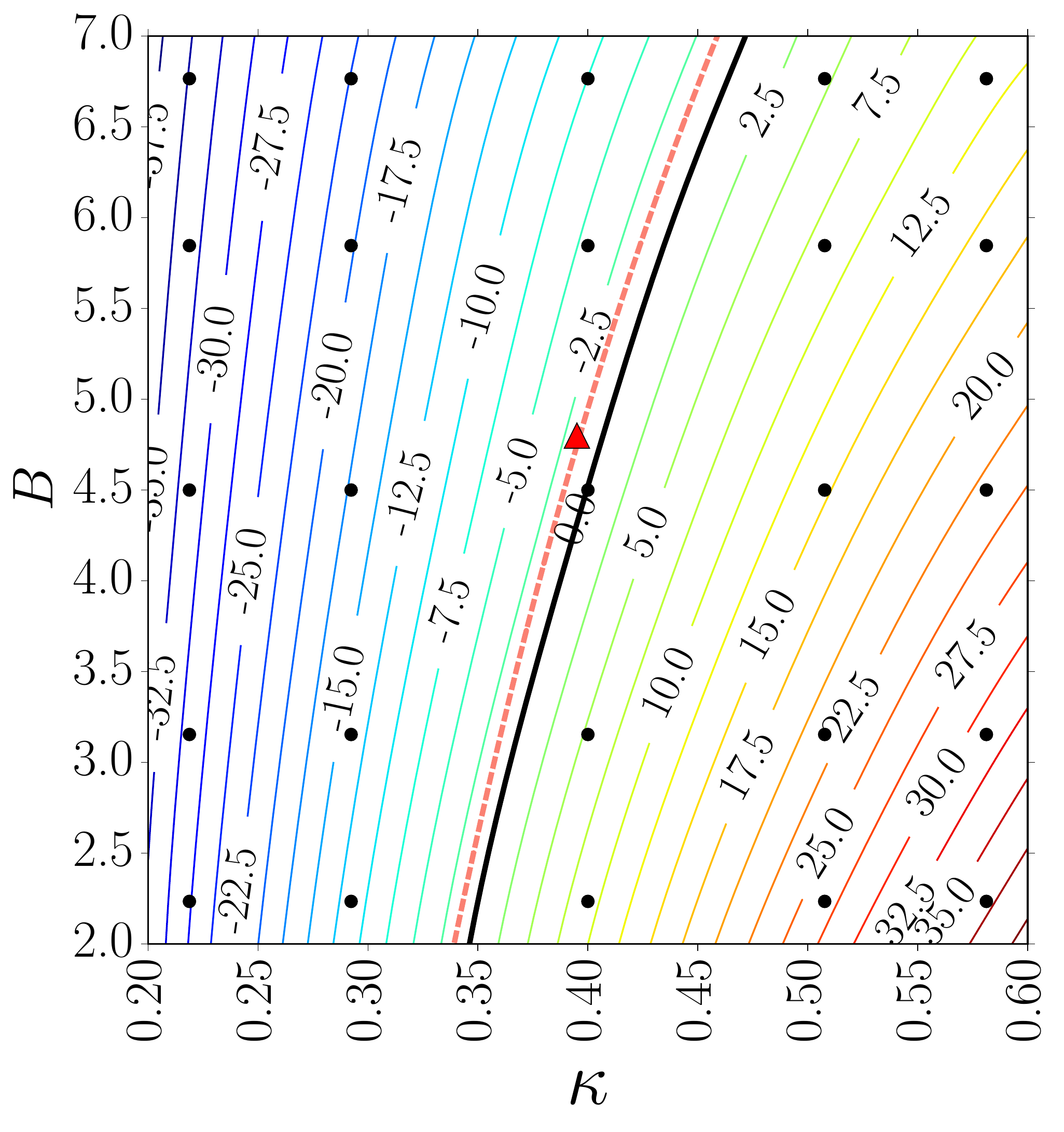} \\   
    {\small{(a)}}  &  {\small{(b)}}  &  {\small{(c)}}  \\
    \end{tabular}
    \caption{Isolines of $\epsilon[\lbut]\,\%$ in the $\kappa \dash B$ plane \revCom{belonging to WMLES by the combinations of} LUST-WALE (a), Linear-WALE (b), and Linear-Smagorinsky (c) using grid resolution $\nd=25$. The sampling point is at $\hd=0.1$ ($\thrd$ cell center) corresponding to which the combinations of a-priori estimated $\kappa^\circ$ and $B^\circ$ are shown by the dashed curve. The black solid line represents $\epsilon[\lbut]=0$ resulting from WMLES. Moreover, the red triangle specifies $\kappa=0.395$ and $B=4.8$.}
    \label{fig:DuTau_kapBIso}
\end{figure}

\subsubsection{Seeking optimal parameters}\label{sec:wmParamMotiv}
It is clear that in general, $\fq_\opt$ are dependent on the choice of numerical scheme, SGS model, grid resolution, \rev{and~$\rey$-number. Therefore, $\fq_\opt$ values are not universal.}
\revCom{Now, it is explored if for a given sampling height~$h$, the~$\fq_\opt$ become the same as the a-priori estimated parameters~$\fq^\circ$.}

For this, the reference DNS values of~$\lu^{\circ^+}_h$ at $h^{\circ^+}$, taken from \cite{lee-moser:15}, are plugged into the Spalding law (\ref{eq:spalding}) to find $\fq^\circ=\{ \kappa^\circ,B^\circ\}$.
Clearly, the formulated problem is under-determined, having one data point and two unknown parameters.
To handle the issue, all possible solutions over presumed admissible ranges for the two parameters are computed. 
In particular, $\Qk=[0.2,0.6]$ and $\QB=[2.0,7.0]$ are chosen for this purpose. 
For the specific choice of $\hd=0.1$, the resulting combinations of $\kappa^\circ$ and $B^\circ$ are plotted by dashed line in \fig~\ref{fig:DuTau_kapBIso} in the~$\kappa\dash B$ plane formed by the admissible ranges.
Obviously, $\kappa=0.395$ and $B=4.8$ is only one possibility out of infinitely many which reproduce $\lu^{\circ^+}_h$. 
Also shown in this figure are the isolines of $\epsilon[\lbut]$ obtained by WMLES using the LUST scheme with WALE model and also the Linear scheme with both WALE and Smagorinsky models.
In all cases, the grid resolution $\nd$ is $25$.
The error isolines are constructed using the gPCE-based metamodel (\ref{eq:pce}). 
To construct the metamodel, parameters~$\kappa$ and~$B$ of the Spalding law (\ref{eq:spalding}) are \rev{treated as} uniform random parameters varying over admissible ranges $\Qk=[0.2,0.6]$ and $\QB=[2.0,7.0]$, respectively.
The error in $\lbut$ and other QoIs are assigned to be the responses.
In particular, 5 sample points in each of the $\Qk$ and~$\QB$ are taken resulting in 25 WMLES in total. 
\rev{Each simulation is represented by a black dot in~\fig~\ref{fig:DuTau_kapBIso}.}

\begin{figure}[!t]
\centering
    \begin{tabular}{ccc}
    \includegraphics[scale=0.26]{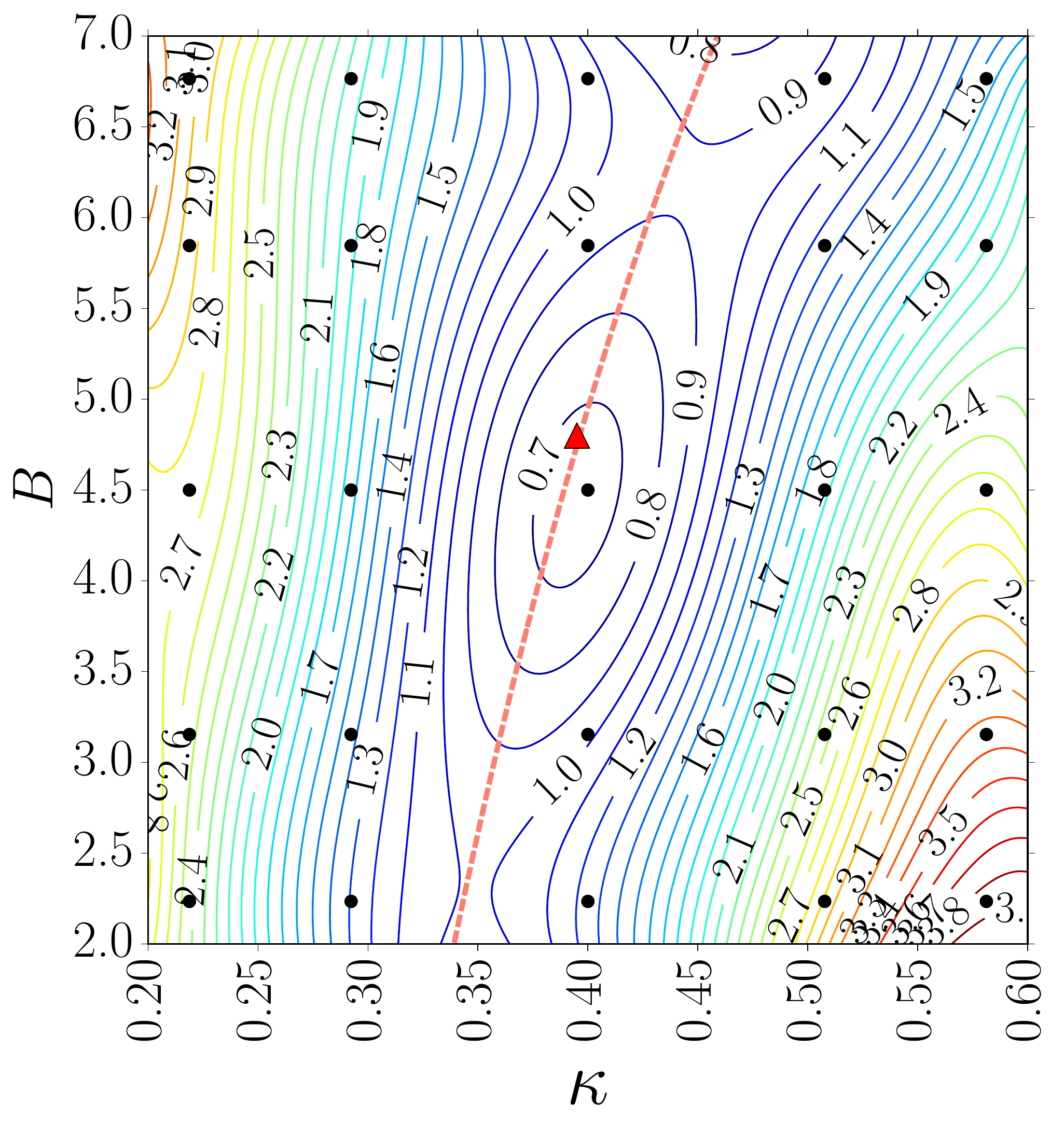} &  \hspace{-0.4cm}
    \includegraphics[scale=0.26]{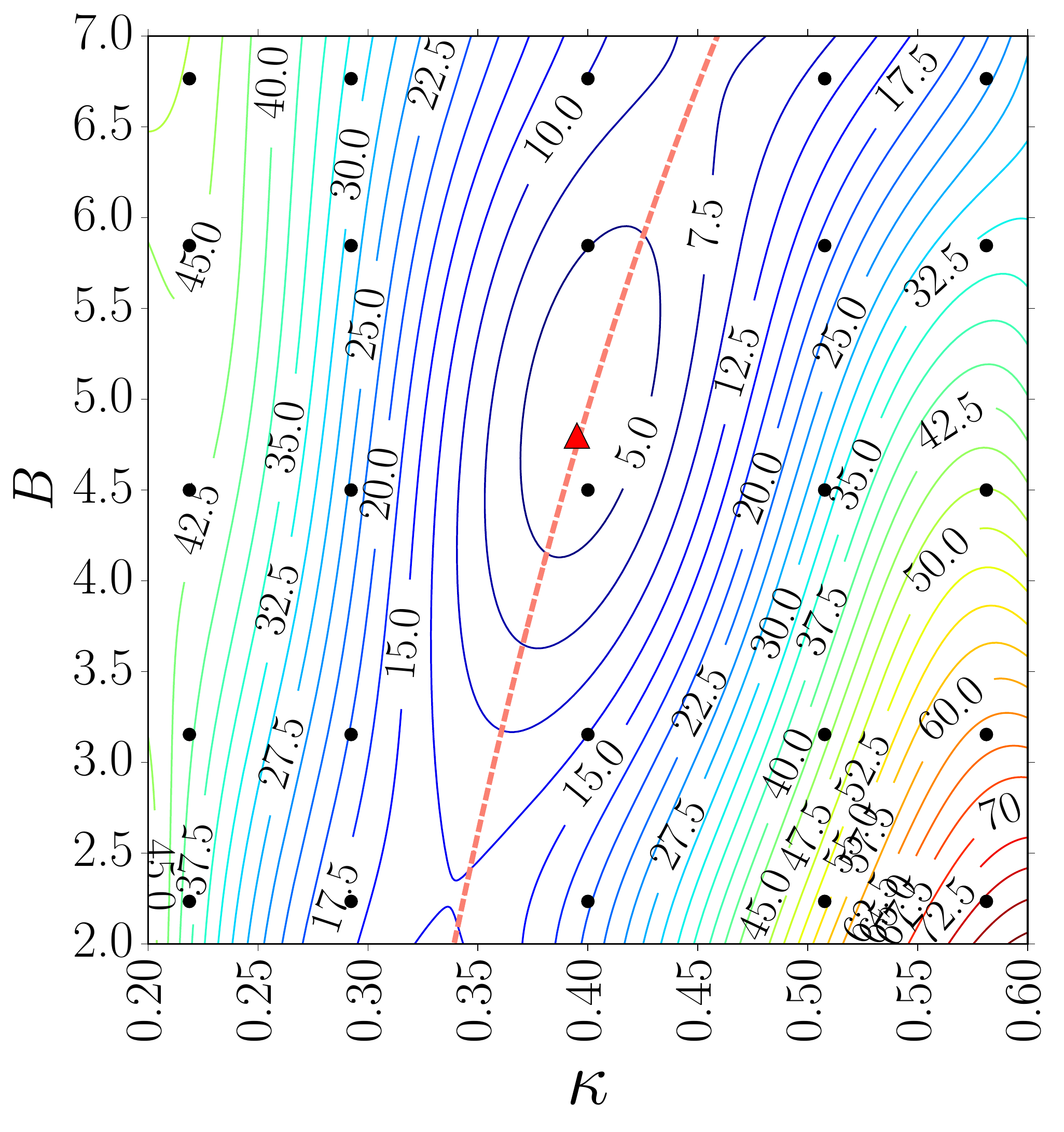} &  \hspace{-0.4cm}
    \includegraphics[scale=0.26]{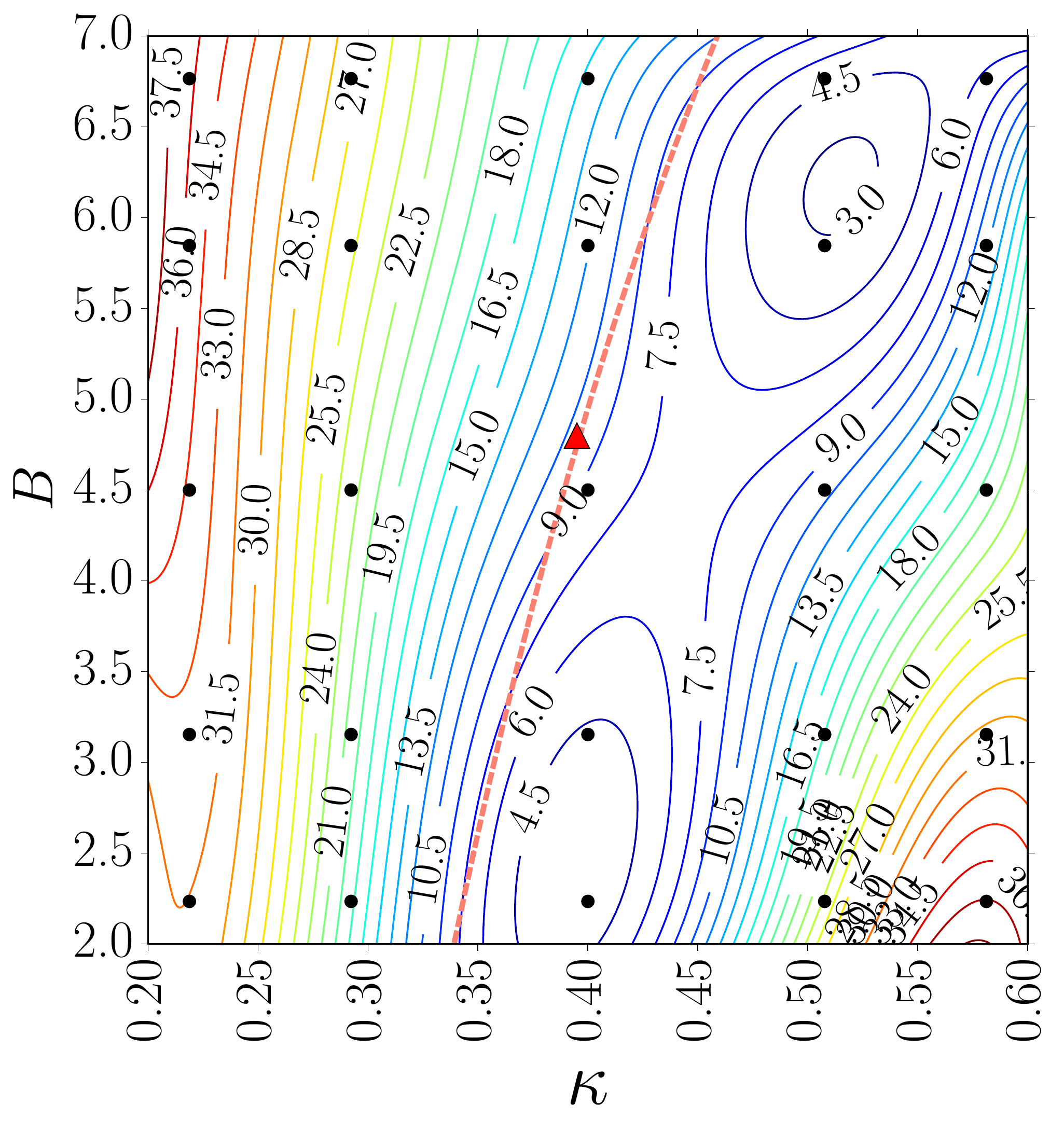} \\
   {\small{(a)}} &    {\small{(b)}}&    {\small{(c)}} \\      
    \includegraphics[scale=0.26]{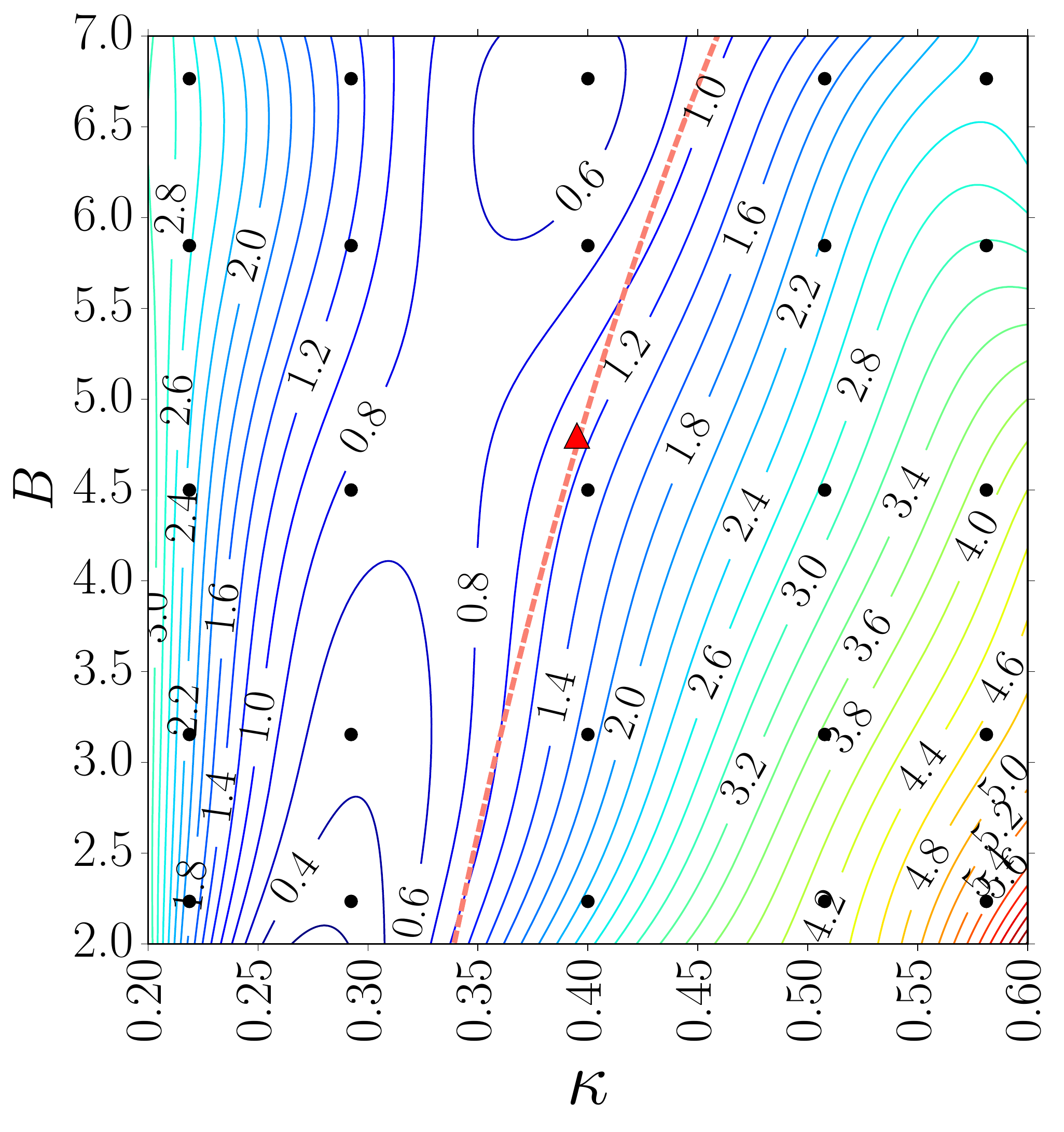} &   \hspace{-0.4cm}
    \includegraphics[scale=0.26]{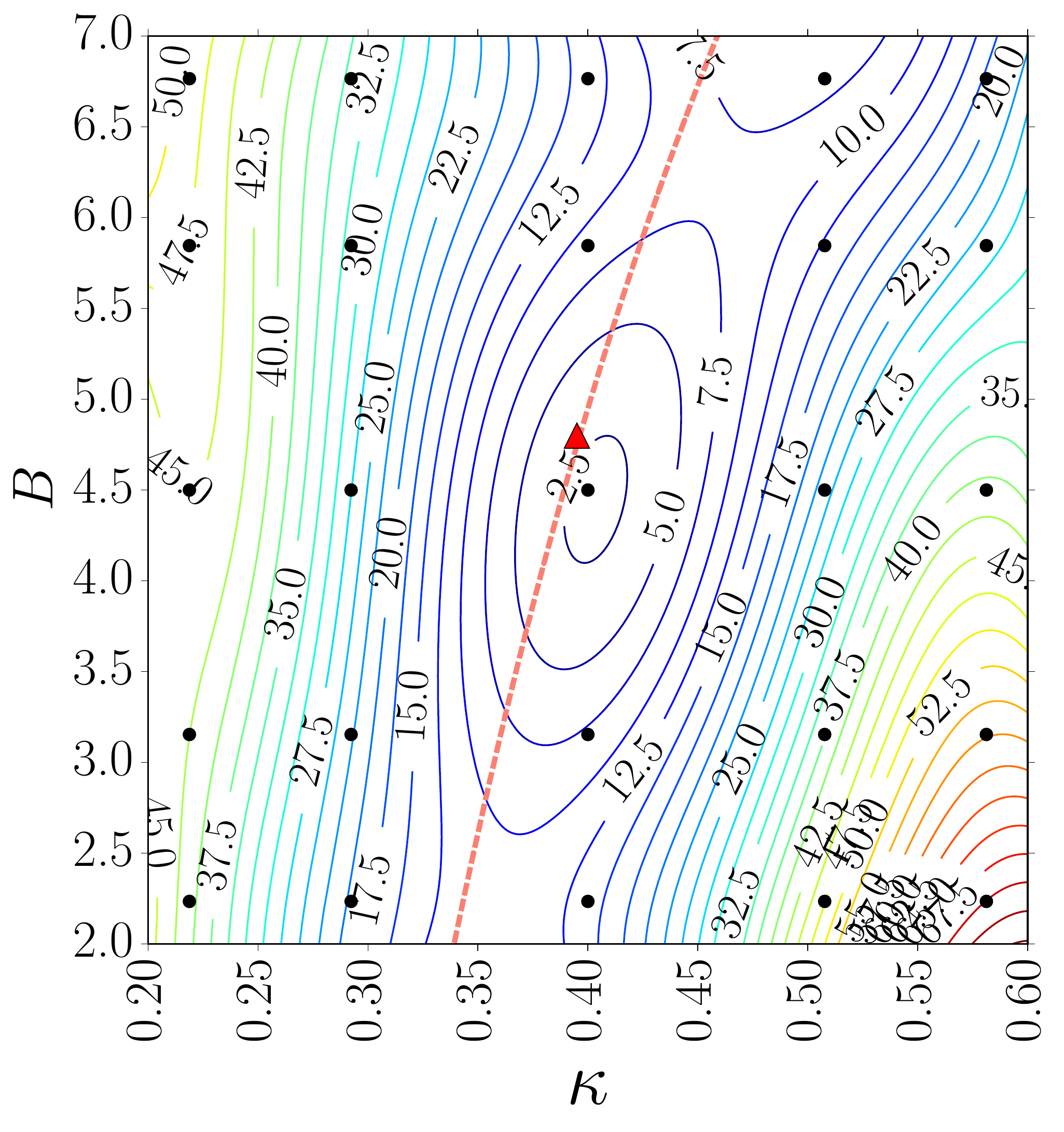} &   \hspace{-0.4cm}
    \includegraphics[scale=0.26]{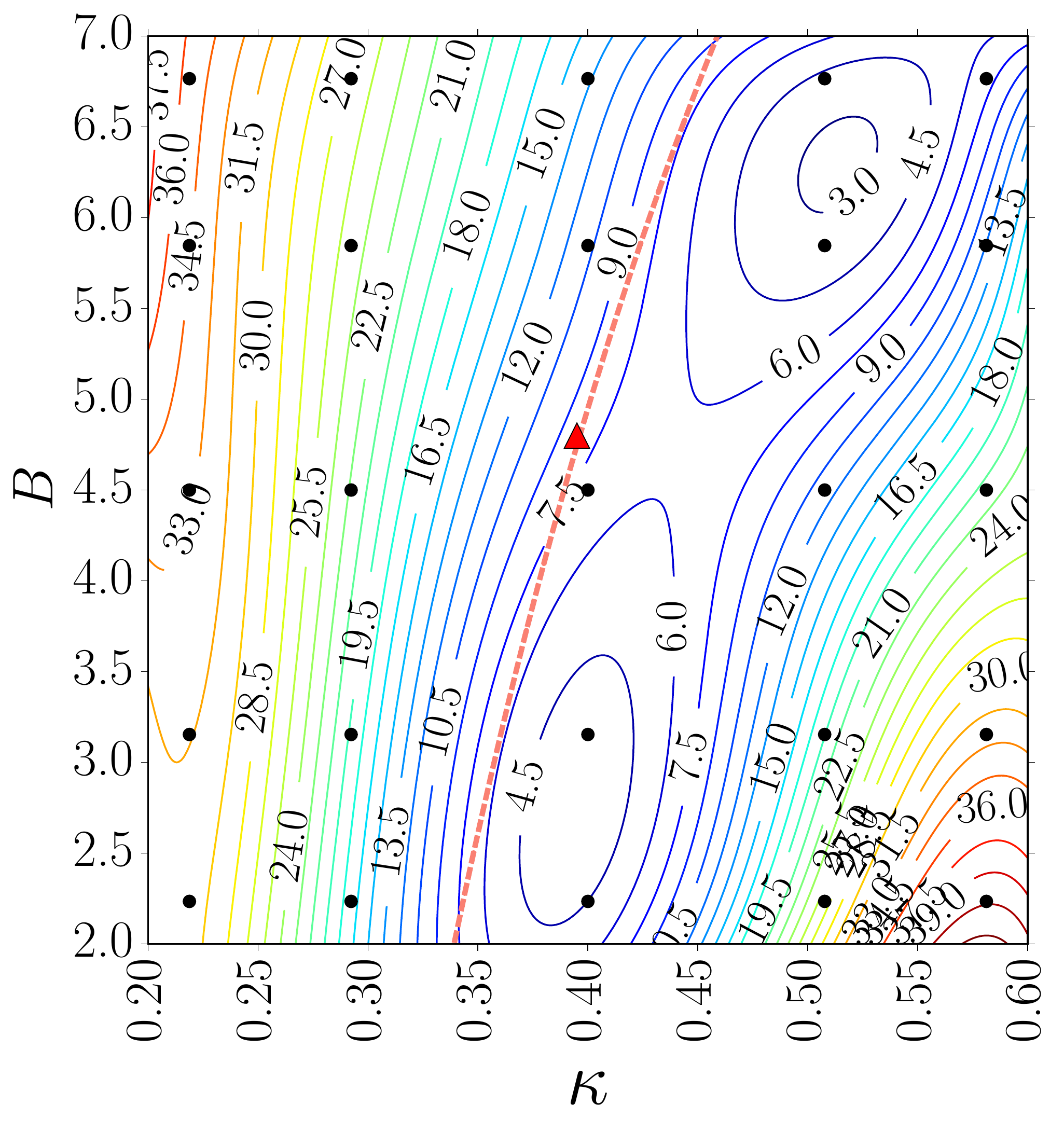} \\
   {\small{(d)}} &    {\small{(e)}}&    {\small{(f)}} \\    
    \end{tabular}
    \caption{Isolines of $\einf[\U]\,\%$ (a,d), $\einf[\buv]\,\%$ (b,e), and $\einf[\bk]\,\%$ (c,f) in the $\kappa \dash B$ plane for WMLES by LUST-WALE (top row) and Linear-Smagorinsky (bottom row). The grid resolution is $\nd=25$ and the Spalding law is used with velocity sampled from the $\thrd$ cell center located at $\hd=0.1$. For associated isolines of~$\epsilon[\lbut]$,~see~\fig~\ref{fig:DuTau_kapBIso}(a,c). The combinations of a-priori estimated $\kappa^\circ$ and $B^\circ$ are shown by the dashed curves. Also, the red triangle specifies $\kappa=0.395$ and $B=4.8$.
    }\label{fig:isoErrors_kapB}
\end{figure}

Based on the isolines of $\epsilon[\lbut]$, it is inferred that both positive and negative LLM can be obtained for different combinations of wall model parameters. 
For the simulations with the LUST scheme and WALE model, the optimal $\kappa$ and $B$ which yield zero error in $\lbut$ coincide with the a-priori estimated~$\fq^\circ$.
In contrast, for the combination of Linear-WALE, there is a considerable difference between $\fq^\circ$ and the loci of zero $\epsilon[\lbut]$ in the $\kappa\dash B$ plane.
This returns to the error in the velocity samples imported to the wall model when the Linear scheme is used with low-dissipative SGS model, as discussed in the previous sections. 
As expected, substituting WALE with \tim{the} Smagorinsky model, reduces the deviation of a-priori estimated $\fq^\circ$ from the a-posteriori optimal parameters $\fq_{\opt}$, see \fig~\ref{fig:DuTau_kapBIso}(c).
Based on this discussion, the superiority of the LUST scheme over the Linear one is further clarified.

The isolines of $\einf[\U]$, $\einf[\buv]$, and $\einf[\bk]$ belonging to the same simulations are illustrated in \fig~\ref{fig:isoErrors_kapB}.
First of all, \rev{as seen in} the pattern of the isolines, $\kappa$ is more influential than $B$ on the error in different QoIs. 
This is in agreement with the observations made in the local and global sensitivity analyses in \sect~\ref{sec:LSAGSA}.
In particular, for the simulations with the LUST scheme and \tim{the} WALE model, the Sobol indices of $\epsilon[\lbut]$ with respect to $\kappa$ and $B$ are respectively equal to $0.916$ and $0.1$. 
Corresponding indices for the error in the velocity statistical moments are $\approx 0.92$ and $\approx 0.30$.

The error in the mean velocity profile far from the wall, measured by $\einf[\U]$, \tim{appears to be} \rev{quite} sensitive to the values of the wall model parameters and hence, to the predicted wall shear stress.
This \tim{may} seem to \tim{contradict} what \tim{was} observed in the previous sections where the error in $\U$ far from the wall was found to be mainly driven by the numerical scheme, SGS model and grid resolution. 
However, note that only for those values of the wall model parameters that the resulting error in $\lbut$ is very large (say $|\epsilon[\lbut]|>7.5\%$), the value of $\einf[\U]$ exceeds $1\%$. 
It is emphasized that such extreme values of the wall model parameters are adopted here only to construct a clear picture of the error isolines in the $\kappa\dash B$ plane.

Based on \fig~\ref{fig:isoErrors_kapB}\rev{(top)}, for the simulations with the LUST scheme, if the wall model parameters are chosen in close vicinity of the optimal values, the values of $\einf[\U]$ are also small.
Contrary to this, when the Linear scheme is used,~\rev{see \fig~\ref{fig:isoErrors_kapB}(bottom),} the loci of small $\einf[\U]$ do not coincide with those of accurate~$\lbut$.
This is more significant when WALE model is used (not shown here).
Independent of the numerical interpolation scheme used with Smagorinsky model, the values of $\epsilon[\lbut]$ and $\einf[\buv]$ are related, following the discussion in \sect~\ref{sec:qoiRelation}. 
Finally, the resolved velocity fluctuations are also substantially affected by changing the wall model parameters. 
Unfortunately, the most accurate $\bk$ profiles far from the wall are obtained for the wall model parameters different than those yielding accurate $\lbut$, $\U$ and $\buv$.
This is consistent with the effect of other controllers of WMLES discussed in the previous~sections. 

\rev{Finally, the following interesting fact is observed concerning the set of optimal wall model parameters,~$\fq_\opt$.}
Based on the plots in \fig~\ref{fig:DuTau_kapBIso}, for a given numerical scheme, SGS model, grid resolution, and sampling height, an infinite number of $\fq_\opt$ exists.
However, for all of these parameters in each case, similar levels of $\einf[\U]$, $\einf[\buv]$, and $\einf[\bk]$ is predicted by the metamodel (\ref{eq:pce}), see \fig~\ref{fig:isoErrors_kapB}.
The validity of this observation has been further confirmed by conducting WMLES using several optimal parameter combinations (results not shown here).

\subsubsection{Possibility of removing the dependency on $h$}\label{sec:hRemove}
\rev{Regardless of the sampling height, the same results for WMLES are obtained as long as associated $\fq_\opt$ are employed.}
To show this, consider cases for which the velocity samples are taken from the first few off-wall cells, where the deviation of the WMLES mean velocity from the reference data is large. 
For such simulations the a-priori estimated parameters, $\fq^\circ$, are definitely not the choice for the optimal parameters~$\fq_\opt$. 
Therefore, to find $\fq_\opt$, WMLES must be performed.
Adopting the same approach as \rev{in \sect~\ref{sec:wmParamMotiv}}, let $\kappa\sim\cU[\Qk]$ and $B\sim \cU[\QB]$, and take 5 samples for each parameter to construct metamodel (\ref{eq:pce}).

\begin{figure}[!htbp]
\centering
   \begin{tabular}{ccc}
   \includegraphics[scale=0.45]{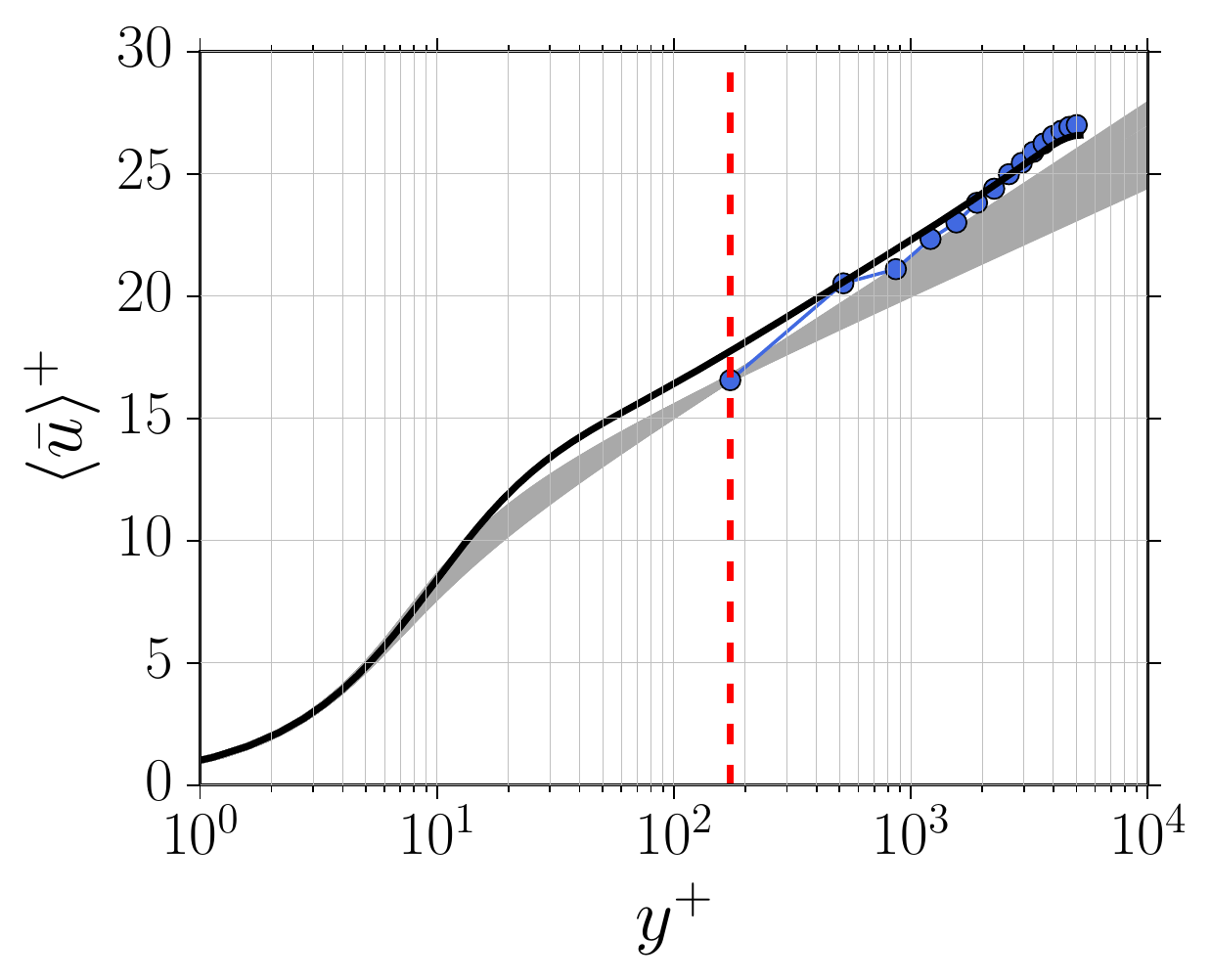} & \hspace{-0.67cm}   
   \includegraphics[scale=0.45]{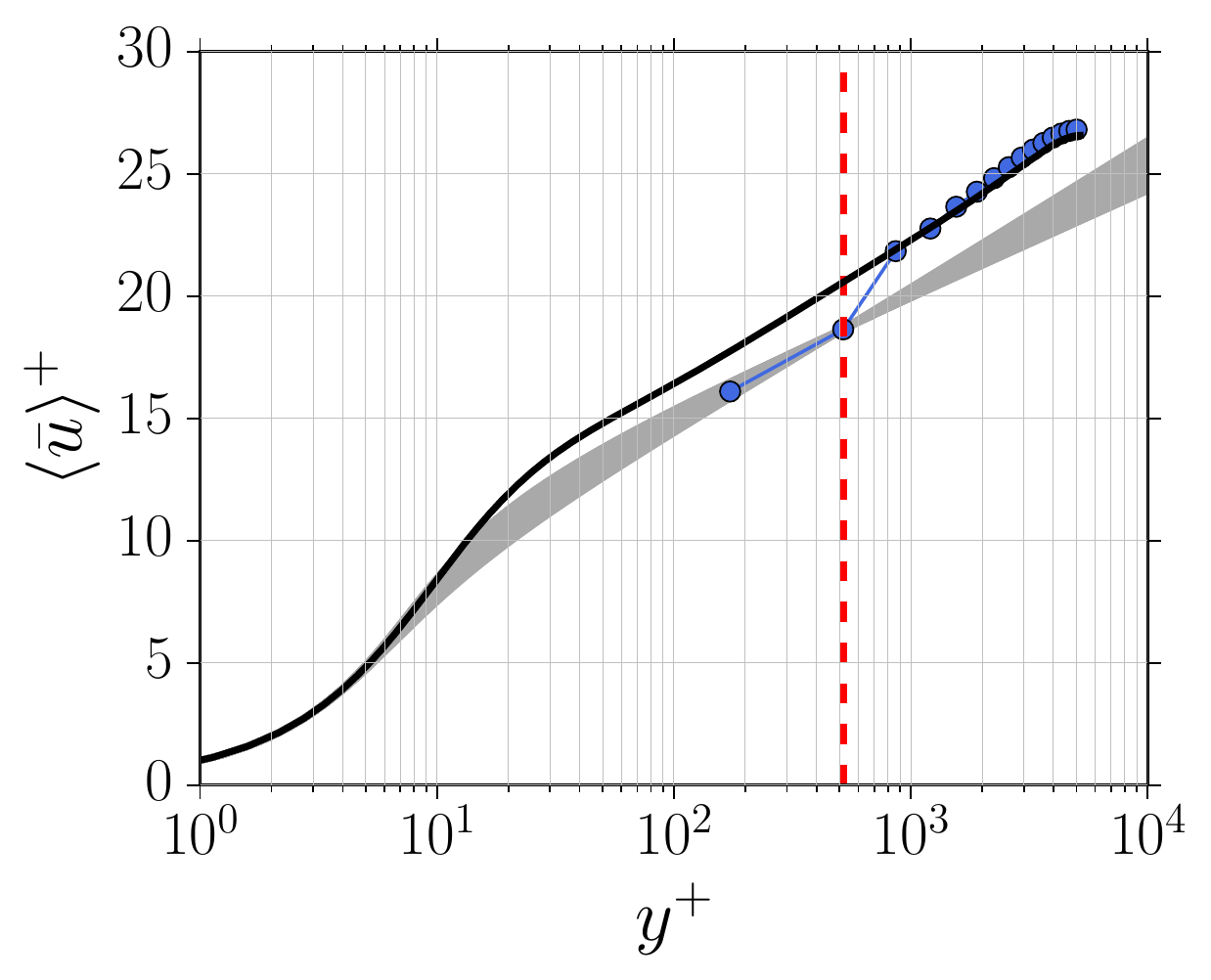} &  \hspace{-0.67cm} 
   \includegraphics[scale=0.45]{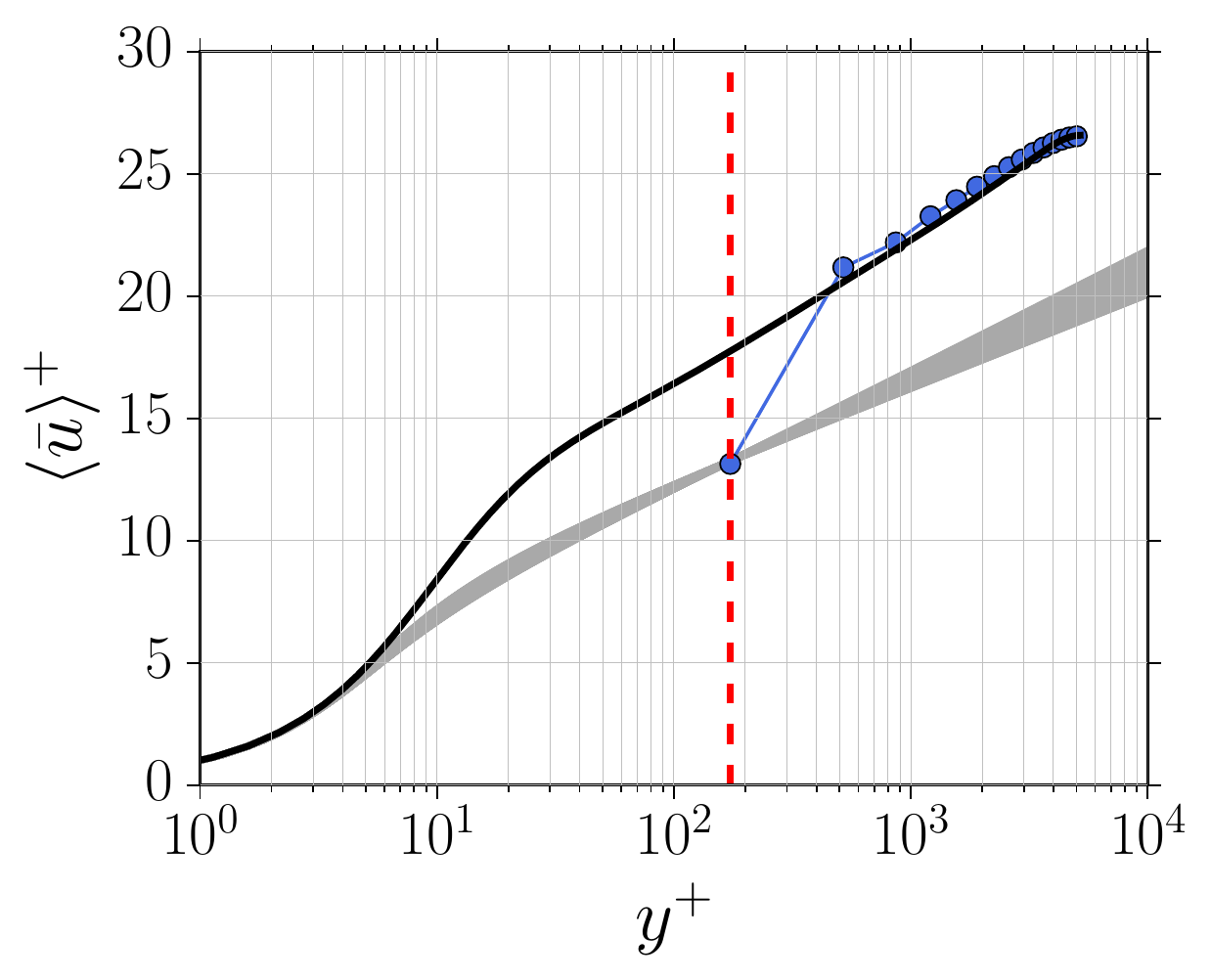} \\      
   {\small{(a)}} &    {\small{(b)}}&    {\small{(c)}} \\   
   \end{tabular}
\caption{Inner-scaled mean velocity of WMLES (symbols) obtained by using \revCom{a single set of} optimal parameters $\fq_\opt$ for the Spalding law, as compared to DNS~\cite{lee-moser:15} (solid black line). 
\rev{In each plot, the gray curves represent the Spalding law (\ref{eq:spalding}) evaluated at \revCom{several sets of} associated $\fq_\opt$.} The WMLES results belong to Linear-WALE, $\kappa=0.4115$, $B=4.1271$ (a), Linear-Smagorinsky, $\kappa=4.105$, $B=4.1271$ (b), and LUST-WALE, $\kappa=0.5218$, $B=3.357$ (c). In all simulations, $\nd=15$. The~vertical~line~represents~the sampling~height.}
\label{fig:spalding_optimalKB_n15}
\end{figure}

\begin{figure}[!h]
\centering
   \begin{tabular}{ccc}
   \includegraphics[scale=0.43]{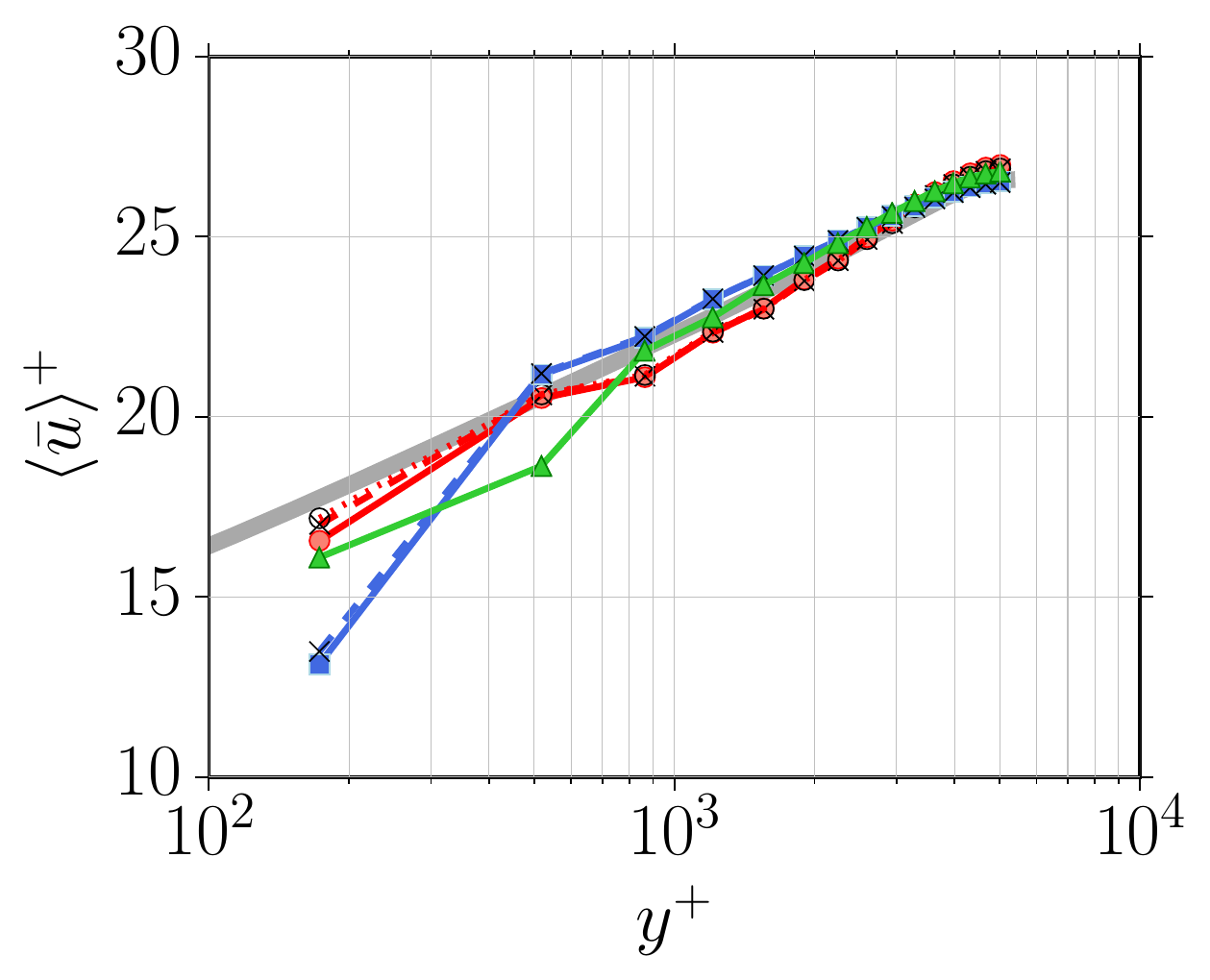} & \hspace{-0.55cm}  
   \includegraphics[scale=0.43]{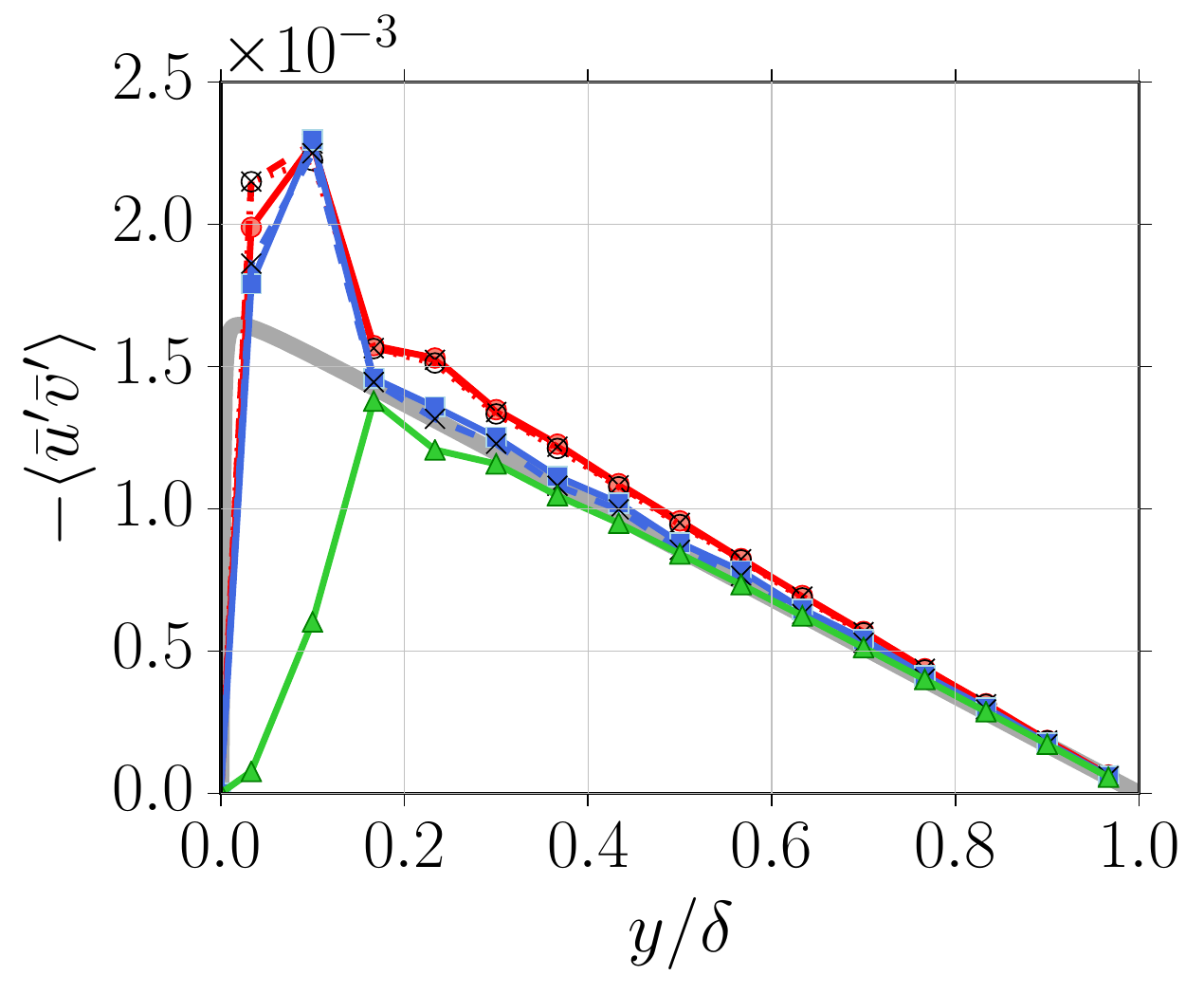} & \hspace{-0.55cm}
   \includegraphics[scale=0.43]{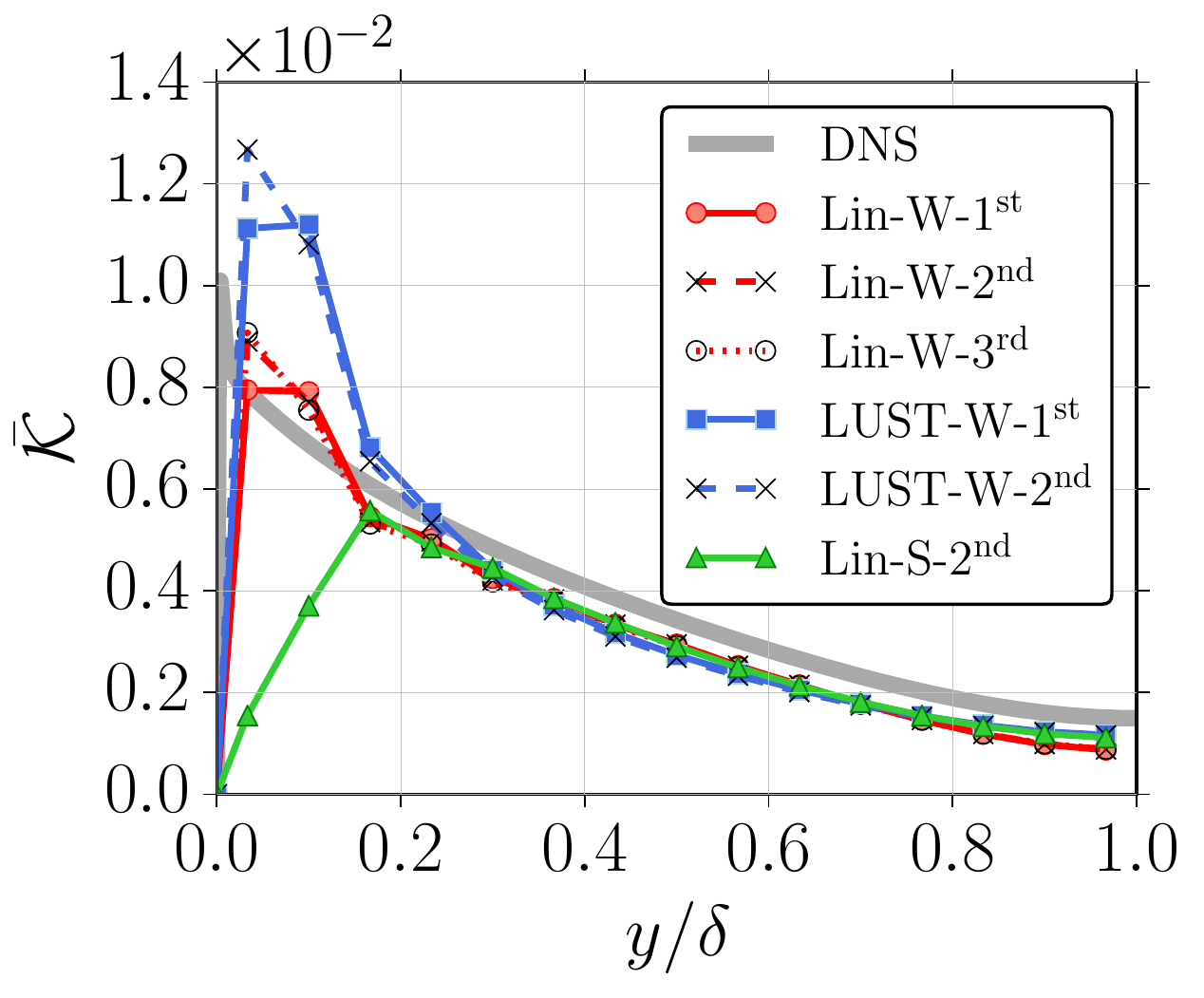} \\
   {\small{(a)}} &    {\small{(b)}}&    {\small{(c)}} \\   
   \end{tabular}
   \caption{WMLES profiles of $\U^+$ (a), $-\buv$ (b), and $\bk$ (c) for grid resolution $\nd=15$. The Spalding law (\ref{eq:spalding}) is used for wall modeling with~$\fq_\opt$ associated with the numerical convective scheme, SGS model, and sampling height \revCom{(in terms of the cell number), see the legend}. Optimal parameters $(\kappa,B)$ are equal to $(0.412,4.127)$ for Linear-WALE-$\frst$, $(0.402,5.093)$ for Linear-WALE-$\scnd$, $(0.402,4.335)$ for Linear-WALE-$\thrd$, $(0.522,3.357)$ for LUST-WALE-$\frst$, $(0.410,6.022)$ for LUST-WALE-$\scnd$, $(0.411,3.381)$ for Linear-Smagorinsky-$\scnd$. }\label{fig:profs_optimalKB_n15}
\end{figure}

For different combinations of numerical scheme, SGS model, and sampling height, the results of the WMLES performed with one \revCom{arbitrary} set of optimal parameters~$\fq_\opt$ (\ie~located on the associated loci of $\epsilon[\lbut]=0$ in the $\kappa\dash B$ plane) are shown in \figs~\ref{fig:spalding_optimalKB_n15} and \ref{fig:profs_optimalKB_n15}.
Also shown in \fig~\ref{fig:spalding_optimalKB_n15} are the \rev{families of}~$\lu^+ \dash y^+$ curves obtained from the Spalding law (\ref{eq:spalding}) using \revCom{several sets of} optimal parameters corresponding to each simulation condition. 
All of these realizations that result in accurate $\lbut$, coincide at the sampling point.
The conclusion is that even when the velocity is sampled from the near-wall cells, the adjustment of the parameters in the law of the wall can remove the LLM.

To motivate these more, \tim{consider} the plots in \fig~\ref{fig:profs_optimalKB_n15} obtained by using $\fq_\opt$.
For all simulations, \tim{the} predicted~$\lbut$ is accurate, hence, it is the footprint of the numerical scheme and SGS model which determines the accuracy of the profiles. 
For instance, see the particular pattern in the under-predicted velocity in the first few cells off the wall. 
\tim{It is} interesting to note that for all simulations, almost the same profile of resolved TKE is obtained, excluding the spurious overshoots in the near-wall region.

\subsection{Influence of the grid anisotropy}\label{sec:gridAnisot}
In all the simulations investigated up to this point, the grid was assumed to be isotropic with \rev{grid cell side length~$\delta/n$} in all directions. 
\rev{Now, the effect of grid anisotropy on WMLES QoIs is investigated.}
The \rev{cell side lengths} in the streamwise, wall-normal, and spanwise directions, respectively \tim{denoted} by $\dx=\delta/n_x$ , $\dy=\delta/n_y$, and $\dz=\delta/n_z$, are allowed to be independently chosen.
To get an idea about the possible range of variation of the grid densities, literature can be consulted. 
In his pioneering work, Chapman \cite{chapman:79} estimated $\nx=\nz=10$ and $\ny=25$ to be adequate for resolving the outer part of a TBL.
For WMLES of channel flow, different ranges of grid resolutions have been recommended, including $5\leq \nx\leq 32$, $16\leq\ny\leq 50$, and $5\leq\nz\leq 50$, based on Refs.~\cite{cabotMoin:00,nicoud:01,lee:13,park:14,yang:17,patil:12,pantano:08}.
It is worth noting that \tim{the} majority of those simulations have been performed using anisotropic grids.
\rev{Also note that number of cells per unit $\delta^3$ in the mentioned studies varies between $830$ and $64\,000$.}

\tim{The non-intrusive gPCE technique of~\sect~\ref{sec:uq} is used to reduce the amount of simulations necessary to quantify the effect of the studied parameters~$\nx$, $\ny$, and $\nz$.
All three are considered to be uniformly distributed and} varying over associated admissible ranges $\BQ_{n_x}$, $\BQ_{n_y}$, and $\BQ_{n_z}$. 
In particular, $\BQ_{n_x}=\BQ_{n_z}=[6.43,43.57]$ and $\BQ_{n_y}=[12.09,37.91]$ are chosen to cover a wide range of possible grid resolutions. 
For \tim{the} target $\reyt=5200$, these ranges correspond to $119.3\leq (\dxp,\,\dzp)\leq 808.7$ and~$137\leq \dyp\leq 430$.
The resulting parameter space $\BQ_{n_x}\times\BQ_{n_y}\times\BQ_{n_z}$ is sampled by $5\times 3 \times 5$  Gauss quadrature points. 
If the sampled $\nx$ or $\nz$ has a decimal part, then the default domain lengths $l_x$ and~$l_z$, see \sect~\ref{sec:simDetails}, are slightly modified, so that they are discretized by \tim{an} integer number of cells.
Since in the wall-normal direction such modification is not possible, $\BQ_{n_y}$ is taken differently from the $\BQ_{n_x}$ and~$\BQ_{n_z}$ with three deterministic samples~$\ny=15,\,25$, and $35$.
\rev{In each grid, all cells have the same size.}

\begin{figure}[!t]
\centering
    \begin{tabular}{ccc}
    \includegraphics[scale=0.26]{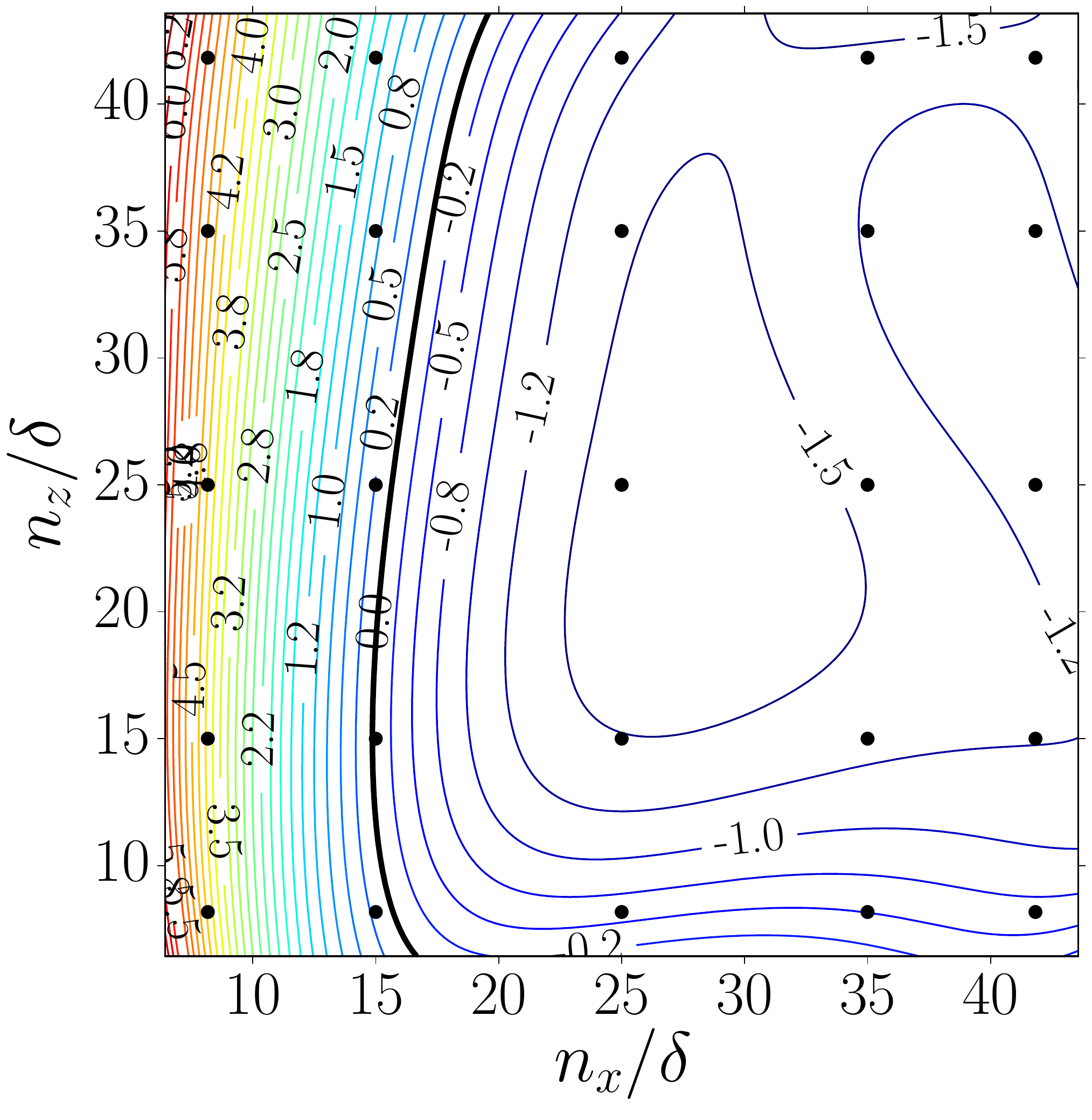} &   \hspace{-0.35cm}    
    \includegraphics[scale=0.26]{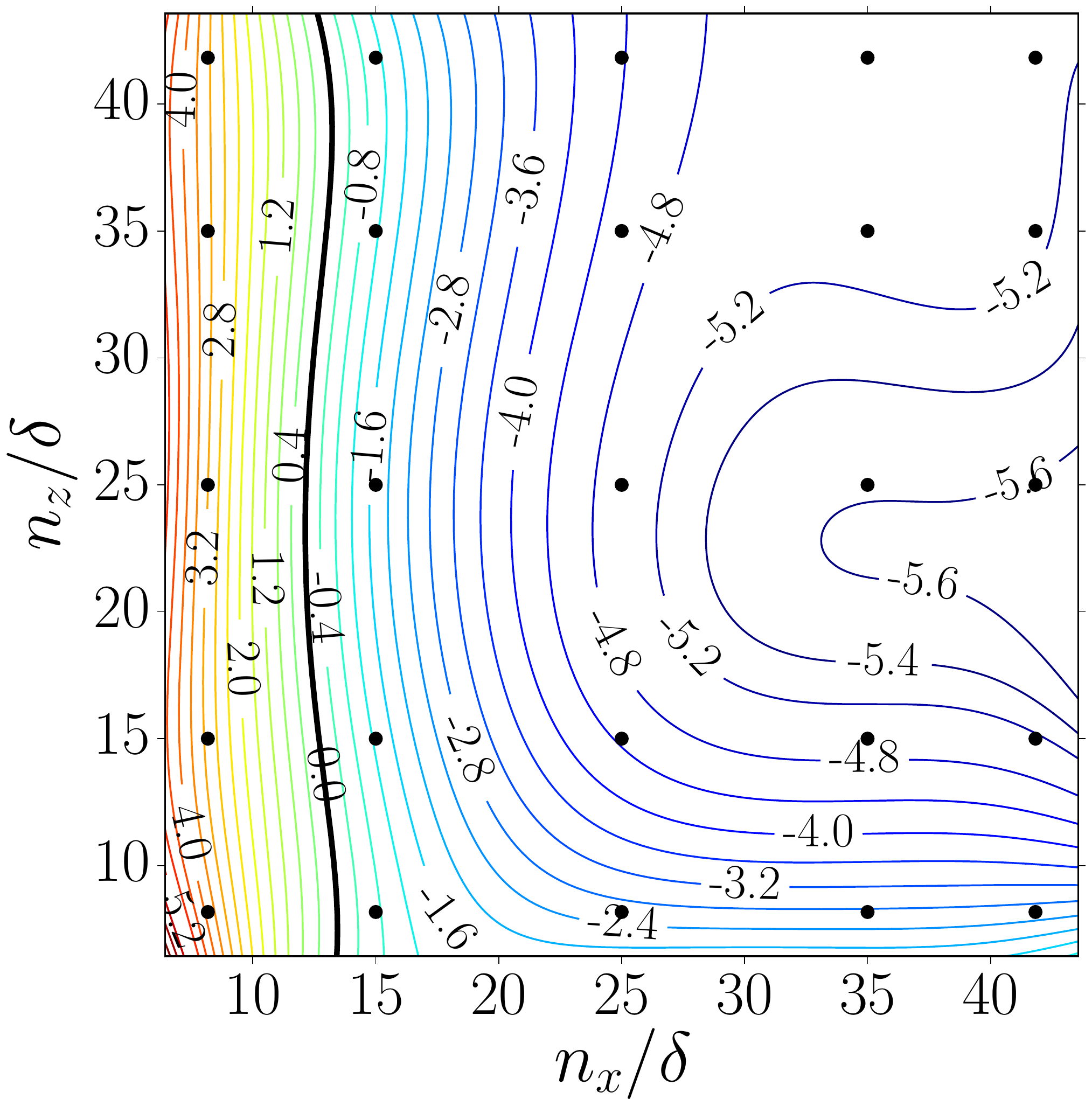} &    \hspace{-0.35cm}  
    \includegraphics[scale=0.26]{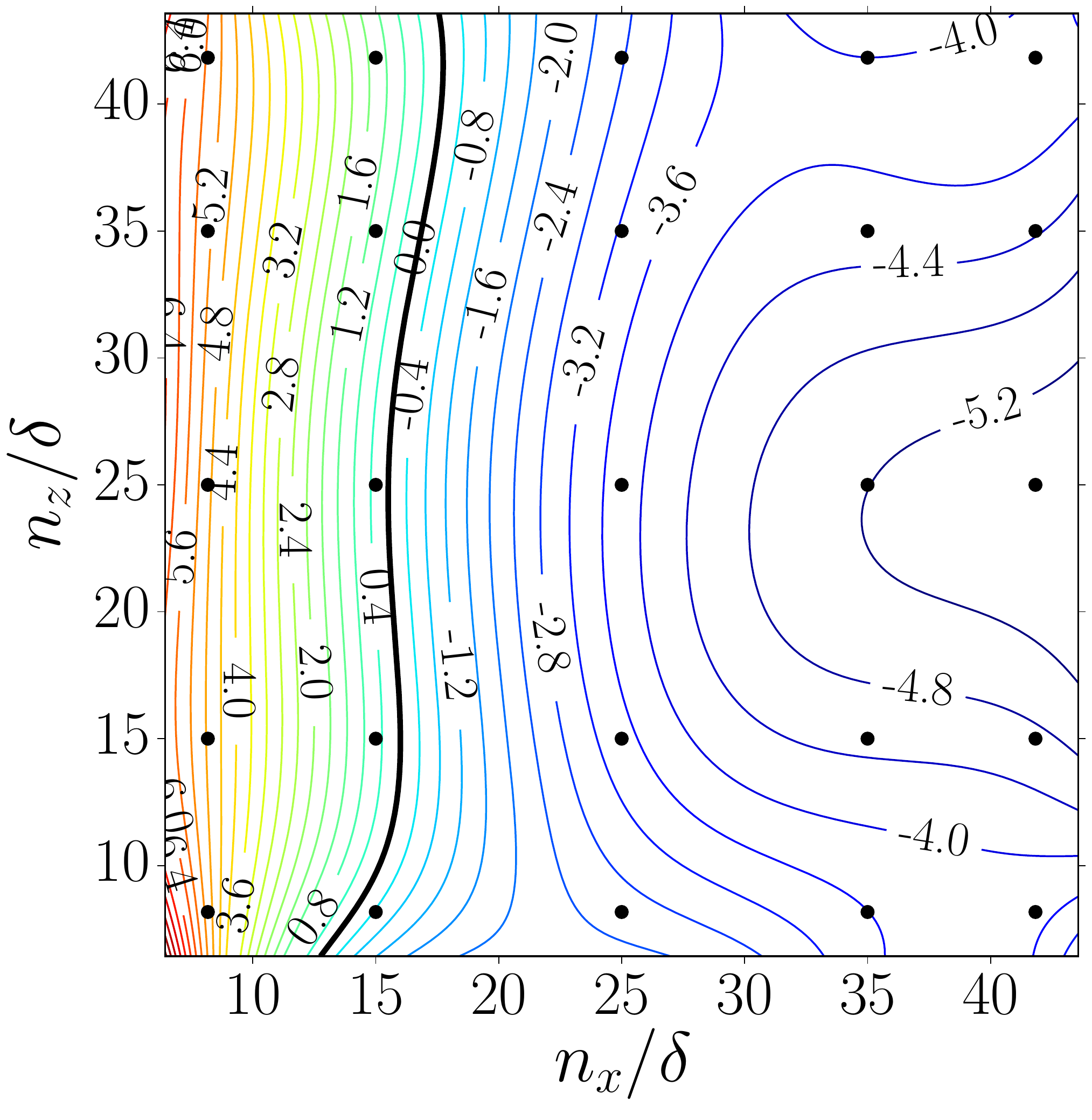} \\
   {\small{(a)}} &    {\small{(b)}}&    {\small{(c)}} \\      
    \includegraphics[scale=0.26]{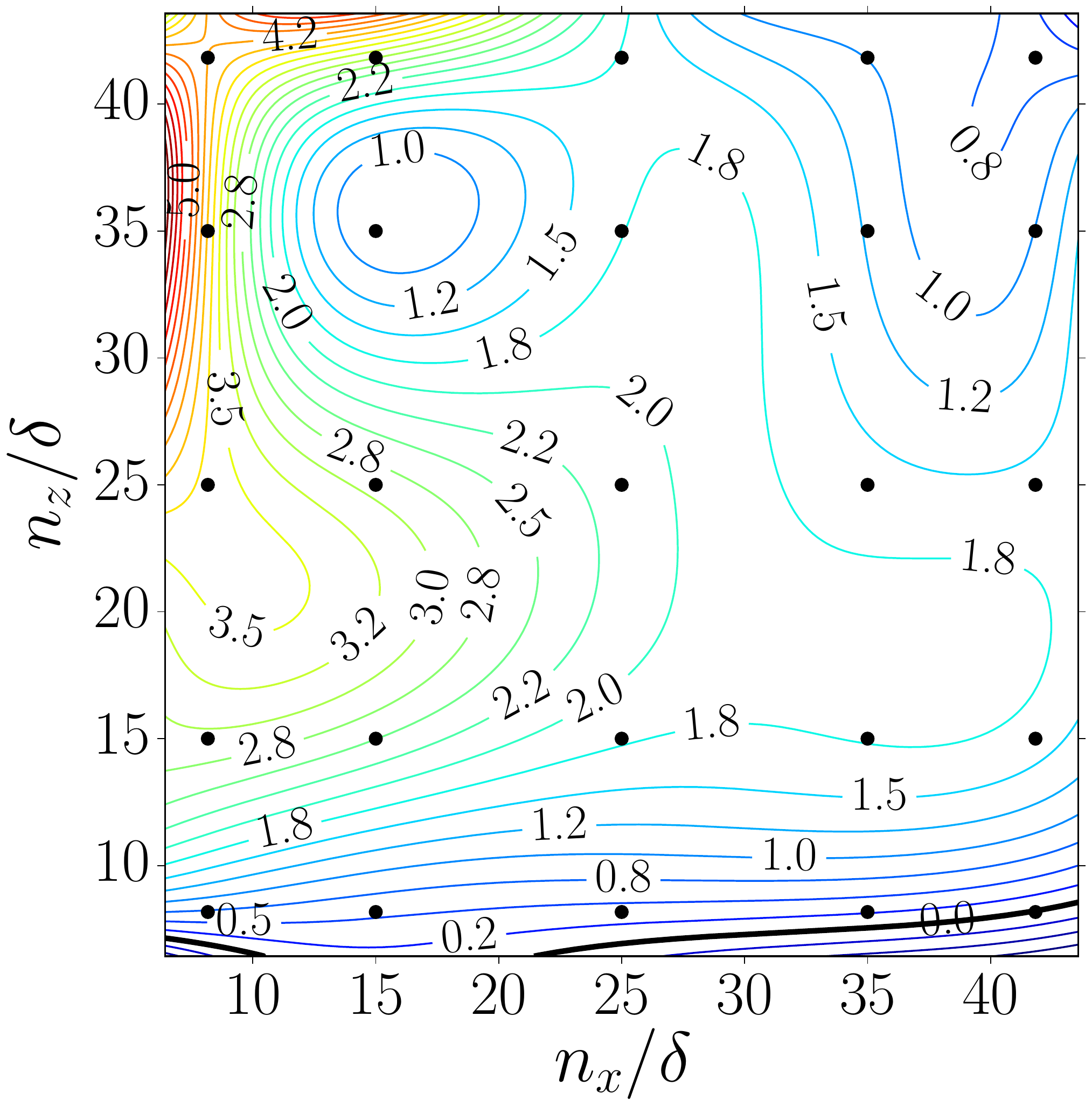} &     \hspace{-0.35cm}
    \includegraphics[scale=0.26]{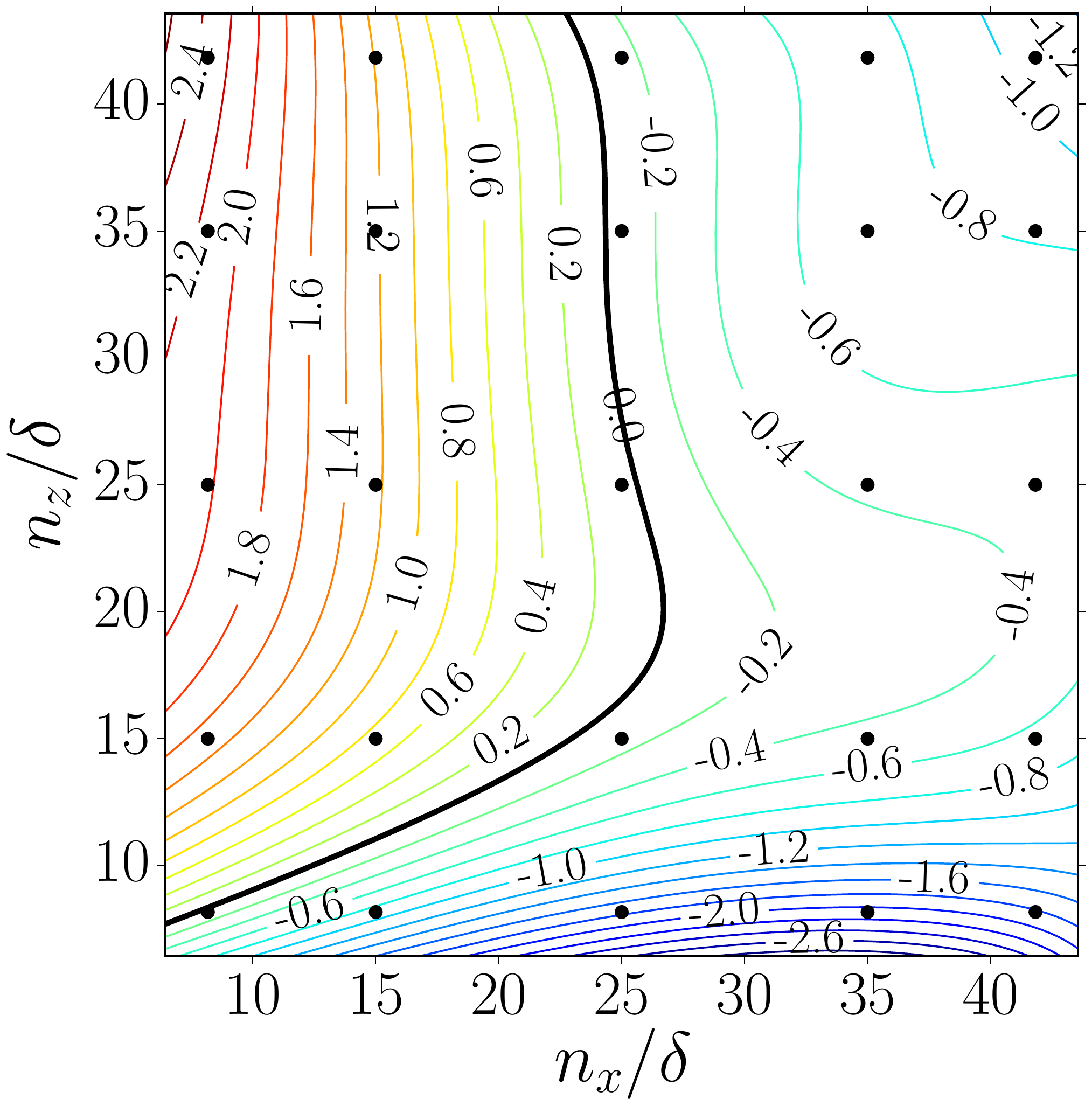} &      \hspace{-0.35cm}
    \includegraphics[scale=0.26]{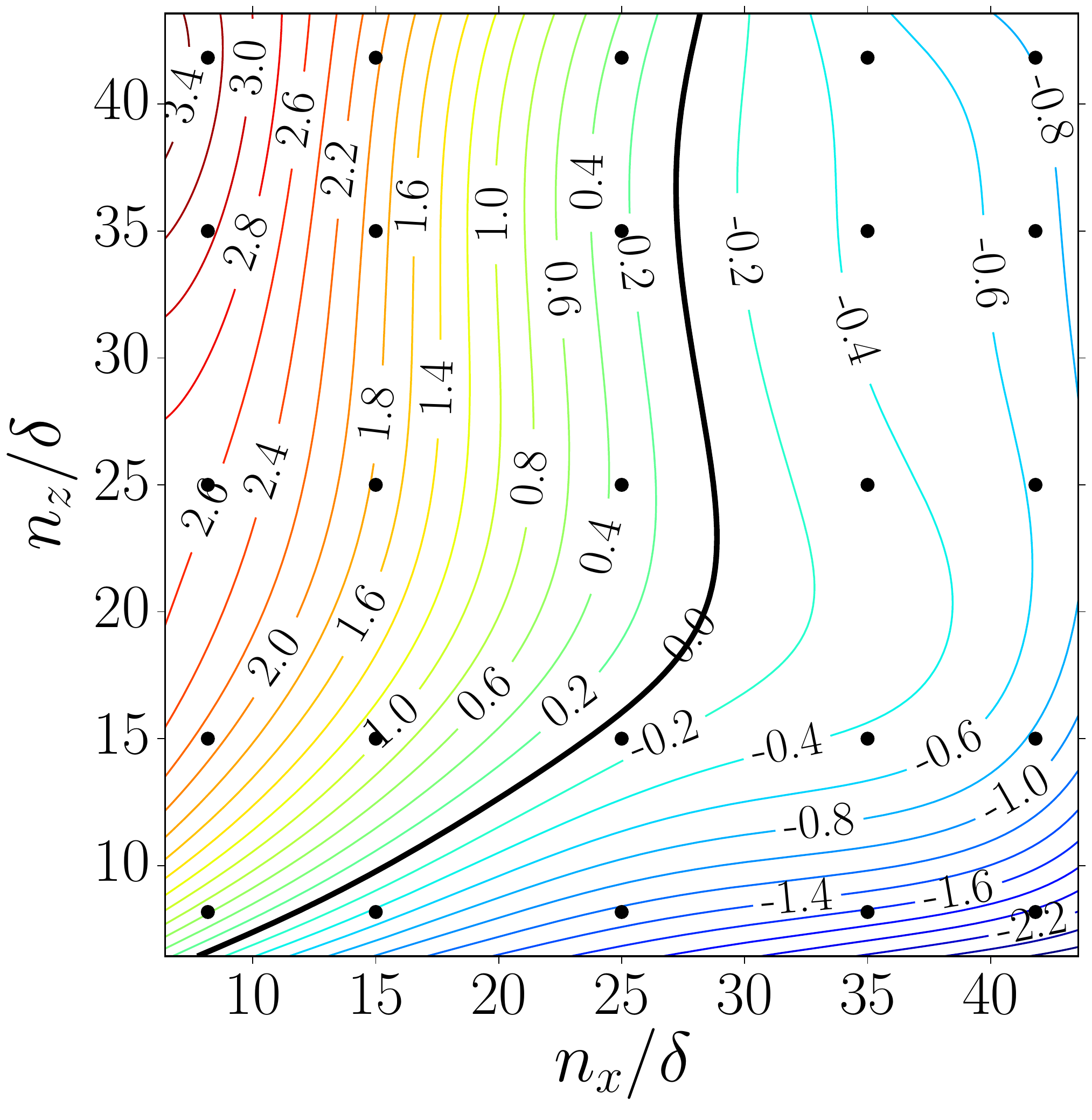} \\      
   {\small{(d)}} &    {\small{(e)}}&    {\small{(f)}} \\         
    \end{tabular}    
    \caption{Isolines of $\epsilon[\lbut]\,\%$ in the $\nx \dash \nz$ plane for the Linear (top row: a,b,c) and the LUST (bottom row: d,e,f) schemes. The wall-normal resolution $\ny$ is equal to $15$ (left: a,d), $25$ (middle: b,e), and $35$ (right: c,f). WALE SGS model and the Spalding law ($\kappa=0.395$ and $B=4.8$) are used with sampling velocity from $\hd=0.1$.}
    \label{fig:duTau_uqGrid}
\end{figure}

The resulting $75$ simulations are carried out using the combinations of the Linear and LUST schemes with \tim{the} WALE and Smagorinsky SGS models. 
The wall modeling is performed \tim{using} the Spalding law (\ref{eq:spalding}) with $\kappa=0.395$ and $B=4.8$.
For this purpose, velocity is sampled from~$\hd=0.1$ that corresponds to the $\scnd$, $\thrd$, and $\frth$ cell centers in the grids with $\ny=15,\,25$, and $35$, respectively.

An important observation based on this simulation set is that grid anisotropy has insignificant effect on the particular pattern of the mean velocity \tim{profile in} the first few cells off the wall.
For instance, the large under-prediction of $\U$ in the first off-wall cell when using the LUST scheme with isotropic grid resolution, see \figs~\ref{fig:numSGSfoot_uPls} and~\ref{fig:WMfoot_uPls}, is also seen for the anisotropic grids.

\fig~\ref{fig:duTau_uqGrid} displays the isolines of $\epsilon[\lbut]$ in the $\nx\dash\nz$ plane for the three adopted values of $\ny$ and using the WALE model. 
The relatively complex pattern of the isolines proves that grid anisotropy can substantially modify the predictions of the wall model. 
It is also noted that in all plots, the rate of changing of the error with grid resolution is inversely proportional to the distance between the isolines. 
Therefore, for both numerical schemes, the rate of variation of $\epsilon[\lbut]$ reduces with increasing $\nx$ and $\nz$, given $\ny$.

Both negative and positive LLM can be produced by the grid anisotropy.
As expected, the numerical scheme has a significant impact on the patterns. 
For the simulations with the Linear scheme, the loci of zero $\epsilon[\lbut]$ are almost insensitive to the variation of the resolution in the spanwise direction.
The $\nx$ at which accurate $\lbut$ are obtained is around $ 14\dash 15$ and slightly varies with $\ny$.
According to the isolines, at each $\ny$, simultaneously choosing high resolutions in the streamwise and spanwise directions leads to \tim{a} large under-prediction of $\lbut$.
This increase in the magnitude of $\epsilon[\lbut]$ can be attributed to large errors in the velocity samples used for wall modeling.
When the LUST scheme is used, the resulting $\epsilon[\lbut]$ at any resolution has smaller values compared to that by the Linear scheme.
Also, contrary to the Linear scheme, the isolines of $\epsilon[\lbut]=0$ at $\ny=15$ are different from those at higher $\ny$.
This originates from the difference in the velocity profiles.

\begin{figure}[!t]
\centering
    \begin{tabular}{ccc}
    \includegraphics[scale=0.26]{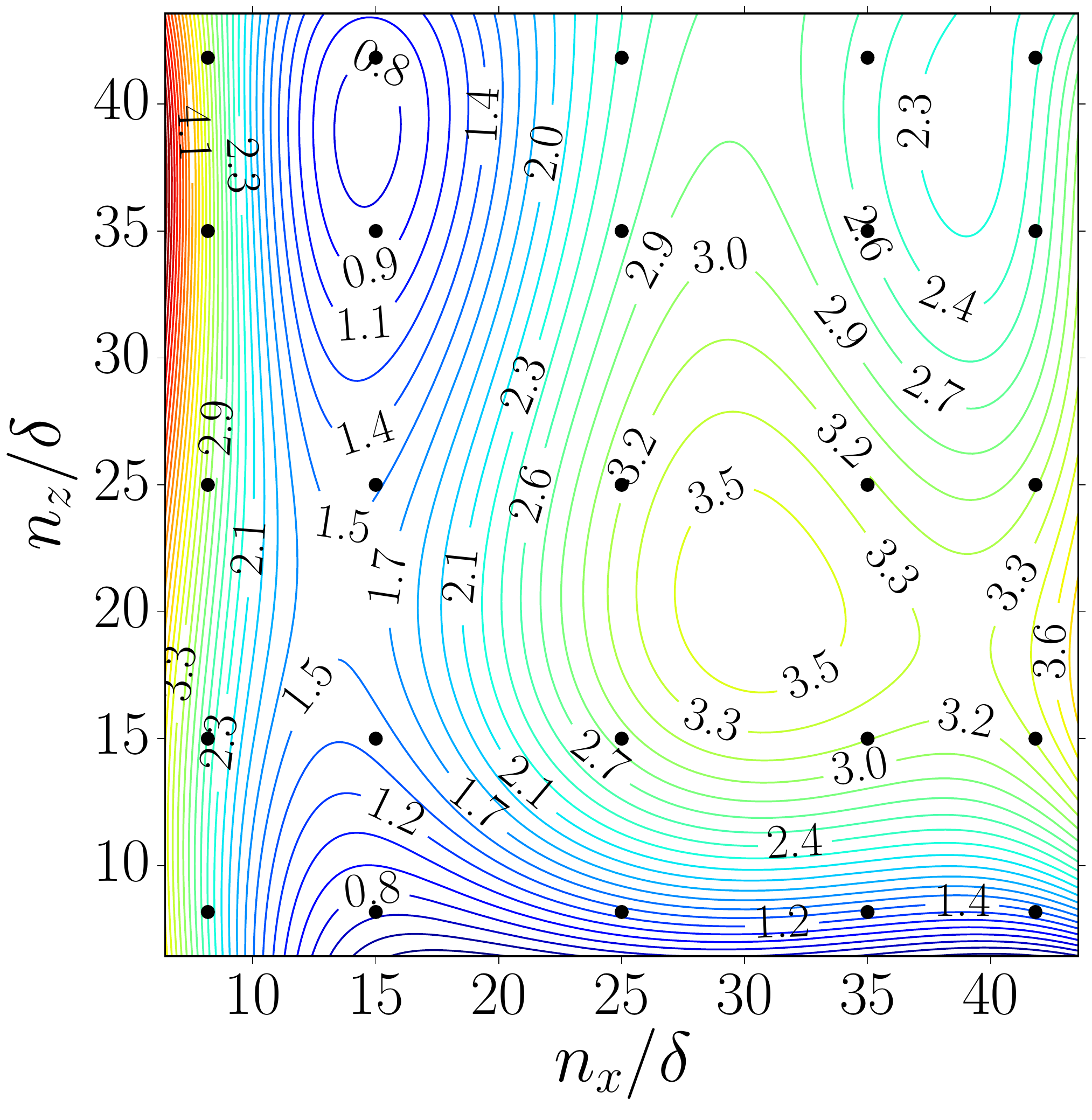} &      \hspace{-0.35cm}    
    \includegraphics[scale=0.26]{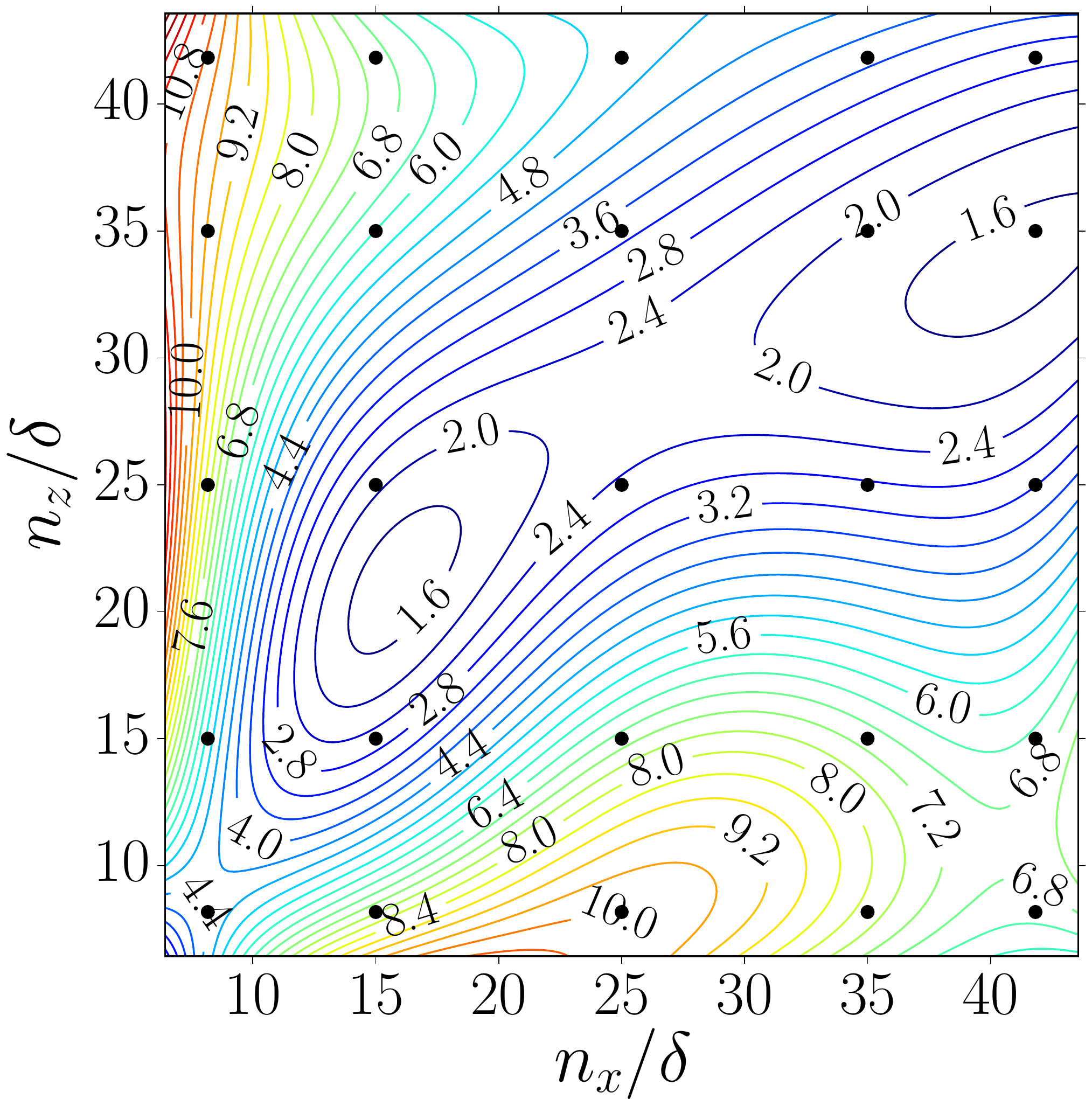} &      \hspace{-0.35cm}    
    \includegraphics[scale=0.26]{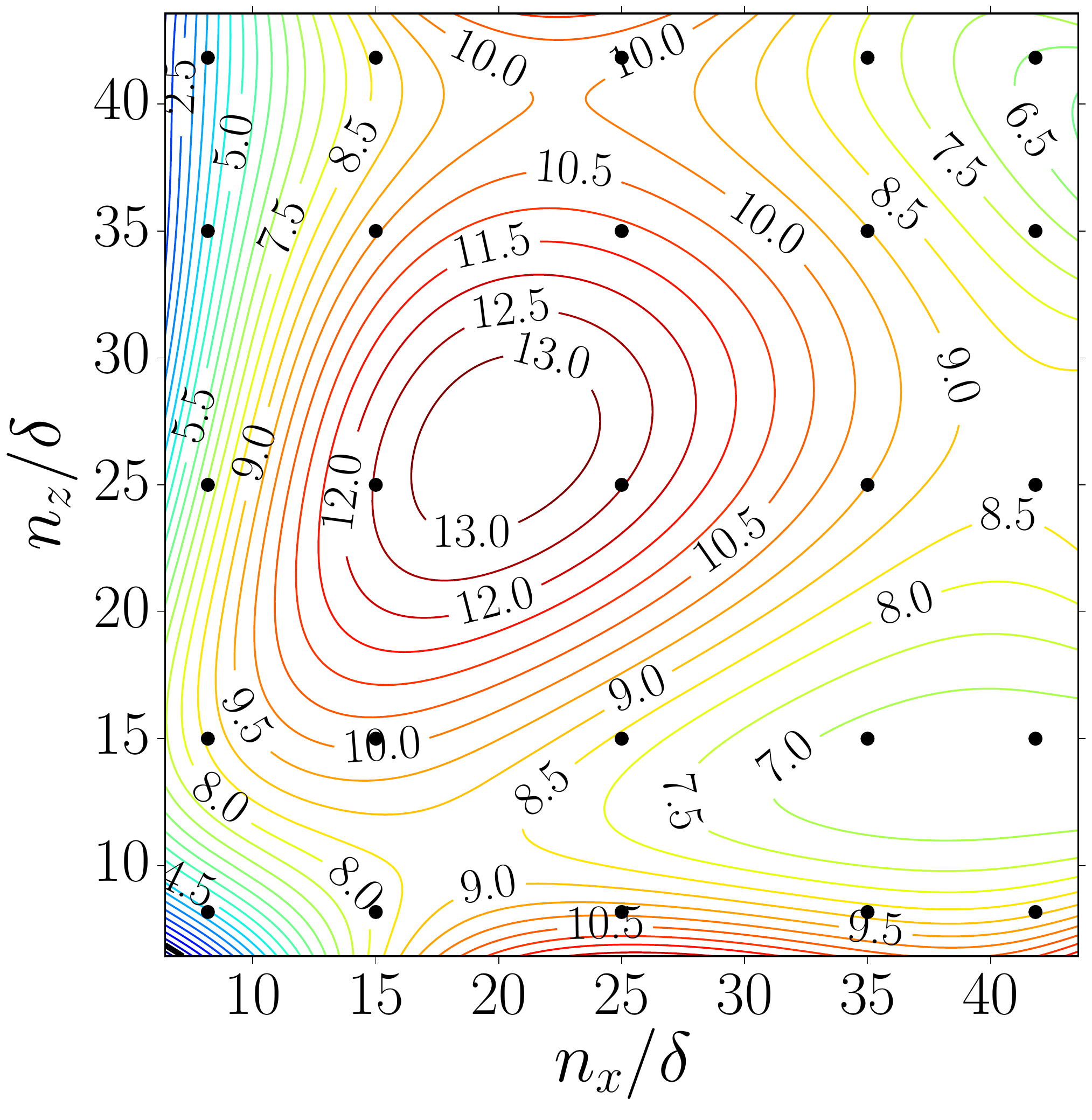} \\
   {\small{(a)}} &    {\small{(b)}}&    {\small{(c)}} \\      
    \includegraphics[scale=0.26]{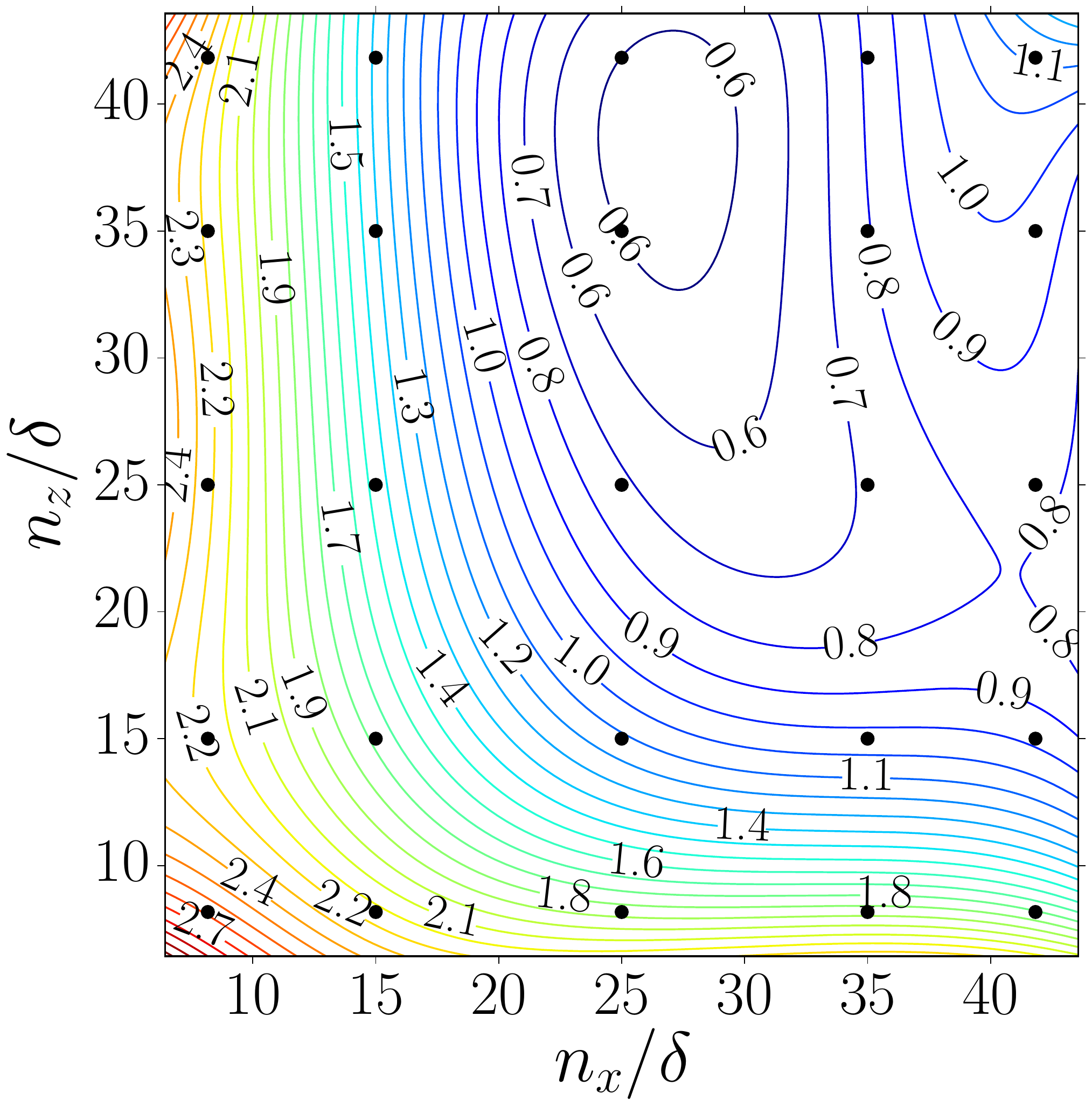} &      \hspace{-0.35cm}    
    \includegraphics[scale=0.26]{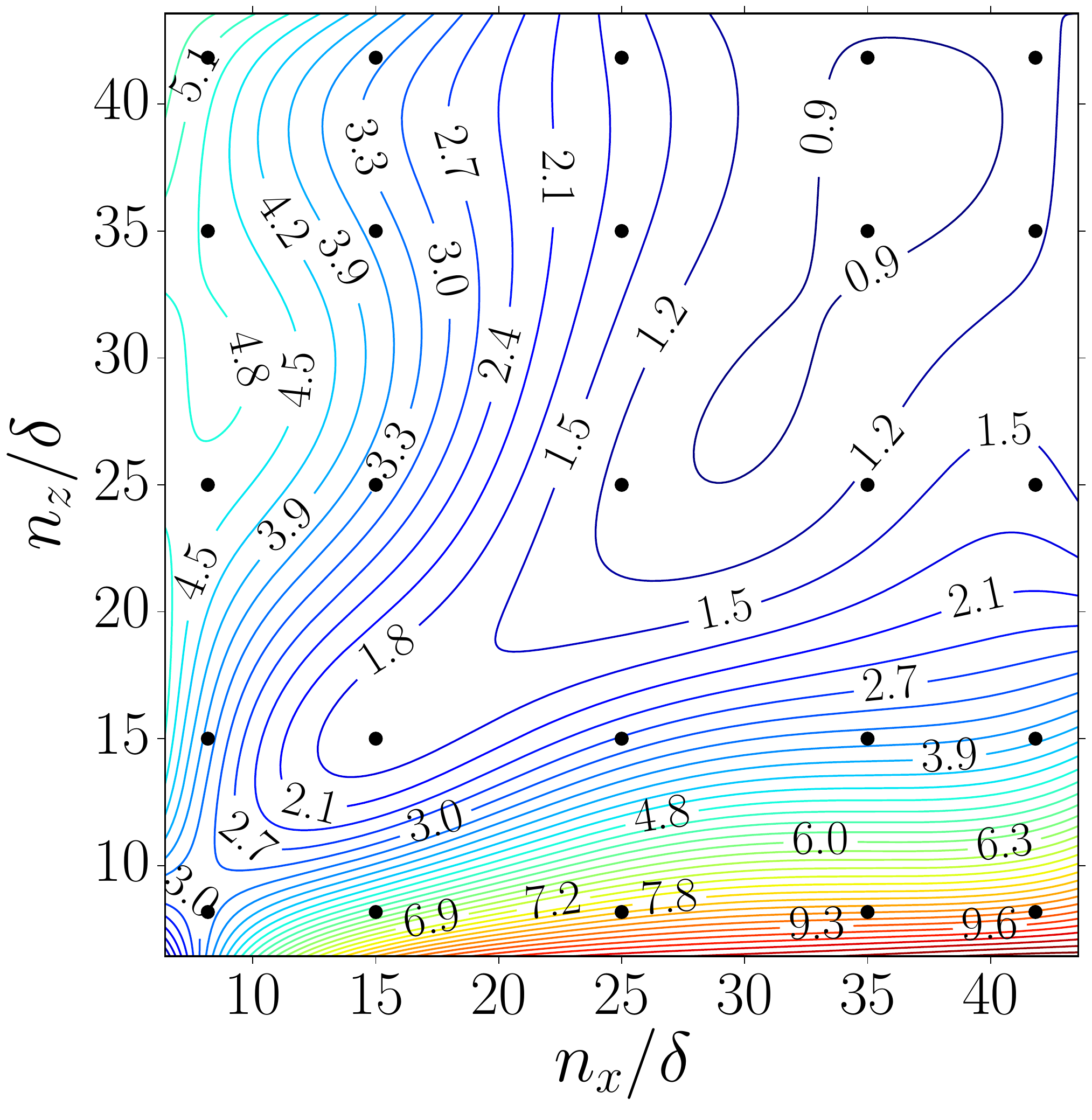} &       \hspace{-0.35cm}    
    \includegraphics[scale=0.26]{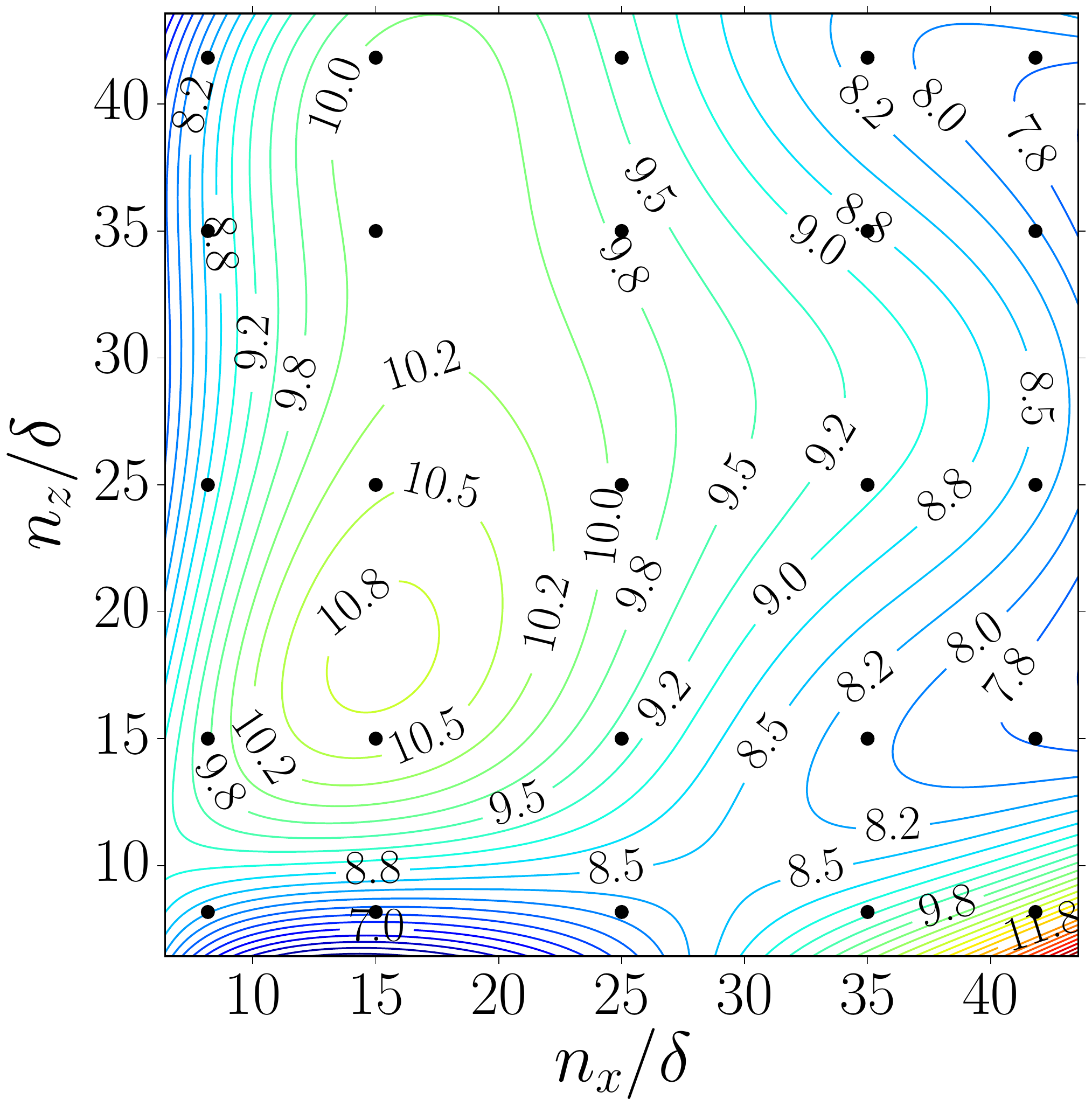} \\
   {\small{(d)}} &    {\small{(e)}}&    {\small{(f)}} \\         
    \end{tabular}
    \caption{Isolines of $\einf[\U]\,\%$ (a,d), $\einf[\buv]\,\%$ (b,e), and $\einf[\bk]\,\%$ (c,f) in the $\nx \dash \nz$ plane for $\ny=25$. The Linear (top~row) and LUST (bottom row) schemes are used. The plots correspond to \fig~\ref{fig:duTau_uqGrid}(b,e).}
    \label{fig:isolinesUny25}
\end{figure}

In \fig~\ref{fig:isolinesUny25}, the isolines of $\einf[\U]$, $\einf[\buv]$, and $\einf[\bk]$ for the simulations in \fig~\ref{fig:duTau_uqGrid} with $\ny=25$ are shown in the $\nx\dash\nz$ plane.
\revCom{For a set of the results represented in these two figures, the convergence of gPCE metamodels is shortly discussed in~\ref{app:gPCEConvergence}.}
By increasing the resolution $\ny$ to 35, small differences appear in the pattern of the error isolines \revCom{of \fig~\ref{fig:isolinesUny25}}, however, at $\ny=15$ considerable differences exist (not shown here).
\tim{Unfortunately,} for neither of the schemes \tim{do} the loci of small errors in the velocity moments occur at the grid resolutions corresponding to $\epsilon[\lbut]=0$. 
When the Linear scheme is adopted, two regions of small~$\einf[\U]$ appear in the $\nx\dash\nz$ plane, one at $\nz\lesssim 15$ and the other one around $\nx\approx 15$. 
Over neither of these regions, the values of $\einf[\buv]$ and $\einf[\bk]$ simultaneously become small.
For the LUST scheme, the coarsest resolution in the $\nx\dash\nz$ plane results in the highest value of $\einf[\U]$. 
By simultaneous refining the grid in the streamwise and spanwise directions, this error reduces, with comparable sensitivities to $\nx$ and $\nz$. 
Eventually the smallest $\einf[\U]$ and $\einf[\buv]$ is obtained for the simultaneous adoption of resolutions $\nx\gtrsim 20$ and $\nz\gtrsim 25$.
This is not followed by the reduction of \tim{the} error in $\bk$.
For both Linear and LUST schemes, $\einf[\bk]$ may decrease at the coarse resolutions $\nz$. 
Moreover, for a fixed resolution $\nx$ and $\nz$, increasing $\ny$ increases $\einf[\bk]$.
These trends are similar to what formerly observed in \fig~\ref{fig:kProfs_nTest} in case of isotropic grid.

\begin{figure}[!t]
\centering
   \begin{tabular}{cc}
   {\small(a)} &
   \includegraphics[scale=0.5]{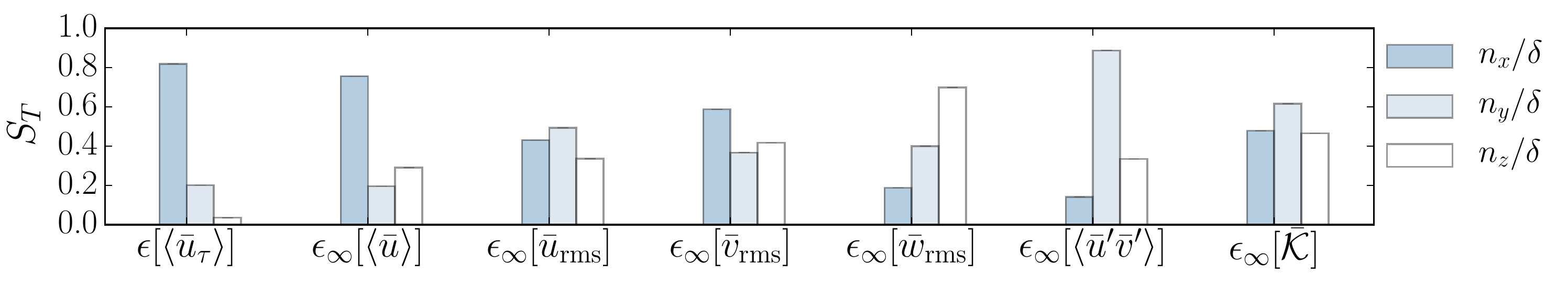}    \\
   {\small(b)} &
   \includegraphics[scale=0.5]{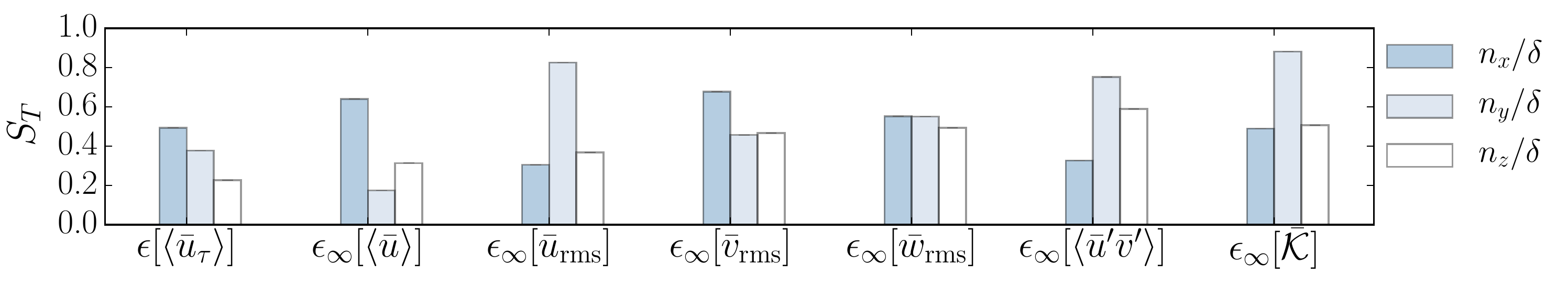}    \\   
   {\small(c)} &
   \includegraphics[scale=0.5]{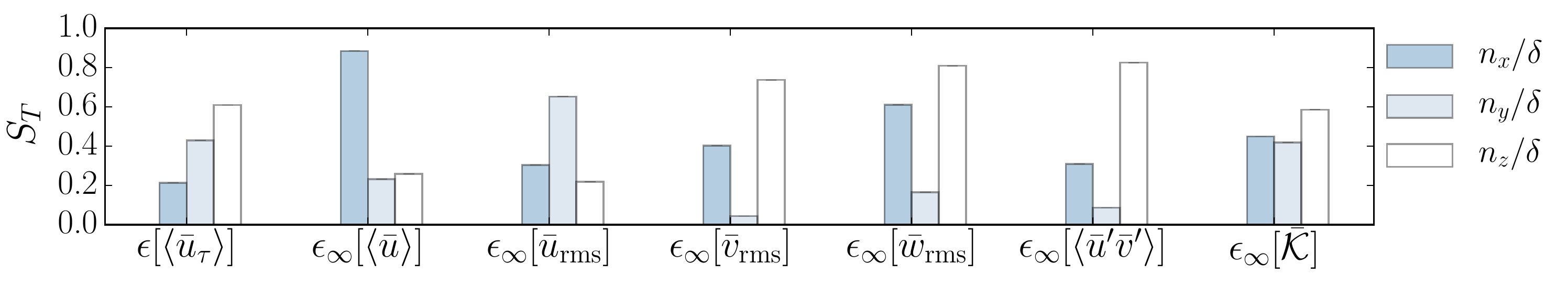}    \\
   {\small(d)} &
   \includegraphics[scale=0.5]{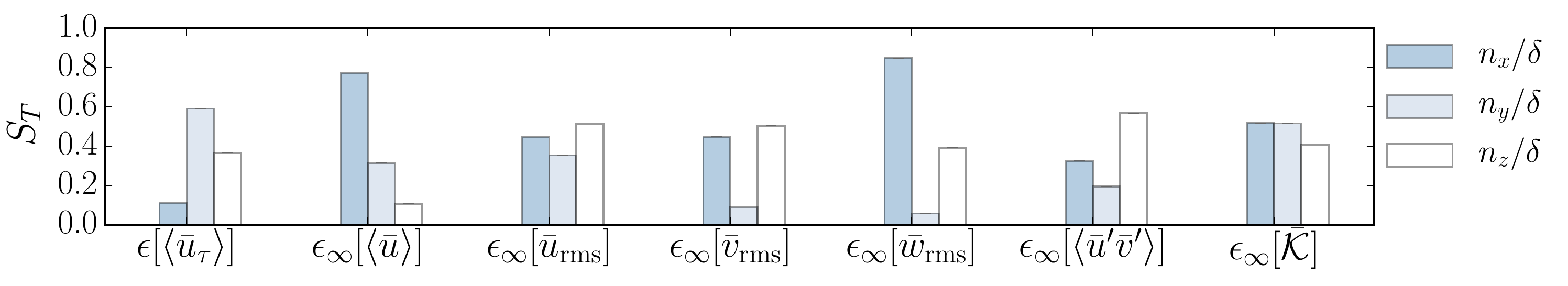}    \\     
   \end{tabular}
\caption{Total Sobol indices (\ref{eq:sobol}) representing the sensitivity of the measured errors with respect to the variations in grid resolutions $\nx$, $\ny$, and $\nz$, for Linear-WALE (a), LUST-WALE (b), Linear-Smagorinsky~(c), and LUST-Smagorinsky~(d). The indices in (a) and (b) belong to the simulations illustrated in \figs~\ref{fig:duTau_uqGrid}~and~\ref{fig:isolinesUny25}. In all simulations, the Spalding law~(\ref{eq:spalding}) with $\kappa=0.395$ and $B=4.8$ and $\hd=0.1$ is used.}
\label{fig:sobolGridAnisot}
\end{figure}

To \tim{further} quantify how sensitive the errors in different QoIs are with respect to grid anisotropy, global sensitivity analysis is conducted.
This analysis is a complement to the above UQ forward problem, \ie~the propagation of the uncertainty in choosing the grid resolution. 
The results are reported in terms of total Sobol indices (\ref{eq:sobol}) for different errors, as shown in \fig~\ref{fig:sobolGridAnisot}.
\rev{Note that, to show the influence of the SGS model, the indices for WMLES similar to those in \fig~\ref{fig:duTau_uqGrid} and~\ref{fig:isolinesUny25} but with the Smagorinsky model are also provided.}

For both numerical schemes and both SGS models, $\einf[\U]$ are most sensitive to the variations of the grid cell size in the streamwise direction. 
This is also the case for $\epsilon[\lbut]$ when the WALE model is used. 
The relative magnitude of the indices with respect to $\nx$, $\ny$, and $\nz$ for the velocity fluctuations and velocity's second-order moments depends on the choice for the SGS model. 
For instance, the highest sensitivity of $\einf[\buv]$ is with respect to $\ny$ \tim{when} using WALE, and to~$\nz$ for using Smagorinsky. 
There are error responses, such as $\einf[\bk]$ when using the Smagorinsky model, for which the Sobol indices are more or less the same.

\tim{The conducted analysis clearly shows that the response to grid anisotropy varies among the QoIs, making it difficult to recommend a bias in the grid resolution along a particular direction.
Moreover, the effect of anisotropy also varies depending on the other modeling parameters employed.
Consequently, adopting an anisotropic grid is not generally recommended.
}

\subsection{Best practice guidelines for WMLES}\label{sec:guidelines}
Based on the extensive study of different influential factors on the accuracy of   WMLES results, \tim{a set of guidelines for obtaining accurate QoIs is proposed here.} 
To obtain \tim{an} accurate $\U$ in the outer layer, the proper choice of the numerical scheme, SGS model, and grid resolution is important.
Between the two employed schemes used for constructing the convective flux of (\ref{eq:LESmomentum}), the LUST scheme \tim{performs better}.
In fact, even without the use of \tim{a} wall model, see \sect~\ref{sec:footprint}, the simulations with \tim{a} coarse grid resolution performed \tim{using} this scheme can lead to \tim{a} low-error mean velocity profile in the outer layer. 
Moreover, the behavior of the LUST scheme is found to be more consistent and predictable. 
For instance, in case of grid refinement, see \figs~\ref{fig:noWM_conv}(b) and~\ref{fig:convtest_n}(b), the values of $\einf[\U]$ \tim{in} the simulations with the LUST scheme decrease with increasing the grid resolution.
In contrast, for the Linear scheme, $\einf[\U]$ may increase when refining the grid.

The LUST scheme provides enough dissipation so that using different SGS models does not \tim{significantly} influence the outcome.
The only improving effect of the WALE model compared to \tim{the} Smagorinsky model, is the slight improvement of $\einf[\U]$, as it can be inferred from \figs~\ref{fig:convtest_n}(b) and \fig~\ref{fig:convtest_h}(b). 
Another significant role of the SGS model is its impact on the balance of the shear stresses, see \eq~(\ref{eq:xMomChan}), and hence its ability to control the spurious overshoots in the velocity's second-order moments. 
In particular, the Smagorinsky model is found to be capable of removing the overshoots in $-\buv$ and $\bk$, see \sect~\ref{sec:spuriousOvershoot}.
However, since the \tim{primary objective is obtaining accurate} profiles of the second-order moments in the outer layer, the overshoots can be neglected.
\tim{Hence, using the WALE model is recommended.}

\tim{As shown in} \sect~\ref{sec:nTests}, for the simulations with the LUST scheme \tim{and an} isotropic grid,~$\einf[\U]$ becomes less than $1\%$ for resolutions $\nd \geq 25$.
This level of error is also found to be enough for obtaining accurate~$\lbut$.

Sampling velocity from any point in the outer part of the TBL, where $\U$ has \tim{a} small deviation from the reference data and also the law of the wall is valid, \tim{is} suitable to get accurate predictions for $\lbut$.
\tim{In the particular case of using the LUST scheme} \revCom{with any SGS model but Smagorinsky}, \tim{any cell excluding the wall-adjacent one can be used.}
To make these predictions more accurate, employing \revCom{adjusted} parameters for the law of the wall is \tim{recommended}. 
These parameters \tim{can be} obtained by calibrating the law of the wall against the \revCom{reference} DNS data for the inner-scaled mean velocity profile. 
\tim{For channel flow at $\reyt=5200$}, the values of $\kappa=0.395$ and $B=4.8$ for the Spalding law of the wall (\ref{eq:spalding}) \tim{are a good starting point for a wide range of~$h/\delta$}.

\rev{For channel flow,} once an accurate $\lbut$ prediction is obtained using a slightly dissipative scheme and enough resolution, an accurate prediction of the resolved Reynolds shear stress profile,~$-\buv$, \rev{in the outer layer follows.}
\rev{This is based on the discussion in \sect~\ref{sec:qoiRelation} and the observations in \sects~\ref{sec:WMgain} and \ref{sec:nTests}.}

The resolved velocity fluctuations and hence the resolved TKE, $\bk$, are the QoIs for which no intuitive procedure for controlling associated accuracy could be proposed. 
\rev{However, if the mentioned best practice guidelines for WMLES are followed, the predicted $\bk$ profiles in the outer layer are relatively acceptable since they resolve at least $ 80\%$ of the total TKE.}

\tim{An important question is whether the proposed guidelines, which are based on a study of a single flow at a single $\rey$-number, are applicable for WMLES of other flow configurations.
Currently, this has been shown to be the case for simulations of the zero-pressure-gradient (ZPG)-TBL flow \cite{timECCOMAS:18} and of the flow over a backward-facing step \cite{mukha:18}.
Considering a wider range of flows is a subject for future work.
The applicability of the guidelines for channel flow at other Reynolds numbers is discussed in the section below.}

\subsection{Extension to other $\rey$-numbers}\label{sec:ReEffect}
\tim{Results from a set of channel flow simulations at $\reyt$ other than $5200$ are presented here in order to demonstrate that the modeling choices recommended in the section above still lead to high accuracy.}
\tim{Following the proposed guidelines}, the LUST scheme, \tim{the} WALE model \tim{and an isotropic grid with resolution $\nd=25$} are used \tim{in all the simulations}. 
The domain size is the default one, see \sect~\ref{sec:simDetails}, that is consequently discretized by $1\,125\,000$ cells.
For wall modeling, the Spalding law (\ref{eq:spalding}) is used for which the velocity samples are taken from $h/\delta=0.1$ which corresponds to the $\thrd$ cell center off the wall.
The values $\kappa=0.395$ and $B=4.8$ are adopted in all the simulations.

\begin{table}
\centering
\caption{Summary of the WMLES of channel flow at different $\reyt$ using the LUST scheme and WALE model with $\nd=25$, $\hd=0.1$ ($\thrd$ cell center), and the Spalding law with $\kappa=0.395$ and $B=4.8$.}\label{tab:otherRe}
   \begin{small}
   \begin{tabular}{ccc|cccc}
   \toprule\toprule
   \multicolumn{3}{c}{$\reyt$} & \multicolumn{4}{c}{Error \% in WMLES QoIs} \\
   Target & Reference & WMLES & $\epsilon[\lbut]$ & $\einf[\U]$ & $\einf[\buv]$ & $\einf[\bk]$ \\
   \hline
   2\,000 & 1\,994.756, DNS \cite{lee-moser:15} & 2\,009.348 & 0.73 & 1.15 & 1.59 & 8.77 \\
   5\,200 & 5\,185.897, DNS \cite{lee-moser:15} & 5\,190.375 & 0.09 & 0.72 & 1.21 & 10.02 \\
   8\,000 & 8\,016, DNS \cite{yamamoto:18} & 7\,957.014 & -0.74 & 0.71 & 1.57 & 10.81 \\
   20\,000 & 20\,000, \eq~(\ref{eq:cess}) & 19\,792.421 & -1.04 & 0.98 & 2.33 & --\\
   100\,000 & 100\,000, \eq~(\ref{eq:cess}) & 98\,262.119 & -1.74 & 0.96 & 3.52 & --\\   
   \bottomrule
   \end{tabular}
   \end{small}
\end{table}

\begin{figure}[!htbp]
\centering
   \begin{tabular}{c}   
   \includegraphics[scale=0.35]{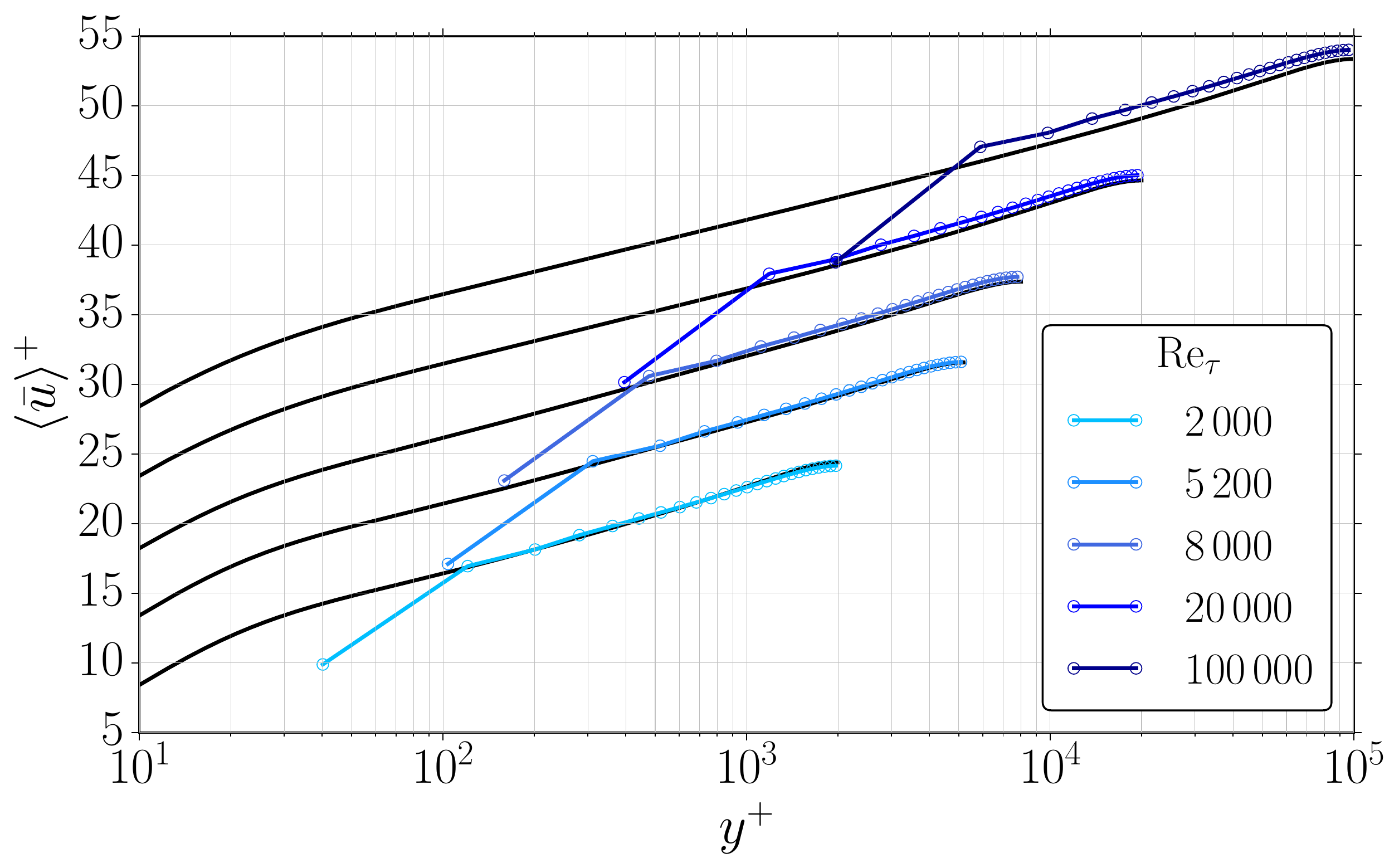} \\
   {\small{(a)}} \\   
   \end{tabular}
   
   \begin{tabular}{cc}   
   \includegraphics[scale=0.38]{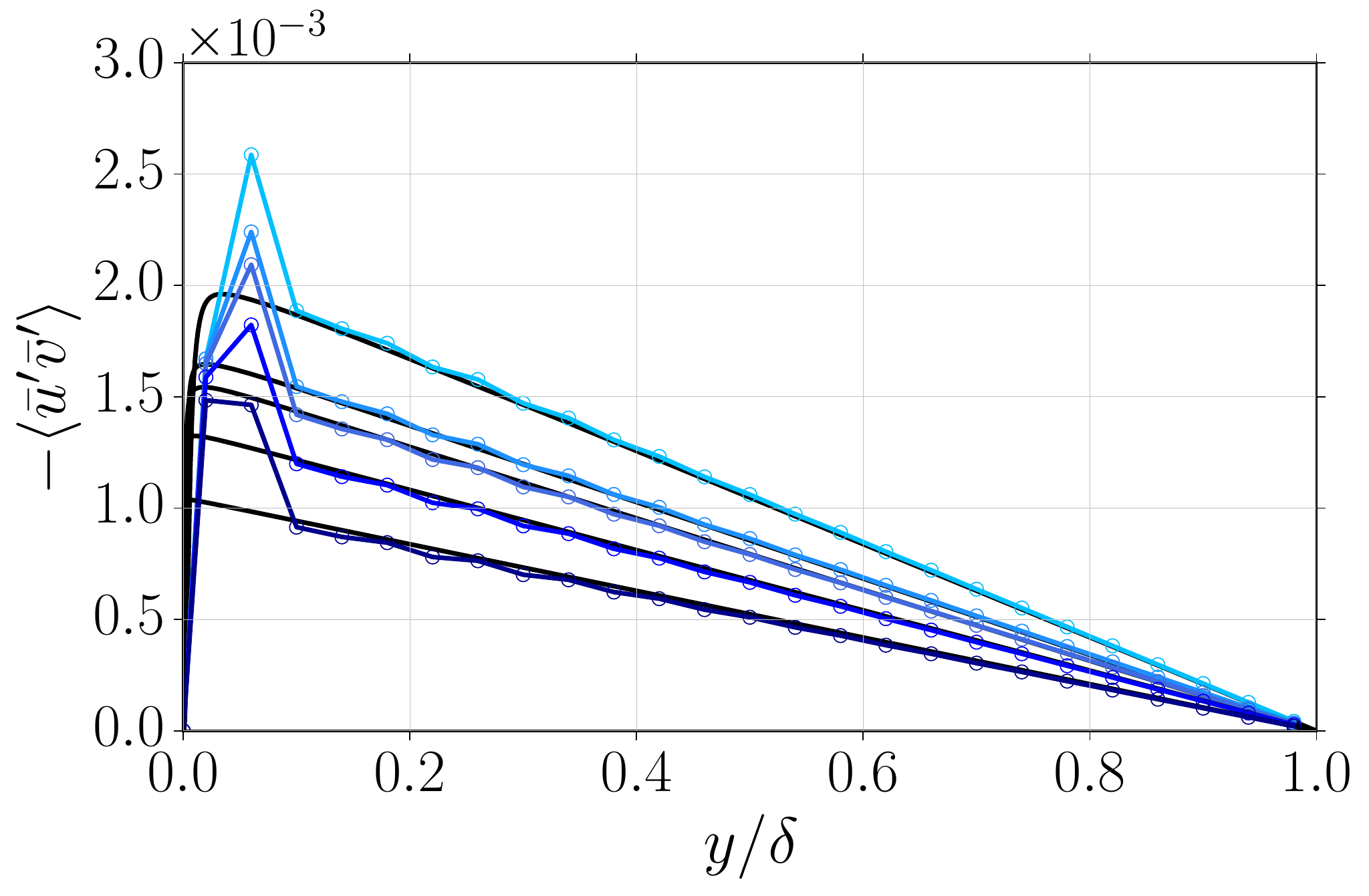}&
   \includegraphics[scale=0.38]{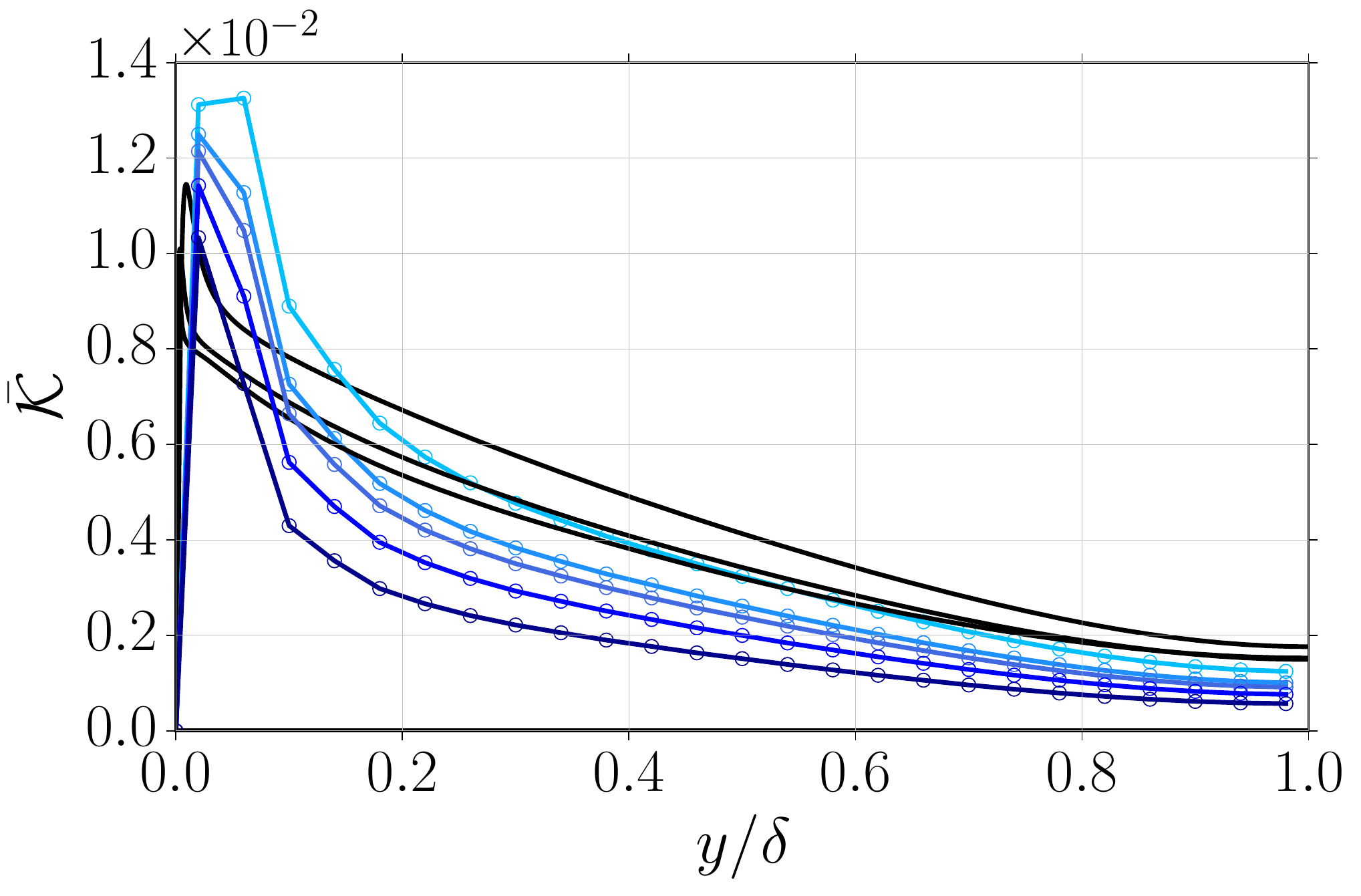} \\
   {\small{(b)}}&    {\small{(c)}} \\         
   \end{tabular}
   \caption{Inner-scaled velocity (vertically shifted by $5$ wall units) (a), outer-scaled $-\buv$ (b), and outer-scaled $\bk$ (c) profiles of WMLES of channel flow at $\reyt=2\,000,\,5\,200,\,8\,000,\,20\,000$, and $100\,000$. Reference data are represented by solid black lines. \revCom{For details of WMLES, see the caption of}~\tab~\ref{tab:otherRe}. }
   \label{fig:allReProfs}
\end{figure}

\begin{figure}[!htbp]
\centering
\includegraphics[scale=0.60]{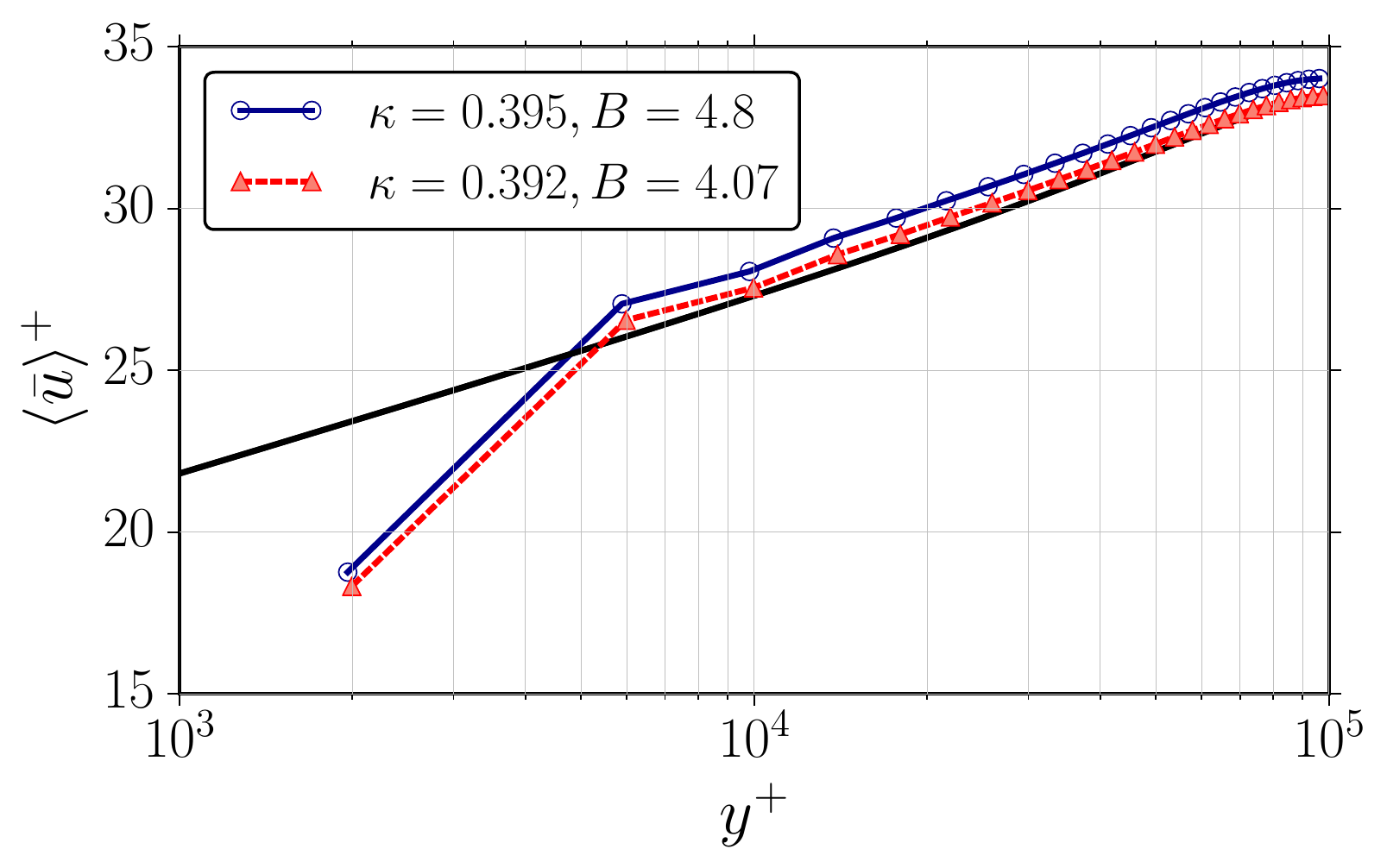}
\caption{Influence of the parameters of the Spalding law (\ref{eq:spalding}) on the inner-scaled velocity profile of channel flow at $\reyt=10^5$. Reference data are represented by solid black line.}
\label{fig:Ret10000KapB}
\end{figure}

The resulting errors in the QoIs are summarized in \tab~\ref{tab:otherRe} and the associated profiles of~$\U$, $-\buv$, and~$\bk$ are plotted in \fig~\ref{fig:allReProfs}. 
For $\reyt$ up to $8000$, the available DNS datasets~\cite{lee-moser:15,yamamoto:18} are used as the reference. 
For higher $\reyt$, an empirical expression suggested by Cess \cite{cess:58} is \rev{calibrated using the DNS data of \cite{hoyas:08,lee-moser:15}} and then employed to create the profiles of the mean velocity and~$\buv$, see\app~\ref{app:CessProfs}. 
\tim{For} these cases, no reference data for the $\bk$ profile is available.
The low error in the QoIs is approximately independent of the Reynolds number, confirming the appropriateness of the suggested guidelines. 
This is despite the fact that \tim{for} the fixed grid resolution $\nd=25$, \tim{the size of the grid cells in wall units is relatively small, and thus not optimal, in the case of $\reyt=2000$}.
This might be the reason that $\einf[\U]$ for~$\reyt=2000$ is slightly larger compared to higher $\rey$-number flows.
The magnitude of the error in $\lbut$ slightly increases with $\reyt$-number, \tim{as} evident from the $\U^+\dash y^+$ plots in \fig~\ref{fig:allReProfs}(a).
This is mainly because of using the values of $\kappa=0.395$ and $B=4.8$ that are optimal choices for $\reyt=5200$, \tim{and} not necessarily for the other~$\reyt$.

To confirm this, the simulation at $\reyt=10^5$ is repeated \revCom{using corresponding} optimal parameters \mbox{$\kappa=0.392$} and $B=4.07$. 
The resulting $\lbut$ and hence $\buv$ are improved significantly.
In fact, the values of $\epsilon[\lbut]$ and~$\einf[\buv]$ become respectively equal to $-0.24\%$ and $0.66\%$. 
In contrast, the value of~$\einf[\U]$ reduces \tim{only slightly,} to $0.92\%$.
The improved inner-scaled velocity profile can be observed in \fig~\ref{fig:Ret10000KapB}.

\section{Summary and Conclusions}\label{sec:conclusions}
This work presents the results of a systematic investigation of the influence of different modeling choices on the predictive accuracy of wall-modeled large eddy simulation.
In particular, algebraic wall-stress modeling is considered and a model based on the Spalding law is employed.
The factors the influence of which is studied are the resolution and anisotropy of the computational grid, the parameters of the wall model ($\kappa$ and $B$), the distance $h$ to the wall model's sampling point, the interpolation scheme for convective cell-face fluxes, and SGS modeling.
To quantify the effect of these parameters on WMLES results, simulations of fully developed turbulent channel flow at~$\reyt = 5200$ are performed.
For each considered combination of modeling choices, errors in several QoIs are evaluated, including the mean wall shear stress as well as first- and second-order statistical moments of velocity.
DNS data~\cite{lee-moser:15} is used as reference.

To reduce the necessary number of simulations \rev{when studying the influence of wall model parameters and grid anisotropy,} uncertainty quantification techniques are employed.
In particular, generalized polynomial chaos expansions~\cite{xiu_gPCE} with collocation method~\cite{xiu:07}, are used to construct metamodels for the responses of the QoIs.
Furthermore, sensitivity analysis (both local and global) is \rev{performed} to complement the investigation, see Section~\ref{sec:LSAGSA} and also Section~\ref{sec:gridAnisot}.

The analysis in Section~\ref{sec:footprint} focuses on the influence of SGS modeling, wall modeling, and  the interpolation scheme for the convective fluxes.
It is shown that both the SGS model and the interpolation scheme have a significant effect on the mean velocity profile, and also on the accuracy of the wall modeling, via its dependency on the sampled velocity signal.
The results reveal the importance of having enough \rev{numerical} dissipation in the \rev{simulation} in order to properly redistribute the streamwise momentum across the channel.
This can be achieved by using a slightly dissipative interpolation scheme, such as LUST, or a dissipative SGS model, such as Smagorinsky.
The combination of the WALE SGS model and the LUST scheme was observed to result in the best accuracy.
As expected, wall modeling is shown to dramatically improve the predictions of the mean wall shear stress, regardless of the other modeling choices.
On the other hand, the effect \rev{of wall modeling} on the mean velocity profile is found to be limited and generally does not change the error pattern in the outer-scaled $\U$ profile typical for a particular combination of SGS model and interpolation scheme.
It is noted, however, that for flows where the thickness of the TBL, $\delta$, is not defined by the geometry (as it is in channel flow), the wall shear stress predictions will also have a significant influence on this quantity.

The effect of grid resolution is studied in Section~\ref{sec:nTests}.
Contrary to what could be expected, grid refinement does not always lead to decreased errors in the QoIs.
The behavior depends significantly on the other modeling choices, and consistent improvement in the accuracy with $n/\delta$ is found only when the LUST scheme is employed, further supporting it as a suitable choice for WMLES.
Based on that, resolution $n/\delta \gtrsim 25$ is recommended.
It is noted that no convergence could be obtained for the \rev{resolved} turbulent kinetic energy, its profile being only slightly affected by the change in the resolution of the grid.
Nevertheless, the observed level of accuracy for the TKE can be considered satisfactory for WMLES.
Further investigation of the behavior of this quantity is necessary.

Addressing the influence of $h$ (Section~\ref{sec:hTests}), it is clear that the optimal value of $h$ is driven by the error pattern in the mean velocity profile.
In particular, $h$ should correspond to a \rev{cell center} where the velocity is predicted accurately, since that is a necessary condition for accurate prediction of~$\lbut$ by the wall model.
Based on the error patterns in $\U$ observed here, it is recommended to avoid associating $h$ with the cell centers of the wall-adjacent and second consecutive off-wall cell.
For the particular case of using the LUST scheme \revCom{with any SGS model but Smagorinsky}, \rev{sampling from any cell center} but that of the wall-adjacent cell is acceptable.

Apart from the accuracy of the sampled velocity values, the performance of the wall model also depends on the accuracy of the underlying law of the wall.
The latter can be adjusted by modifying the parameters of the law ($\kappa$ and $B$ for the Spalding law~(\ref{eq:spalding})).
Using existing reference data, optimal values for the parameters can be found, i.e.~those that lead to perfect agreement between the law and the reference data at the sampling point.
The associated gain in the accuracy of~$\lbut$ is, perhaps, larger than what could be anticipated, see Section~\ref{sec:motivational}.
Considering the model parameters to be freely adjustable coefficients instead of physical constants, it is possible to use them to compensate for any systematic errors in the input velocities, thus eliminating the dependency of the wall model accuracy on $h$, see Sections~\ref{sec:wmParamMotiv} and~\ref{sec:hRemove}.
Unfortunately, it is difficult to know a-priori\rev{, without reference data,} what parameter values should be adopted to that end.
\revCom{Studying the possibility of designing a procedure for dynamic adjustment of the parameters remains as a future work.}

To further investigate the role of grid resolution, anisotropic grids are considered in Section~\ref{sec:gridAnisot}.
The simulations reveal a complicated pattern in the dependency of the QoIs on the anisotropy.
Unfortunately, these patterns not only depend on the other modeling choices, but also vary among the considered QoIs.
Therefore, it is impossible to choose a particular grid resolution bias which would not lead to a deterioration in accuracy for at least some of the QoIs.
Consequently, it is recommended to use isotropic grids.

Summarizing the results from the performed simulation campaigns, best practice guidelines for WMLES were formulated in Section~\ref{sec:guidelines}.
Concisely, the following is advised.
Use the WALE SGS model, the LUST scheme for interpolating convective fluxes, an isotropic grid with resolution~\mbox{$n/\delta \gtrsim 25$}, sample from the second off-wall cell or higher, and, when possible, \rev{adjust} the parameters of the law of the wall to match reference data at the sampling point.
\revCom{
These guidelines are expected to be useful for WMLES by similar nominally second-order accurate finite-volume solvers other than~\of. 
For WMLES by other numerical methods, we recommend that the procedure demonstrated here is applied to derive corresponding best practice guidelines.
}

To see whether the proposed guidelines remain valid for channel flow at other Reynolds numbers, simulations for several $\reyt$ in the range between $2\,000$ and $1\,000\,000$ were performed.
The results are shown in Section~\ref{sec:ReEffect} and the conclusion is that the guidelines are not Re-number dependent.

Application of the proposed WMLES methodology to other, more complicated, wall-bounded flows is a subject for future work.
However, initial experiments on simulating a flat-plate TBL~\cite{timECCOMAS:18} and the flow over a backward-facing step~\cite{mukha:18} strongly indicate that the proposed practices are applicable to these flows as well.
Considering flows separating from a curved surface is of~high~priority.

\section*{Acknowledgments}   
All channel flow simulations were performed on resources provided by the Swedish National Infrastructure for Computing (SNIC) at PDC Centre for High Performance Computing (PDC-HPC).
The study was supported by grant No 621-2012-3721 from the Swedish Research Council.

\appendix
\section{Sensitivity analysis of the algebraic wall models}\label{app:LSA}
\revCom{This section details the local and global sensitivity analyses of the algebraic wall models, in general, and the Spalding law (\ref{eq:spalding}), in particular. 
The quantity of interest is the wall shear stress, sensitivity of which is studied with respect to the variations in the model parameters and sampled velocity.}
Consider the general form of a law of the wall that is written~as, 
\begin{equation}\label{eq:LoTgen}
F(y,\lut,\lu,\fq_{m}) = 0 \,,
\end{equation}
in which, $\fq_m$ \revCom{are the model parameters and} $\fq_m = \{\kappa, B\}$ for the Spalding law (\ref{eq:spalding}).
Perturbing the \rev{parameters}~$\fq_m$, \rev{input} $\lu$ (\tim{the} mean velocity at height $y=h$ from the wall), and the resulting $\lut$ about associated nominal states, \ie,
\begin{equation}\label{eq:lsaDeriv}
\frac{\dd}{\dd \varepsilon} F\left(
   h ,
   \langle {u}_{\tau_0}\rangle + \varepsilon \Delta{\lut} ,
   \langle {u}_0\rangle + \varepsilon \Delta {\lu} , 
   \fq_{m0}+\varepsilon \Delta {\fq_{m}} 
   \right) 
   |_{\varepsilon=0} 
   =0 \,,
\end{equation}
leads to,
\begin{equation}\label{eq:lsaWM}
\frac{\Delta \lut}{\langle u_{\tau_0}\rangle} = X_{\lu} \frac{\Delta \lu}{\langle u_0 \rangle} +\sum_{i=1}^2 X_{q_{m_i}} \frac{\Delta q_{m_i}}{q_{m0_i}} \,.
\end{equation}
Here, $X$ represent the local sensitivity indices and $\Delta$ specifies the variation in each quantity. 
Moreover, the subscript $0$ stands for the nominal state. 
For the Spalding law (\ref{eq:spalding}) the following analytical expressions for local sensitivity indices are derived, 
\begin{eqnarray*}
X_{\lu}&=&\langle u_0 \rangle^+ \left[1+\kappa_0 \mathcal{T}_0  \right] /\mathcal{S}_0 \,, \\
X_\kappa &=& \kappa_0\left[(\langle u_0 \rangle^+-B_0)\mathcal{T}_0 +B_0 e^{-\kappa_0 B_0} {(\kappa_0 \langle u_0 \rangle^+)^{3}}/{3!} \right] /\mathcal{S}_0 \,, \\ 
X_B&=&-\kappa_0 B_0\left[ \mathcal{T}_0 -e^{-\kappa_0 B_0}{(\kappa_0 \langle u_0 \rangle^+)^{3}}/{3!}\right]/\mathcal{S}_0 \,,  
\end{eqnarray*}
in which,
\begin{eqnarray*}
\mathcal{T}_0&=&e^{-\kappa_0 B_0}\left[e^{\kappa_0 \langle u_0 \rangle^+} - \sum_{m=0}^{2} \frac{(\kappa_0 \langle u_0 \rangle^+)^m}{m!}\right] \,,\\
\mathcal{S}_0&=&\left(h^+_0+ \langle u_0 \rangle^+ +\kappa_0 \langle u_0 \rangle^+ \mathcal {T}_0 \right) \,.
 \end{eqnarray*} 
For evaluating these indices \revCom{in \fig~\ref{fig:SAIndices}}, the values of $\langle u_0\rangle$ are taken to be the DNS mean velocity data~\cite{lee-moser:15}, and, $\fq_{m0}=\{\kappa_0,B_0\}=\{0.395,4.8\}$.

The relative importance of the LSA indices, $X$, is only meaningful when the normalized variations in the mean velocity,~\ie~$\Delta \lu/\langle u_0 \rangle$, and the parameters in (\ref{eq:lsaWM}) are of the same order. 
To avoid this shortcoming and also the underlying linearization, global sensitivity analysis (GSA) can be carried out.
Another advantage of the GSA is that the algebraic wall models are analyzed imitating the conditions in the practical WMLES, that is, instantaneous velocity samples are imported to the wall models, see~\sect~\ref{sec:wallModels}. 
To synthesize these inputs, both the fluctuations and mean values of the velocity samples should be taken into account.

To conduct the GSA, at a given target $\reyt$ and any $h$, both $\bu$ and $\fq_m$ are assumed to be random and allowed to simultaneously vary over the prescribed ranges. 
Let $\bu$ be Normally distributed with the mean and standard deviations set respectively equal to the mean and root mean square (rms) fluctuations of the DNS velocity, \ie~$\bu\sim \cN(\langle u_0 \rangle , u_{\rms,0})$. 
This assumption is accurate enough for the purpose of the analysis, although it has been shown that the velocity fluctuations exhibit sub-Gaussian~(i.e~slightly deviating from Gaussian) behavior, see Refs.~\cite{jimenez:98,meneveau:13}.

The wall model parameters are assumed to be uniformly distributed independent random variables.
In particular \revCom{to produce the graphs in \fig~\ref{fig:SAIndices}(b)}, it is assumed that $\kappa\sim \cU[0.35,0.45]$, where the admissible range is chosen based on the values of the \vk~coefficient suggested in the literature, see \eg~\cite{nagib:08,orlu:10,zanoun:03}, and the references therein. 
Further assume $B\sim\cU[4.0,6.0]$. 
Given joint samples of~$\fq_m$ and~$\bu$, the wall friction velocity~$\but$ is iteratively computed from the wall model
\begin{equation}\label{eq:wmGen}
F(y,\but,\bu,\fq_m) = 0 \,.
\end{equation}
The gPCE technique described in \sect~\ref{sec:uq} is employed to evaluate realizations of the wall models (\ref{eq:wmGen}) for different combinations of the velocity samples and parameters. 
In particular, for each uncertain factor,~5 Gaussian quadrature samples are chosen.

\section{On the accuracy of satisfying \eq~(\ref{eq:xMomChan_inScal}) by WMLES}
\label{app:xMomBalance_Converge}
\begin{figure}[!htbp]
\centering
\includegraphics[scale=0.6]{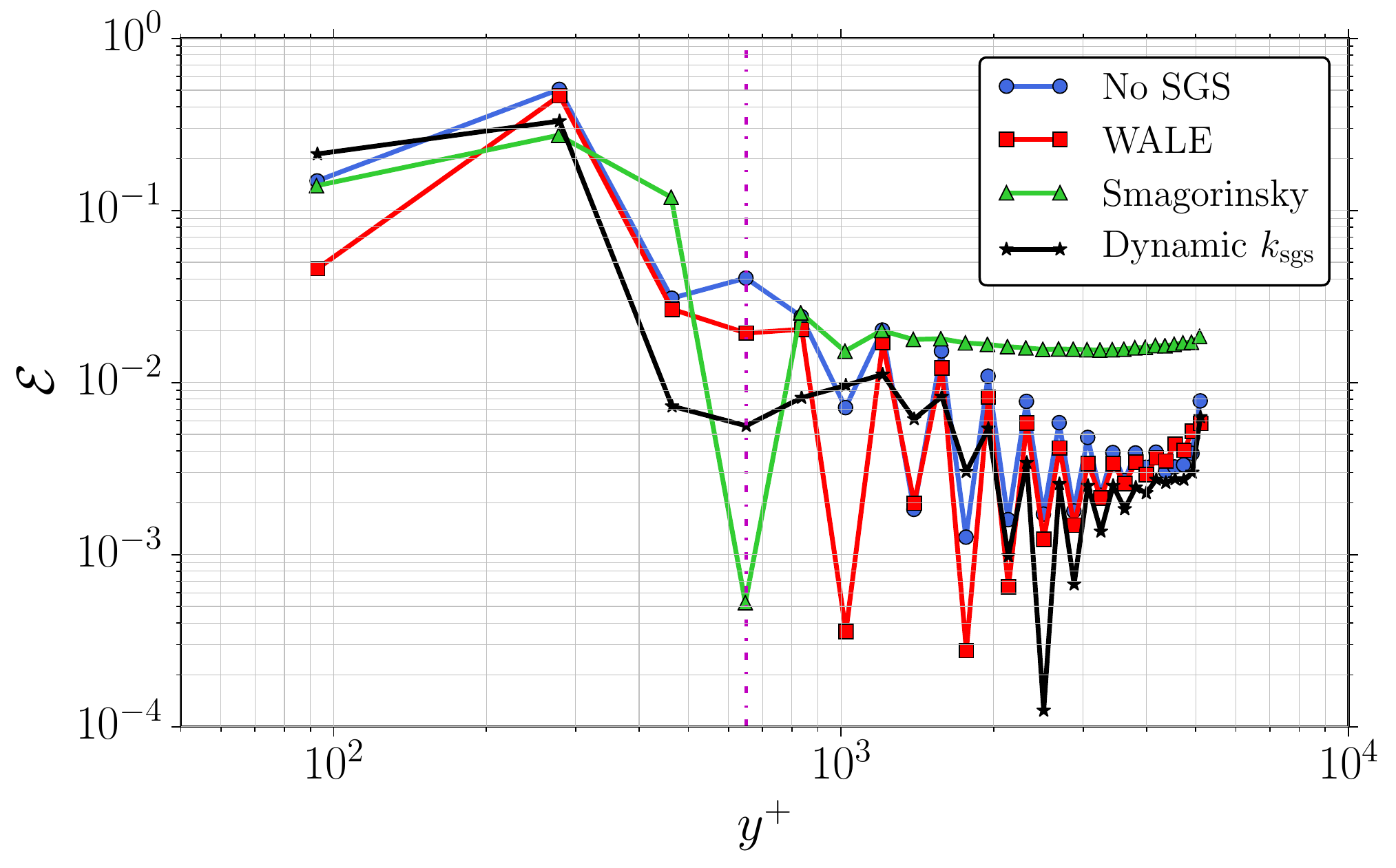}
\caption{\revCom{
Variation of $\mathcal{E}$ defined by (\ref{eq:bigEps}) across the channel for the WMLES of \fig~\ref{fig:uvBalance_lust}. Simulations are performed with the LUST scheme and different SGS models (shown by different colors). 
For details, see the captions of \fig~\ref{fig:uvBalance_lust} and \tab~\ref{tab:WMfootDuTau}.}}
\label{fig:uvBLnce_lust}
\end{figure}

\revCom{
The accuracy to which \eq~(\ref{eq:xMomChan}), or equivalently, \eq~(\ref{eq:xMomChan_inScal}) is satisfied by WMLES of channel flow, is good but depends on the choice of the numerical convective scheme, SGS model, and in any case, on the distance from the wall. 
To quantify such accuracy, the following normalized error measure can be evaluated,
\begin{equation}\label{eq:bigEps}
\mathcal{E}:=\left| 1 - \frac{\frac{\dd \U^+}{\dd y^+} - \buv^+ - \langle B_{xy}\rangle^+ }{(1-\eta)} \right| \,,\quad
0\leq \eta < 1 \,.
\end{equation}
In particular, for the simulations with the LUST scheme represented in \fig~\ref{fig:uvBalance_lust},~$\mathcal{E}$ is plotted across the channel in \fig~\ref{fig:uvBLnce_lust}. 
Similar plots can be obtained for the Linear scheme (not shown here). 
Independent of the SGS model and interpolation scheme, at the first few off-wall cells, $\mathcal{E}$ is the highest compared to the other points across the channel. 
This, however, is not an issue since, \eqs~(\ref{eq:xMomChan}) and (\ref{eq:xMomChan_inScal}) have been employed to support the reasonings associated with the behavior of the QoIs in the outer layer of the TBL, see \eg~\sect~\ref{sec:qoiRelation}.}

\revCom{
For both the Linear and LUST schemes, the use of Smagorinsky model results in similar values of $\mathcal{E}$ in the outer layer. 
At higher~$y^+$, say $y^+ \gtrsim 10^3$, for other employed SGS models there are oscillations in $\mathcal{E}$. 
However, it should be noted that for this range of off-wall distance, $\mathcal{E} \lesssim 0.02 (1-\eta)$ where the magnitude of the normalized total stress, $(1-\eta)$, decreases moving away from the wall. 
}

\section{On the accuracy of the gPCE metamodels}
\label{app:gPCEConvergence}
\revCom{
The gPCE-based metamodels (\ref{eq:pce}) are used in \sects~\ref{sec:wmParamMotiv} and \ref{sec:gridAnisot}.
Here, the convergence of the gPCE which directly affects the accuracy of the predictions made by the metamodels is shortly discussed. 
According to \sect~\ref{sec:uq}, in the non-intrusive use of gPCE the exact $\cR$ is not known over the whole $\BQ$.  Therefore, it is not possible to derive estimates for the error between the exact response and what is predicted by the metamodel (\ref{eq:pce}). 
An alternative to assess the accuracy of the metamodel is to investigate if the order of the polynomials in (\ref{eq:pce}) is high enough. }

\begin{figure}[!htbp]
\centering
    \begin{tabular}{cc}
        \includegraphics[scale=0.33]{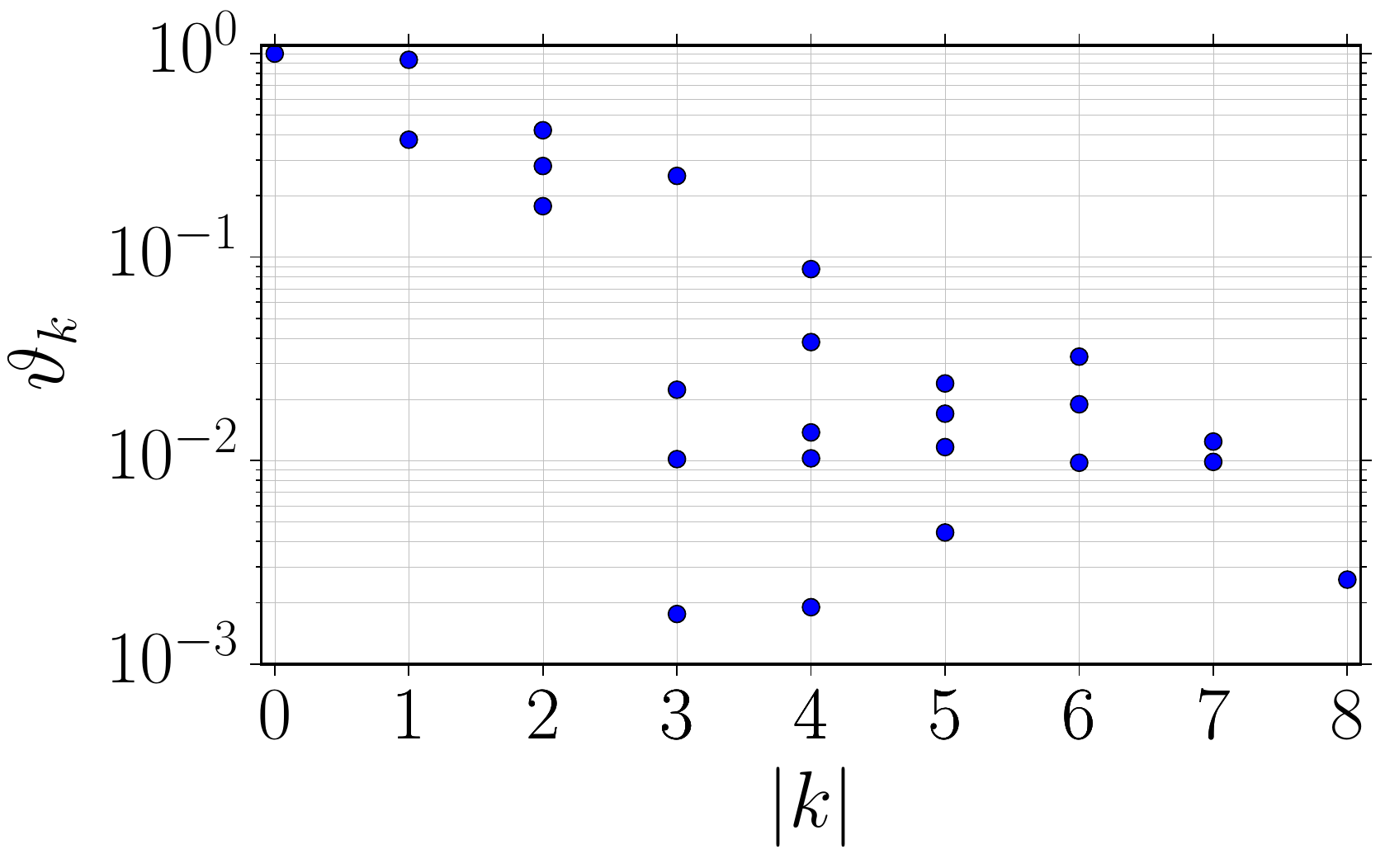} & 
        \includegraphics[scale=0.33]{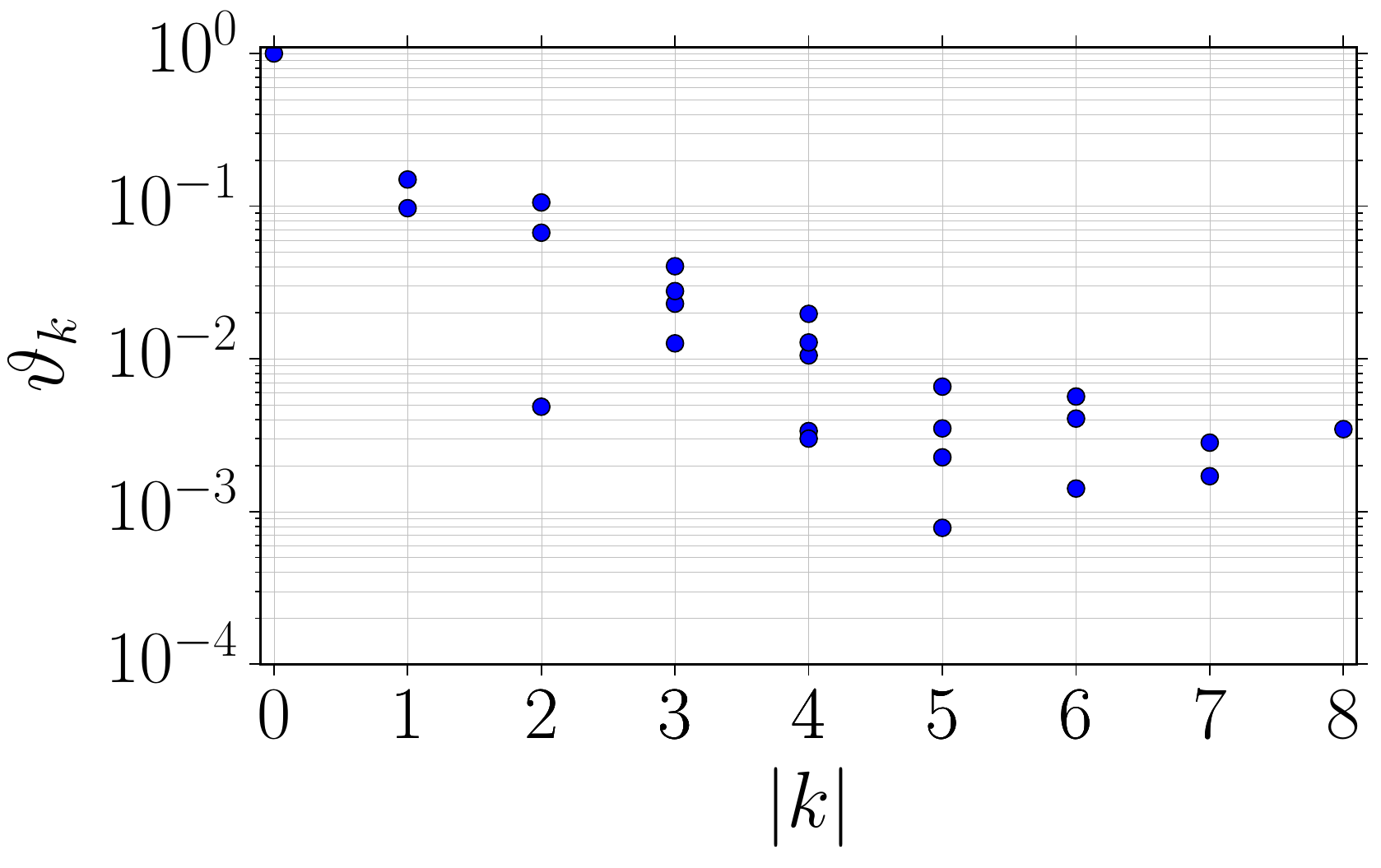} \\
        {\small{(a)}} &    {\small{(b)}} \\        
        \includegraphics[scale=0.33]{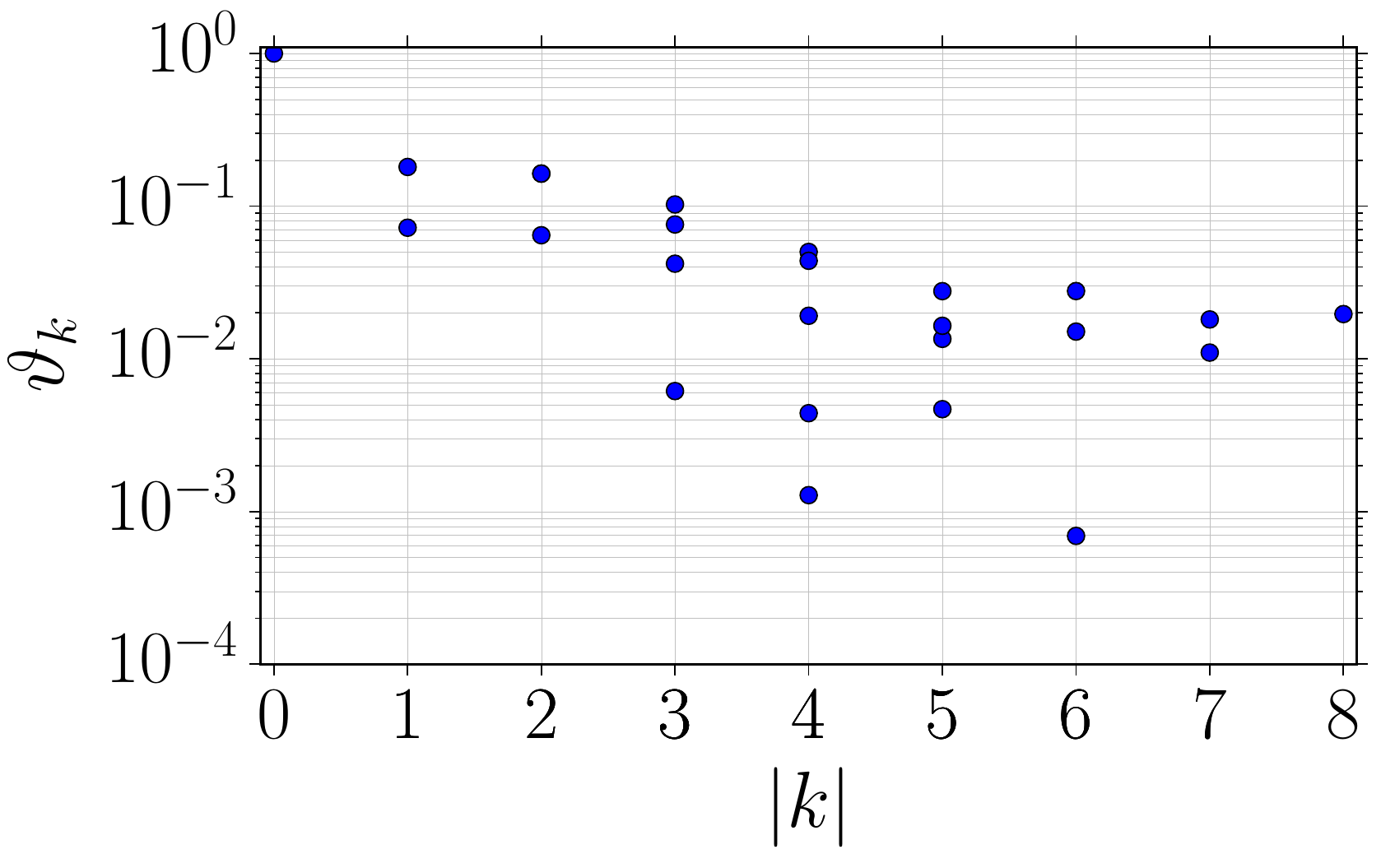} &
        \includegraphics[scale=0.33]{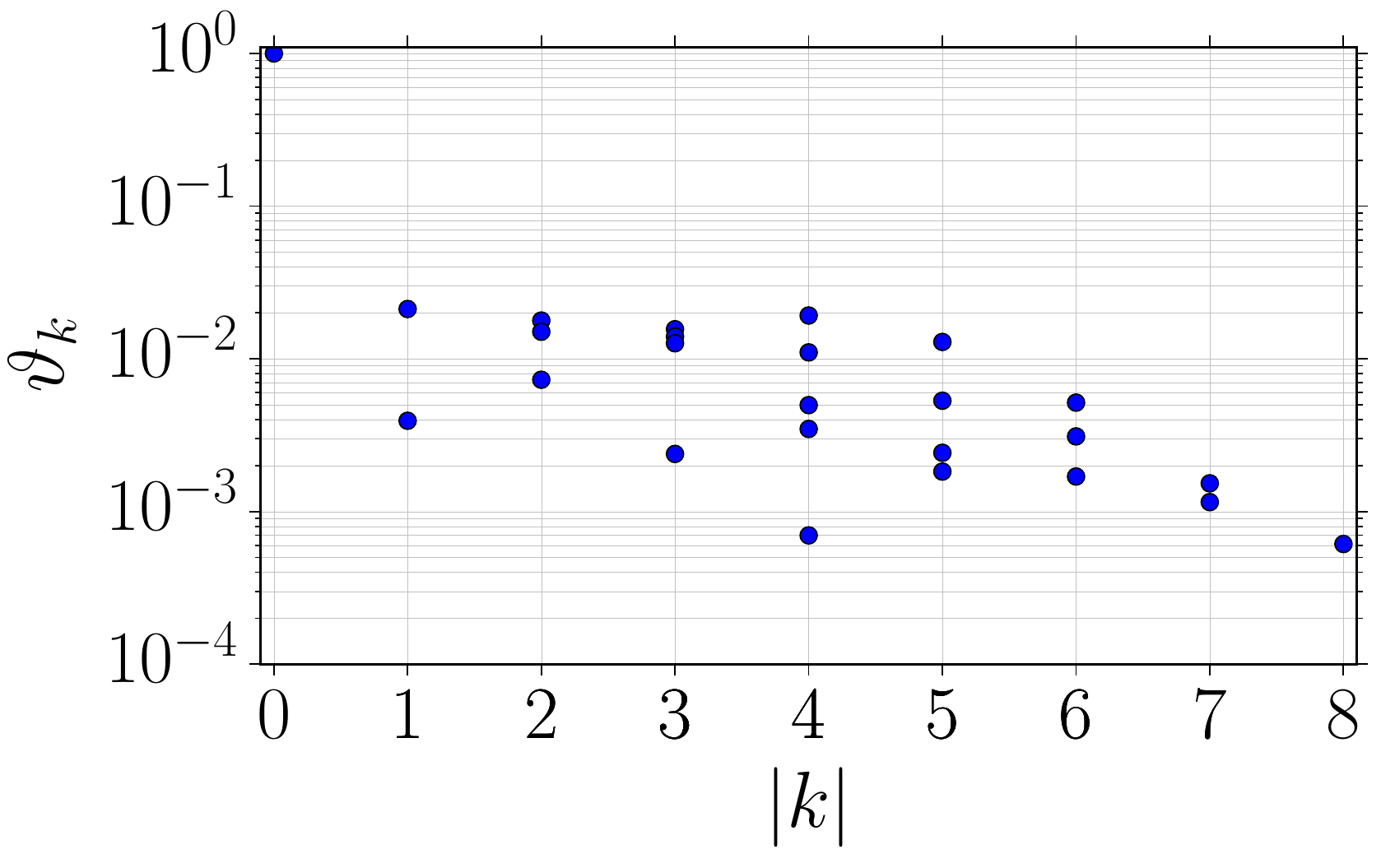} \\                 
        {\small{(c)}} &    {\small{(d)}} \\
    \end{tabular}
    \caption{\revCom{Variation of $\vartheta_\rk:=\| \hat{f}_{\rk} \Psi_{\rk}(\fq)  \|_2 / |\hat{f}_0|$ with $|\rk|=k_1+k_2$. Coefficients $\hat{f}_k$ belong to the metamodel~(\ref{eq:pce}) constructed for $\epsilon[\lbut]$ (a), $\einf[\U]$ (b), $\einf[\buv]$ (c), and $\einf[\bk]$ (d), the isolines of which are plotted in Figure~\ref{fig:duTau_uqGrid}(e), Figure~\ref{fig:isolinesUny25}(d,e,f), respectively.}}\label{fig:pceConv_wmB1}
\end{figure}

\revCom{
To quantitatively examine this, the magnitude of $\vartheta_\rk:=\| \hat{f}_{\rk} \Psi_{\rk}(\fq)  \|_2 / |\hat{f}_0|$ can be evaluated and shown to reduce with the polynomial order. 
This has been done for all the results presented in \figs~\ref{fig:DuTau_kapBIso}, \ref{fig:isoErrors_kapB}, \ref{fig:duTau_uqGrid}, and \ref{fig:isolinesUny25}. 
For instance, for the constructed metamodels for $\epsilon[\lbut]$, $\einf[\U]$, $\einf[\buv]$, and $\einf[\bk]$ illustrated in \figs~\ref{fig:duTau_uqGrid}(e) and~\ref{fig:isolinesUny25}(d,e,f), $\vartheta_\rk$ are plotted in \fig~\ref{fig:pceConv_wmB1} versus $|\rk|=k_1+k_2$ where $k_1,\,k_2=0,\,1,\,\ldots,\,4$. 
It is clearly observed that the relative importance of the terms in the expansion (\ref{eq:pce}) is reduced with increasing the polynomial order. 
In particular for $|\rk|\geq 4$, the value of $\vartheta_\rk$ is less than $0.1$ in the worst case (\ie~for $\epsilon[\lbut]$).
}

\section{A calibrated empirical expression for cross-channel profiles}\label{app:CessProfs}
In the RANS sense, the momentum equation in the streamwise direction for the channel flow reads as, 
\begin{equation}\label{eq:xMomChanRANS}
(1-\eta) =\left(1+\frac{\langle \nu_t\rangle}{\nu} \right) \frac{\dd \lu^+}{\dd y^+} \,,
\end{equation}
where, $\langle \nu_t \rangle$ is the mixing-length-based RANS eddy viscosity.
The inner-scaled mean velocity profile at target~$\reyt$ can be obtained by integrating \eq~(\ref{eq:xMomChanRANS}), 
\begin{equation}\label{eq:cess}
\lu^+(\eta)=\reyt \int_0^\eta \frac{1-s}{1+ \zeta(s) }\dd s \,,
\end{equation}
where, $\zeta = \langle \nu_t \rangle/\nu$.
For this quantity, the following analytical expression suggested by Cess~\cite{cess:58} for turbulent pipe and channel flows can be used, 
\begin{eqnarray*}
\zeta(\eta)=
\frac{1}{2}\left[1+\frac{\kappa^2\reyt^2}{9}\left(1-(\eta-1)^2\right)^2\left(1+2(\eta-1)^2\right)^2\left(1-\exp(\frac{-\eta\reyt}{A^+})\right)^2\right]^{1/2}-\frac{1}{2} \,.
\end{eqnarray*}
The original fixed values suggested for the parameters are $\kappa=0.4$ and $A^+=31.0$ for channel flow, see~\cite{rt:67}.
Using the available DNS datasets for channel flow from different sources \cite{hoyas:08,lee-moser:15} covering $180\leq\reyt\leq 5200$, the following expression for the dependency of $\kappa$ and $A^+$ on $\reyt$ (even for $\reyt>5200$) can be proposed, 
$$
\kappa\; {\rm or}\; A^+=c_0+c_1 \rey_\tau^{-m} \,,
$$
where, $c_0,c_1,m$ are estimated to be respectively equal to $0.4326,\, 4.2321 \times 10^3\,, 1.9228$ for $\kappa$, and $28.1539,\, 2.5345\times 10^6,\, 2.2619$ for $A^+$.
Once the profile of the mean velocity is obtained by numerical integration of~(\ref{eq:cess}), the inner-scaled Reynolds shear stress is computed from $-\uv^+=\zeta\,\dd \lu^+/\dd y^+$, where $y^+=\eta \reyt$.

\bibliographystyle{abbrv}  
\section*{References}  
\bibliography{main_uqWMChan}

\end{document}